\documentclass[12pt,a4paper]{article}

\usepackage[a4paper,margin=1.15in]{geometry}
\usepackage{xltabular}
\usepackage[T1]{fontenc}
\usepackage[utf8]{inputenc}
\usepackage{tocbibind}
\usepackage{url}
\usepackage{graphicx}
\usepackage{float}
\usepackage[font=small,labelfont=bf,labelsep=period]{caption}
\usepackage{makecell}
\usepackage{hyperref}
\usepackage{apacite}
\usepackage{mathtools}
\usepackage{commath}
\usepackage{booktabs}
\usepackage{multirow}
\usepackage{adjustbox}
\usepackage{enumitem}
\usepackage{bbold}
\usepackage{xcolor}
\usepackage{amssymb}
\usepackage{amsmath}
\usepackage{mathtools}
\usepackage{dutchcal}
\usepackage{tabularx}
\usepackage{pdflscape}
\usepackage{subcaption}
\usepackage[bottom]{footmisc}

\usepackage{placeins}
\let\Oldsection\section
\renewcommand{\section}{\FloatBarrier\Oldsection}

\RequirePackage{fix-cm} 
\DeclareMathSizes{12}{12}{10}{10}
\usepackage[doublespacing]{setspace}

\usepackage[ruled, vlined, linesnumbered]{algorithm2e}
\usepackage{etoolbox}
\AtBeginEnvironment{algorithm}{\setstretch{1.1}}
\SetCommentSty{\footnotesize\texttt}

\newboolean{showcomments}
\setboolean{showcomments}{true}
\ifthenelse{\boolean{showcomments}}
{ \newcommand{\mynote}[3]{
    \fbox{\bfseries\sffamily\scriptsize#1}
    {\small$\blacktriangleright$\textsf{\textit{\color{#3}{#2}}}$\blacktriangleleft$}}}
{ \newcommand{\mynote}[3]{}}

\newcommand{\ts}{\textsuperscript}

\title{A general framework to quantify the event importance in multi-event contests\footnote{ We are grateful for helpful comments and feedback from Nora Bearth, Enzo Brox, Hannah Busshoff, Christian Deutscher, Riccardo Di Francesco, Michael Knaus, Alex Krumer, Michael Lechner, Fabian Muny, Gabriel Okasa, and Carina Steckenleiter. The usual disclaimer applies.}
}
\author{Daniel Goller$^{1,2}$, Sandro Heiniger$^2$ \\  $^1$Centre for Research in Economics of Education, University of Bern \\ $^2$Swiss Institute for Empirical Economic Research, University of St.Gallen}

\date{\today}
\begin{document}

\maketitle
\thispagestyle{empty}

\begin{abstract}

We propose a statistical framework for quantifying the importance of single events that do not provide intermediate rewards but offer implicit incentives through the reward structure at the end of a multi-event contest. Applying the framework to primary elections in the US, where earlier elections have greater importance and influence, we show that schedule variations can mitigate the problem of front-loading elections. When applied to European football, we demonstrate the utility and meaningfulness of quantified event importance in relation to the in-match performance of contestants, to improve outcome prediction and to provide an early indication of public interest.

\end{abstract}


\clearpage
\pagenumbering{arabic} 
\newpage
\section{Introduction}
Incentives are an important tool for motivating people to exert effort. In many environments, from the workplace to sporting contests, incentives are put in place to ensure that invested efforts are optimised to achieve predefined goals~\cite{Lazear2000,Laffont2002}. Changing incentives directly translates into altered performance or success probability \cite{Rosen1985,Prendergast1999}. In contests where the reward depends solely on the outcome of a single event, incentives are provided directly through the potential rewards. 
In multi-event contests, the translation of contest rewards into incentives for single events is not directly observable.
Moreover, single events may be unequally important for obtaining the final rewards, e.g. performance in an interview is rated higher than the previous assessment centre test, or the results of previous events lead to momentum for subsequent events. \par

As this transmission of multi-event contest rewards into incentives reflects the (personal) expected rewards, incentives vary not only between events but also between participants. Understanding these disparate and possibly asymmetric incentives in multi-event contests is essential as it could lead participants to strategically allocate their efforts \cite{preston2003cheating}. Asymmetric individual incentives may have spillover effects on the outcome probabilities of all other participants, which in turn could lead to potentially unbalanced or unfair contests.\par

In this work, we propose a general statistical framework to quantify the importance, or implicit incentive, of single events in complex multi-event contests for each participant individually -- the \textit{event importance} (EI). The EI measures the capability of an event to change a participant's (expected) reward for a contest. Our approach is based on two steps. First, we calculate the probability distribution for each contestant to reach certain end-of-contest rewards based on an outcome model determining the outcome probabilities for every single event. Second, the importance of a single event is determined through the changes in the end-of-contest reward probabilities with respect to the possible outcomes of this particular event. If the probabilities to reach the final rewards are changed substantially, the importance of this event for the participant is high and our methodology returns a high event importance measure. \par

The proposed framework generalises previous approaches \cite{Schilling1994,Scarf2008,buraimo2022armchair} in that it is suitable for any contest design and any number of participants and is not specific to any particular contest environment. Moreover, it allows for participant-specific reward structures and both the reward structure and the schedule can change dynamically during the contest. Crucial for practical usage is that the proposed statistical procedure can also be used in situations in which the importance of the single event potentially plays a role in the determination of the event's outcome -- this can be accounted for by calculating the specific event importance in an iterative procedure. \par

Our statistical framework can be applied in a variety of practical use cases, like competing pharmaceutical companies developing a drug for the same medical indication, presidential elections which are held in a series of local elections, a job or promotion contest among applicants or workers, or sports tournaments.\par

To showcase our methodology, we use our framework in two applications.
In the first, we apply the framework to the US presidential primaries to examine the problem of front-loading: Earlier elections are known to have a greater impact on the outcome of the nomination process, which is why several states are pushing for earlier election dates.
We analyse the Democrats' electoral schedule for the 2020 primaries and compare it with two alternative hypothetical schedules, one sorted by the number of delegates and one randomised.
In this analysis, we find that the positioning of a state's election in the schedule substantially affects its impact on the outcome of the nomination -- indicated by higher event importance measures. 
A comparison of the different schedules shows that the problem of front-loading can be mitigated by arranging the schedule according to the number of delegates in the states and completely eliminated by a random scheduling.\par

In a second application of our framework to the double round-robin tournament structure in football leagues we provide explicit measures of the EI that express implicit incentives for teams. 
In this setting, the relevance of a particular match with respect to the team's expected rewards varies substantially, even though every match is actually awarded the same number of points. Hence, the importance of a match varies over the season and between teams. 
This leads to pairings between teams with potentially very different incentives that change the presumed probabilities of winning.\par

We demonstrate the meaningfulness of the derived values by analysing their relationship to various observable characteristics of the matches. 
The integration of the EI information into a prediction model improves the accuracy of match outcome forecasts. We show that bookmakers do not fully take into account the team-specific importance of events in their prediction model. Furthermore, a positive interrelation can be drawn between the estimated importance of the match and the public's interest in certain matches in the form of larger stadium attendance and social media engagement for more important matches. For the in-match activity of the players and the outcomes of the match, we observe a comprehensive pattern suggesting that teams approach more important matches with a more aggressive, direct, and successful playing style. \par

Both the event importance values and replication code for the applications are publicly available on
\href{https://doi.org/10.7910/DVN/F3QA9N}{Harvard Dataverse}
\cite{Dataverse}.\footnote{For the US presidential primaries, the EI estimates are published for all states and territories, for the three different scheduling scenarios. The published EI estimates for the European football leagues cover the seasons 2009/10 through 2018/19 and all seven leagues, i.e., the German 'Bundesliga 1', the Dutch 'Eredivisie', the Spanish 'La Liga', the French 'Ligue 1', the English 'Premier League', the Portuguese 'Primeira Liga', and the Italian 'Serie A'.}
The rest of the paper is structured as follows. Section 2 discusses related literature. Section 3 explains the proposed statistical method. Section 4 applies the framework to the front-loading in US primaries and the application to double round-robin tournaments appears in Section 5. Section 6 concludes.

\section{Related literature}



This work mainly refers to two types of literature, (a) the importance of specific events and attempts to quantify them and (b) the literature on incentives in contests. The literature investigating the role of (explicit) incentives on performance generally finds that higher incentives increase performance \cite{Ehrenberg1990,Prendergast1999,Lazear2000}. However, it is important to distinguish between effort- and skill-based tasks; in the latter, strong incentives can lead to performance decrements, a phenomenon known as choking-under-pressure \cite{Ariely2009,Cohen-Zada2017,Goller2021}. While most studies, and all of the studies mentioned above, consider individual incentives, the role of team-based incentives on performance may be different \cite{Alchian1972}.
\par

The importance of specific events in multi-event contests can be found in several fields of literature. The order of action literature finds that the outcome of the contest is influenced by the order of the events which has been shown in musical contests \cite{Ginsburgh2003}, song contests \cite{de2005save}, or judicial decisions \cite{danziger2011extraneous}. 
More specifically, several works focus on potential advantages in the first event in (usually sequential) contests, like in R\&D \cite{harris1987racing}, sports \cite{apesteguia2010psychological}, or elections \cite{klumpp2006primaries}. \par

Research documents the differential importance of sequential elections in US presidential primaries. Surprising wins in early states led to momentum effects in the 2004 primaries \cite{knight2010momentum}. In their work they find an unbalanced influence on the final result for voters in early and late elections. \citeA{klumpp2006primaries} model this first-winner advantage for primaries -- known as the New Hampshire effect -- giving an explanation for the more intense campaigning in early elections. With more influence in the nomination process in the early events front-loading, i.\,e.\ states moving their elections to earlier dates, is well documented \cite{mayer2003front}. \citeA{ridout2008importance} find more attention of the candidates to the states the earlier their elections are. Moreover, they find that scheduling is more important than the delegates' count.\par 


The first approaches to determining an event's importance use rather simplistic measures \cite{Jennett1984} and basic contest structures \cite{Schilling1994}. Most influential was \citeA{Schilling1994}'s general idea of defining and calculating event importance in terms of how the probability to reach a final goal varies for different event outcomes. Recently, more sophisticated approaches have emerged, for instance, \citeA{Scarf2008} simulating probabilities of final contest rewards conditional on two different event outcomes. \citeA{Lahvicka2015} and \citeA{buraimo2022armchair} build on the ideas of \citeA{Jennett1984} and \citeA{Schilling1994} but simulate final standings in the ranking in a Monte Carlo simulation to estimate the importance of single events.
A different objective is followed in \citeA{Geenens2014}: The importance of a match with regard to its influence on the final contest outcome is investigated. This has an interesting use case to investigate contests from the neutral spectator's perspective but is conceptually very different from the importance of a match for the specific contestant. \par

The drawback of all the discussed approaches is that they are specific to a certain type of contest that is prevalent in sports, i.\,e.\ a fixed number of event outcomes and one specific reward (e.g., winner-takes-it-all contests). This does not encompass more complex or dynamic contest designs and reward structures, which are common in society. The approach we propose in the following section provides the flexibility to handle a variety of practical applications with a variety of contest and reward structures.

\section{The event importance}\label{sec:The general framework}
\subsection{Introduction to the general framework}\label{sec:Interpretation of the event importance}
The event importance measures the difference between the contest reward probability distributions induced by the possible outcomes of a single event. If the probabilities for the final rewards vary substantially with the differential outcomes of the examined event, its impact on the tournament reward is large and a high event importance measure is attributed. \par

To quantify the importance of a particular event we hence require the probability distribution of the contest rewards conditional on each possible outcome of the investigated event. 
To determine the probability distributions, our framework sets the outcome of the examined event accordingly and solves the remainder of the contest by successive evaluation of the outcome model. Subsequent to the examined event, whose outcome is set by the framework, all entities (outcomes, covariates, schedule) are subject to the probabilistic outcome model. The successive application of the outcome model until the end of the contest generates the probability distribution for the contest reward conditional on the initial outcome.\par

There are six valuable attributes of our approach: 
First, by evaluating the reward probability from the perspective of every contestant individually, the event importance measure is specific to every participant and not the event itself. Second, we do not impose narrow restrictions on the contest setup. Since the contest reward probability distribution is conditional on each possible event outcome, we only need to assume a finite number of contestants, a finite number of possible outcomes for an event, and a finite schedule for the contest. Moreover, the tournament rewards have to be measurable based on the outcomes of all single events. \par

Third, the reward structure can be any function of all single event outcomes or a final contest ranking if such exists, e.\,g.\ close-by ranks can be grouped together if valued equally. Fourth, the framework is not restricted to a specific schedule. As the probability distributions are calculated through a successive evaluation of all events in the contest, the reward probabilities encompass all the essential features of the schedule as well, e.\,g.\ early elimination of participants in the contest or differences in information sets induced by events held in parallel or sequentially. Fifth, the framework is not tied to a specific distance metric to calculate the difference in the reward probability distributions. Sixth, if the outcome model is not known, it can be estimated on training data using any well-suited statistical method. \par

In the following, the details of how the described framework can be implemented to determine the event importance values in a general case are outlined.

\subsection{Technical implementation}\label{sec:Technical implementation}
\subsubsection{Notation}\label{sec:Notation}
This section defines the notation used to describe the general framework. Table~\ref{tab:notation} summarises the notation as a quick reference. 
Upper case letters denote random variables, lower case letters denote their realisations or other exogenous variables, and calligraphic letters are sets. Multi-character names, such as \textit{EI} or function names, are evident choices.\par
\begin{table}[ht]
	\centering
	\caption{Notation}
	\begin{tabular}{ll}
		\toprule
		$t \in \mathcal{T}$ & Time $t$ in contest schedule $\mathcal{T}$ \\
		$\mathcal{T}_{t^-}=\bigcup_{t'\leq t}t'; \ \mathcal{T}_{t^+}=\bigcup_{t'>t}t'$ & Sub-schedule up to ($^-$) or after ($^+$) time $t$ \\
        $e_{t,i} \in e_t$ & An event $e_{t,i}$ held at time $t$\\	
		$k \in \mathcal{K}_e \subseteq \mathcal{K}$ & Contestant $k$ participates in event $e$  \\ 
		$x_{e}=\bigcup_{k \in \mathcal{K}_e} x_{e,k}$ & Information on all contestants in event $e$\\
		$\mathbcal{Y}_e$ & Set of all possible outcomes for event $e$\\
		$y_e \in \mathbcal{Y}_e$ & Realised outcome of event $e$ \\
		$\text{out}(x_e)= \bigcup_{\mathbcal{Y}_e}P[Y_e=y_e|X_e=x_e]$ & Probabilistic outcome model \\
		$\left\{\mathcal{T_{t^+}},x_{t^+}\right\}=\text{gen}\left(\mathcal{T}_{t^-},x_{t^-},y_{t^-}\right)$ & Outcome-dependent schedule and covariates\\
		$r_k=\text{rew}_k\left(\bigcup_\mathcal{T} y_e\right)$ & Reward for contestant $k$ after contest $\mathcal{T}$\\
        $\text{EI}_{e,k}=\text{dist}\left(\bigcup_{\mathbcal{Y}_{e}}r_{k,y_e},\text{out}(x_e)\right)$ & Event importance for contestant $k$ in event $e$  \\
		\bottomrule
	\end{tabular}
	\caption*{\footnotesize Note: For better readability, we omit the subscripts of $e_{t,i}$ if a notation is not tied to a particular event but holds for any arbitrary event $e$.}
	\label{tab:notation}
\end{table}
The contest is held along a finite schedule $\mathcal{T}$. Because of the implicit chronological ordering of $\mathcal{T}$ we can define the notation $\mathcal{T}_{t^-}=\bigcup_{t'\leq t}\,t'$ and $\mathcal{T}_{t^+}=\bigcup_{t'>t}\,t'$, denoting the sub-schedule up to and after time $t$. Multiple events $e_{t,i}$ can be held simultaneously at time $t$, in this case $e_t= \bigcup_i e_{t,i}$. A finite set of contestants $\mathcal{K}$ participate in the contest of which a subset $\mathcal{K}_e\subseteq\mathcal{K}$ participate in event $e$. For each event, a set of covariates $x_e=\bigcup_{k\in\mathcal{K}_e} x_{e,k}$ and its outcome $y_{e}=\bigcup_{k\in\mathcal{K}_e} y_{e,k}$ is observed.\footnote{In this context, observed refers to recording an outcome within the framework and not to the actual outcome of an event that may have been observed.}
The outcome $Y_e$ of event $e$ is a random variable that follows a conditional probabilistic outcome model $\text{out}(x_e)= \bigcup_{\mathbcal{Y}_e}P[Y_e=y_e|X_e=x_e]$ which, in case out() is not known, is approximated by $\widehat{\text{out}}()$. In the description of the general framework we assume w.l.o.g. that $\text{out}_e()$ is known and uniform. The cases of an approximated outcome model $\widehat{\text{out}_e}()$ or event specific outcome models $\text{out}_e()$ can both be handled in the general framework.\par

The chronological feature of the events further allows to define the sets of covariates $x_{t^-} = \bigcup_{e,t'\leq t} x_{e_{t'}}$ and $x_{t^+} = \bigcup_{e,t'>t} x_{e_{t'}}$ which combine all information on events and participants taking place either up to or after time $t$. The analogous operation on the outcomes defines $y_{t^-}$ and $y_{t^+}$. In settings where parts of the covariates $x$ and/or the schedule $\mathcal{T}$ depend on past outcomes, they are generated at run time based on the previous outcomes by  $\left\{\mathcal{T}_{t^+},x_{t^+}\right\}=\text{gen}\left(\mathcal{T}_{t^-},x_{t^-},y_{t^-}\right)$. After the full contest $\mathcal{T}$, the probability distribution of the final rewards is determined according to the valuation function $r_{k,y_e}=\text{rew}_k\left(\bigcup_\mathcal{T} y_e\right)$ which can be individually specific for every contestant $k$. The event importance $\text{EI}_{e,k}=\text{dist}\left(\bigcup_{\mathbcal{Y}_{e}}r_{k,y_e},\text{out}(x_e)\right)$ for contestant $k$ in event $e$ is the difference between the multiple probability distributions of the final rewards measured by any distance measure dist(). The distance function can incorporate the outcome probabilities $\text{out}_e(x_e)$ of the starting event as weights.

\subsubsection{Algorithm}\label{sec:Algorithm}
Algorithm~\ref{algo: general event importance} describes the computation of the event importance value for a competitor $k$ in event $e_{t,j}$. Readers that are less familiar with the pseudo-code notation can consult the literal translation of the algorithm in Appendix~\ref{app:Details to algorithm}.\par
\vspace{6 pt}
\begin{algorithm}
    \caption{Event Importance\label{algo: general event importance}}
    \DontPrintSemicolon
    \SetKwComment{Comment}{$\triangleright$}{}
    \KwData{$e_{t,i},k,t,\mathcal{T}, x, \mathbcal{Y}_{e_{t,i}}$}
    \KwResult{Event importance for competitor $k$ in event $e_{t,j}$}
    \Begin{
        \ForAll($\; \triangleright$ \small We denote all conditional dependence on $\mathbcal{y}_{e_{t,i}}$ by *){$\mathbcal{y}_{e_{t,i}}$ in $\mathbcal{Y}_{e_{t,i}}$\label{line:loop possible states}}{
            $y_{e_{t,i}}^* \gets \mathbcal{y}_{e_{t,i}}$\;
            \ForAll{$e_{t,j}$ in $e_t$ with $j\neq i$\label{line:loop same time events}}{
                $y_{e_{t,j}}^* \gets \text{out}\left(x_{e_{t,j}}\right)$\;
            }
            $\left\{\mathcal{T}_{t^+}^*,x_{t^+}^*\right\} \gets \text{gen}\left(\mathcal{T}_{t^-},x_{t^-},y_{t^-}^*\right)$\;
            \ForAll{$t'$ in $\mathcal{T}_{t^+}^*$\label{line:future events}}{
                $y_{e_{t'}}^* \gets \text{out}(x_{e_{t'}}^*)$\;
                $\left\{\mathcal{T}_{t'^+}^*,x_{t'^+}^*\right\} \gets \text{gen}\left(\mathcal{T}_{t'^-}^*,x_{t'^-}^*,y_{t'^-}^*\right)$\;
            }
            $r_{k,\mathbcal{y}_{e_{t,i}}} \gets \text{rew}_k\left(\bigcup_\mathcal{T} y_e^*\right)$ \label{line:final ranking reward}\;
        }
        $\text{EI}_{e_{t,i},k} \gets \text{dist}\left(\bigcup_{\mathbcal{Y}_{e_{t,i}}}r_{k,\mathbcal{y}_{e_{t,i}}},\text{out}(x_{e_{t,i}})\right)$\;
      \Return{$\text{EI}_{e_{t,i},k}$}
    }
\end{algorithm}
\vspace{6 pt}

\subsubsection{Approximation of the probability distributions}\label{sec:Approximation of the probability distributions}
By the subsequent evaluation of the outcome model, the probability distribution of the rewards at the end of the season can be determined, independent of the contest design. However, a large amount of sequential events opens an extremely large number of possible outcome paths which causes numerical problems if their probability would be evaluated exactly. For an outcome model that depends on past outcomes, the outcome paths can additionally become very complex. For this reason, it is often appropriate to perform a Monte Carlo simulation to approximate the probability distribution of the final rewards. Each run simulates one path for the remaining contest according to the outcome model and the generated covariates/schedule information at run time. With an adequate number of \textit{$N_\text{MC}$} Monte Carlo runs, the estimated probability distribution and hence the event importance values become sufficiently close to the exact values. \par

In a simulation that estimates the EI values for all events consecutively, a chronologically backward iteration over the events allows for the reuse of already evolved paths as they can be merged with respective previous outcome to longer paths and thus reduce the computational complexity. In this case, \textit{$N_\text{MC}$} is an upper bound for the number of actually performed runs in each step and at the same time, a lower bound for the number of runs the event importance estimate is premised on.

\subsubsection{Iterative approximation of event importance}\label{sec:Iterative approximation of event importance values}
In many applications, the event importance is an integral part of the outcome model, e.\,g.\ when the importance can be interpreted as an incentive for the contestants to provide effort that in turn influences the outcome of the event. Independent of whether the outcome model is known or not, it encompasses the event importance values which are not available beforehand. \par 

To determine the unknown event importance values, Algorithm~\ref{algo: general event importance} is at first executed with an approximate outcome model that does not feature the event importance in the variable set. This returns an initial approximation of the desired EI values. A subsequent iterative application of Algorithm~\ref{algo: general event importance} with the full covariate set including the preliminary EI variables updates the event importance estimates accounting for their own impact through the outcome model. This iterative procedure can be continued until a predefined stopping criterion is reached. The application in Section~\ref{sec:Application} is an example of an outcome model which includes the event importance in the covariates. Algorithm~\ref{algo:specific event importance} in Appendix~\ref{app:Specific framework algorithm} illustrates how the iterative procedure is implemented in the context of the application.

\subsubsection{Distance functions}\label{sec:Distancefunctions}
 
To measure the difference between the probability distributions on the contest rewards, an appropriate distance function needs to be chosen. In simple applications with only binary event outcomes and a binary reward scheme, the difference between the contest-winning probabilities conditional on the event outcome is a straightforward choice as the distance function. \par

More complex cases which feature either multiple possible event outcomes or multiple rewards require a statistical distance function. For most settings, the Jensen-Shannon divergence (JSD) is an appropriate distance function to cope with multiple discrete probability distributions~\cite{Lin1991}. It is a common distance measure~\cite{Nielsen2021} with desirable properties, such as allowing for a weighting of the probability distributions as well as being bounded and symmetrical. The JSD measures the difference in the Shannon entropy between the probability distributions which implies that it does not have an intuitive linear interpretation. If such an interpretation is of relevance, other candidates such as the total variation distance can be applied.\par

\section{Application: Front-loading in US primaries} \label{sec:Pr_application}
\subsection{Introduction}
Presidential primary elections in the United States are held by the Democratic and the Republican party to determine the presidential election nominees. Both parties follow a similar procedure where each state, every permanently inhabited US territory, and party members living abroad\footnote{For the ease of readability, all entities are henceforth labelled as states.} are attributed a certain number of votes (pledged delegates). In addition to the pledged delegates, selected party officials' have additional votes (unpledged delegates) that are not tied to states' election results. 
In order to be nominated for the presidential elections, the candidates in the primaries must receive the majority of the delegate votes.\par

Each state holds its election or caucus on an individually chosen date, and the election results determine how its delegates vote. Several states can vote on the same day, e.g. on "Super Tuesday" about one third of all delegate votes are determined. Due to the partially sequential nature of the primaries, it may happen that later elections become irrelevant to the outcome of the nomination if a candidate has already received more than half of the total delegate votes. Moreover, the first elections are of greater importance as they reveal voters' preferences and influence later elections through their results. These two features of the electoral process lead to a long-known and unresolved problem of front-loading \cite{mayer2003front,ridout2008importance}.\par

From a state's perspective, an early election date can increase its influence in the primaries. If all states are considering moving their elections to earlier dates, a solution must be found to regulate the timing of state elections that takes into account the different importance of the dates. Currently, additional delegates are granted for late election dates, but these do not provide sufficient incentive for states to resolve front-loading.
In the following analysis, we compare the US Democrats' 2020 election schedule with two proposed election schedules, namely randomising the election dates and arranging the states according to their delegate count.
The aim of the analysis is to find out whether this leads to a more balanced distribution of the importance of elections for the individual states that is less driven by the timing of the elections.

\subsection{Setup}\label{sec:Pr_setup}

We utilise the actual schedule and reward structure of the 2020 democratic party presidential primary elections. The ordering of the elections and the number of delegates rewarded by the election are displayed in Table \ref{tab:Primaries summary} in Appendix \ref{App:frontloading}. For ease of exposition, we simplify the model of the election process by discarding unpledged delegates, implementing only winner-takes-it-all elections, and engaging only two candidates $i\in \{0,1\}$. 
\par
The reward function of the contest is given by winning the primaries, i.\,e.\ obtaining the majority of the delegates' votes. We model the state's election as a representative voter facing a binary choice model with random utility. The utility function \eqref{eq:Score} is composed of four components: a) the fixed reputation $\{\eta_0,\eta_1\}=\{0.5,0\}$ of the candidates, b) the match between state preferences $\rho_s\sim \mathcal{N}(0,1)$ and the candidates' positions $\{\rho_1,\rho_2\}=\{-1,1\}$, c) spillover effects $\zeta_{i,s}$ of previous elections. d) a standard Type I extreme value error term $\epsilon_{i,s}$. Under the assumption, that the representative voter always chooses the candidate with maximum utility, the setting describes a conditional logit model~\cite{mcfadden_conditional_1974} with outcome probability $\pi_{i,s}$ that candidate $i$ wins the election in state $s$ as described by Equation \eqref{eq:Probability}.\par 
Spillover effects from the results of early states on future elections occur if the share of obtained delegates' votes in prior states differs from the expected share solely based on the candidates' reputation. The spillover effect as defined in \eqref{eq:Spillover} is an example of dynamic covariates in the outcome model that have to be re-evaluated when new election results are determined.
\begin{align}
    \psi_{i,s}&=\eta_i+\zeta_{i,s}-\frac{\lvert\rho_i-\rho_s\rvert^2}{2} + \epsilon_{i,s} \label{eq:Score} \\
    \pi_{i,s}&=\frac{\exp(\psi_{i,s})}{\exp(\psi_{i,s})+\exp(\psi_{1-i,s})} \label{eq:Probability} \\
    \zeta_{i,s}&=\frac{\sum_{s'<s} y_{i,s'}}{\sum_{s'<s} (y_{i,s'}+y_{1-i,s'})}-\frac{\exp(\eta_{i})}{\exp(\eta_i)+\exp(\eta_{1-i})} \label{eq:Spillover}
\end{align}
\par
Based on the model for the election process given by equations \eqref{eq:Score}-\eqref{eq:Spillover}, the probability for each candidate to win the primaries conditional on the outcome of a single state's result is determined. Because the number of states and territories is too large to allow an exact numerical calculation, the winning probabilities are approximated by a Monte Carlo simulation using 5'000 simulation runs. As suggested in Section \ref{sec:Distancefunctions}, we choose the difference in the winning probabilities as the distance function for this contest with binary reward structure, i.\,e.\ to be nominated as a presidential candidate or not. This distance function leads to symmetric EI estimates for both candidates. To eliminate the dependency on a particular set of states' preferences, we simulate 1'000 random draws of state preferences.\par

The three schedules we evaluate are defined as follows: The \textit{regular} schedule is according to the actual election dates\footnote{Because of the Covid-19 pandemic, several states have postponed their election such that the actual dates can differ from the initially planned schedule.} and the allocated number of delegates by state. In the \textit{random} schedule, we randomly permute the ordering of the states keeping the framework of the schedule fixed. The \textit{rank increase} scheme ranks the states increasing by their number of delegate votes and applies the ordering to the actual schedule framework.

\subsection{Results}\label{sec:Pr_results}
\begin{figure}[ht]
        \centering
    \begin{subfigure}{.48\textwidth}
      \centering
      \includegraphics[width=\textwidth]{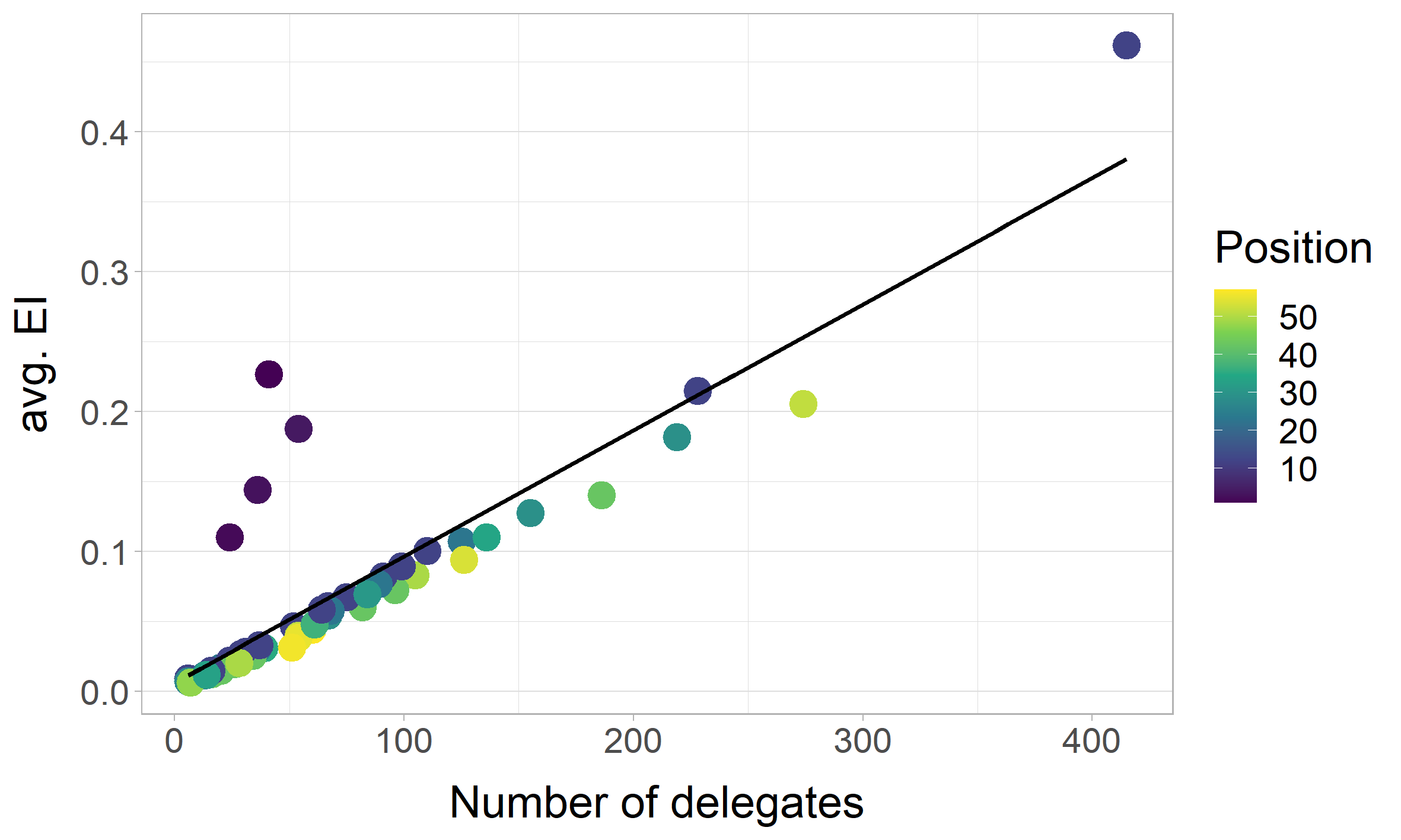}
      \caption{Regular}
      \label{fig:front_loading_scatter_regular}
    \end{subfigure}%
    \hfill
    \begin{subfigure}{.48\textwidth}
      \centering
      \includegraphics[width=\textwidth]{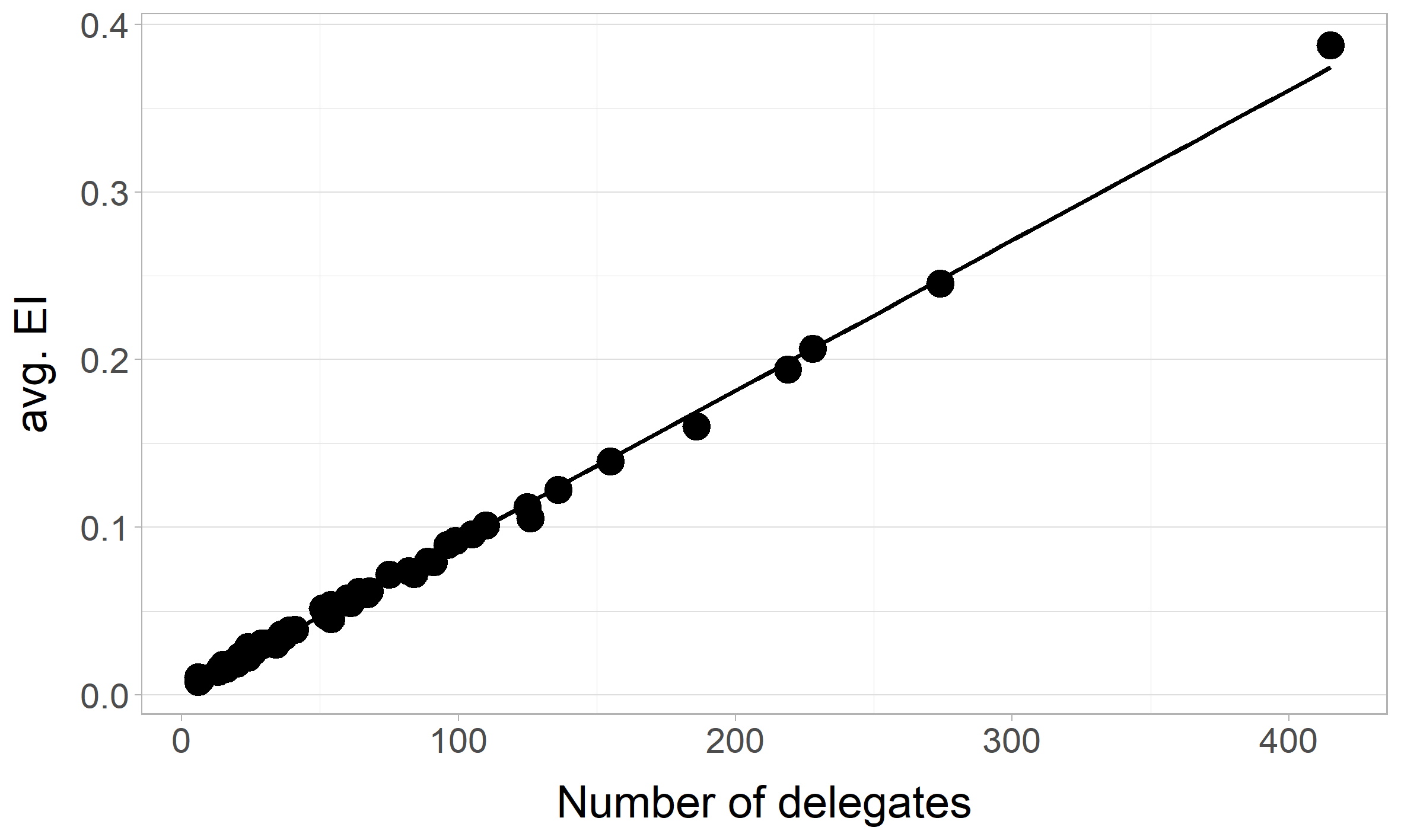}
      \caption{Random}
      \label{fig:front_loading_scatter_random}
    \end{subfigure}
  
\vspace{8pt}  
    
    \begin{subfigure}{.48\textwidth}
      \centering
      \includegraphics[width=\textwidth]{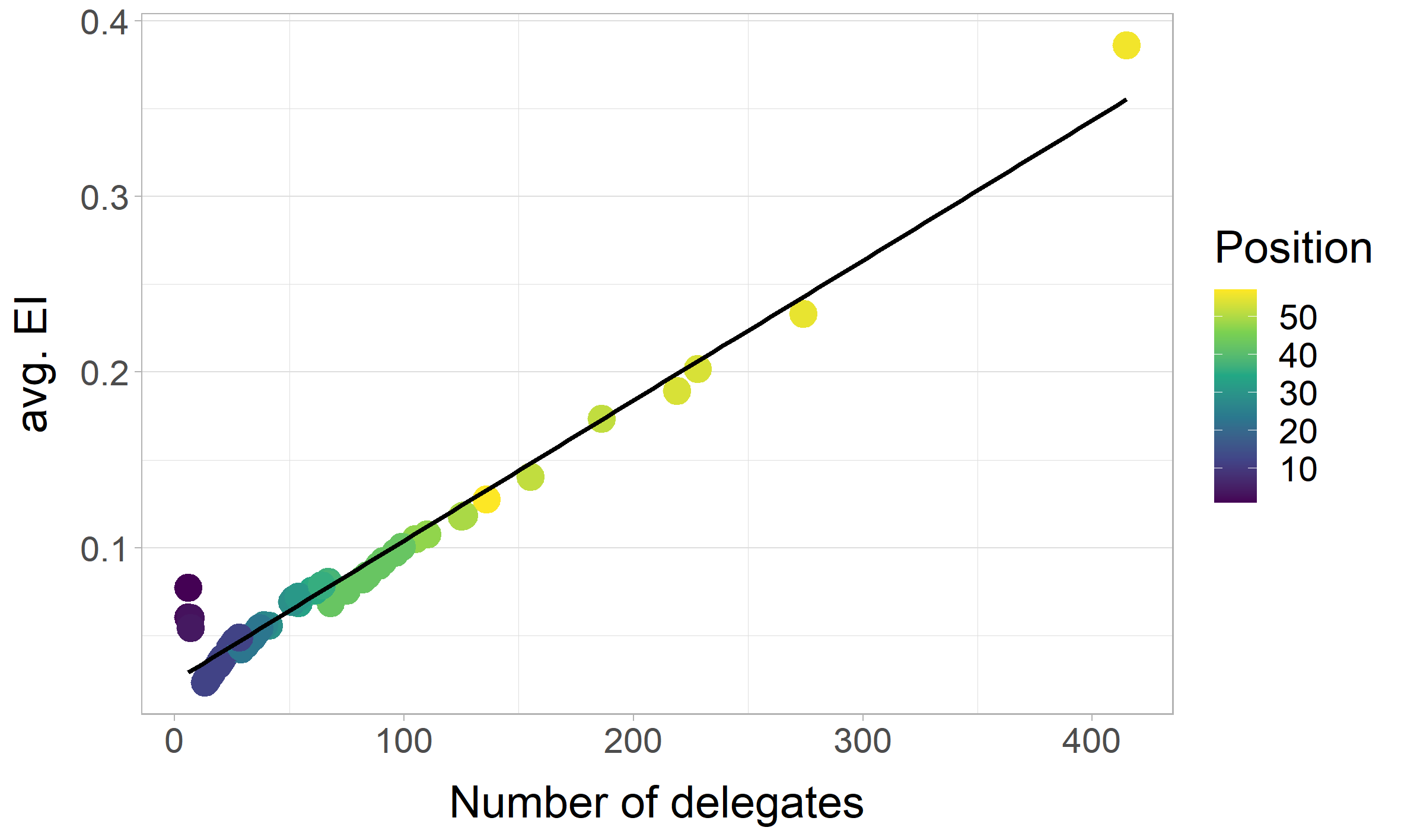}
      \caption{Rank increase}
      \label{fig:front_loading_scatter_rank_increase}
    \end{subfigure}%
    \caption{Average event importance estimates over 1'000 states' (preferences) samples with linear fit.}
    \label{fig:front_loading_scatter}
\end{figure}

Figure \ref{fig:front_loading_scatter} shows the average EI estimates over all samples for the three schedule types. The regular schedule of the 2020 democratic party primaries (\ref{fig:front_loading_scatter_regular}) displays the increased importance of the early elections, as the respective states have a higher average EI estimate than the number of delegates allocated to them would suggest. The randomised schedule (\ref{fig:front_loading_scatter_random}) reveals a linear relationship between the ability of a state to change the outcome of the primaries and its number of delegates. Since all states will eventually benefit from an early position in the schedule, the positive spillover effects are spread across all states and territories and balance each other out. \par

Ordering the states by their number of delegates (\ref{fig:front_loading_scatter_rank_increase}) cannot entirely eliminate the first-winner effect but substantially alleviates it. The increased importance of states due to the early election date can be compensated by a smaller number of delegates, which is an option already considered in the allocation of delegates. Because an ex-ante randomisation requires many repetitions to balance the positional effects, the ordered schedule seems to be a good compromise between practicability and fairness.\par

\begin{figure}[ht]
    \centering
    \includegraphics[width=0.67\textwidth]{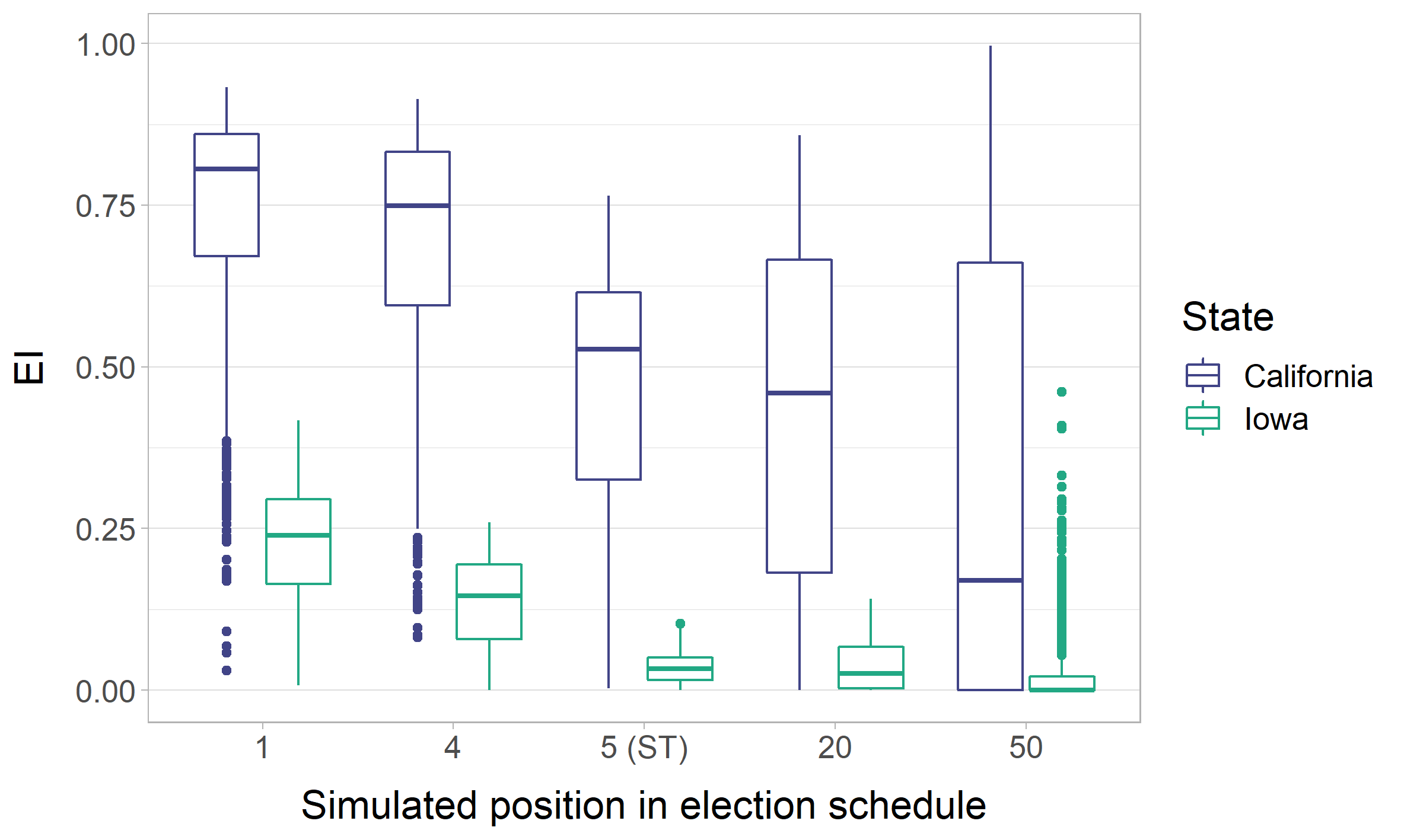}
    \caption{Estimated event importance values for Iowa and California at different hypothetical positions in the election schedule for 400 random samples of states' preferences.}
    \label{fig:state_boxplot_positions}
\end{figure}

To illustrate the importance of both the first-winner effect and the state's size we show in Figure \ref{fig:state_boxplot_positions} the estimated EI values for two states, Iowa (41 delegates) and California (415 delegates), at different hypothetical positions in the election schedule. For the very early positions (1 and 4), both states are of considerable importance for the final nomination of the presidential candidates. From position 5, the "Super Tuesday" on which numerous states hold their elections, the capability of the elections to influence the final nomination decreases considerably. \par

The importance of elections in small states thrives on the fact that there are spillover effects through the first-winner effect. Because of the substantial amount of delegates, California remains considerably important for the nomination result in late stages of the schedule, while the low number of Iowa's delegates become irrelevant in many realisations. For Iowa in particular, if the election would be held in the middle (20) or at the end (50) of the schedule, the importance of the election would be determined only by the possibility that the state's delegates could act as tiebreakers in the nomination if the election race is close.

\section{Application: European football leagues} \label{sec:Application}
\subsection{Introduction}
As with many analyses of contests, sports data provide a suitable and well-structured framework for applications because it features accurate observational data \cite{kahn2000sports,bar2020ask}. We apply the EI framework to the seven major European football leagues. Those contests have a non-trivial schedule of multiple events, held sequentially or in parallel, between pairings of the participants and a non-linear reward structure that can vary individually or change throughout the season -- all of which can be handled naturally with the proposed framework. With, for example, postponed games leading to changes in the schedule, or supplementary rewards achieved by national cup tournaments changing the reward structure individually, this application is a good showcase to demonstrate the flexibility of the framework. \par


Quantifying EI in this context is interesting for several reasons. Contest designers should avoid match-ups that pit contestants with unequal incentive levels against each other. Such matches are potentially more susceptible to bribery and a lower engagement of certain participants could give an unfair (dis)advantage to participants not even involved in the event itself. Other valuable use cases of the EI in football tournaments include (a) selecting intense or interesting matches for prime-time broadcast, (b) improving the prediction of winning probabilities, and (c) detecting or avoiding unfair match schedules.\footnote{E.\,g.\ pairings with unequal EI levels may result in unfair schedules for involved or non-involved teams.}

\subsection{Setup}\label{sec:Setup}
\subsubsection{Data}\label{sec:Data}
We analyse data from the 2006/07 through 2018/19 seasons of seven major European football leagues, namely the German 'Bundesliga 1', the Dutch 'Eredivisie', the Spanish 'La Liga', the French 'Ligue 1', the English 'Premier League', the Portuguese 'Primeira Liga', and the Italian 'Serie A'. These leagues were the major leagues in Europe in terms of sporting and financial success throughout the studied period. All leagues are designed as double round-robin tournaments, which means that each team plays each other twice - once at each home venue. The rewards are distributed after the season. With the seven analysed leagues, we cover a variety of different reward structures. A detailed description of the tournament design, league format, and reward structure of the considered European football leagues can be found in Appendix~\ref{app:Appendix_European_football leagues}.\par
For each individual match, we record a long list of characteristics: describing the match setting, such as the time or day of the week, characterising the participating teams, as their success in recent matches, whether they play in international competitions and metrics of the squad players, e.\,g.\ age, height, estimated market value, and preferred foot. The full set of all 133 covariates is listed in Appendix~\ref{app: List of covariates}.

\subsubsection{Specific application framework}\label{sec:Sp_Ap_Fw}

In this section, we describe how we implement the general framework from Section~\ref{sec:The general framework} and elaborate on all generic functions outlined in Algorithm~\ref{algo: general event importance}. Algorithm~\ref{algo:specific event importance} in Appendix~\ref{app:Specific framework algorithm} presents the pseudo-code of the specific framework tailored to this application.
At the end of a football season, rewards are allocated to the teams based on their final rank. The areas in the ranking which denote the championship title, qualification for international competitions, and relegation are stated by strict thresholds. We use those boundaries to group all the ranks between two thresholds as a single reward.\footnote{Financial rewards, i.\,e.\ money from broadcasting rights, are determined by the final league table. However, this gradation is less relevant for the team, the coaches, and the players. Becoming champions, qualifying for next year's European Cup or not being relegated to a lower league is what we assume is more in the focus of the involved entities. Strict thresholds implicitly assume that it's not particularly relevant for teams to finish 10\ts{th} or 11\ts{th}, but that the potential implications around crucial positions in the ranking -- which guarantee participation in next year's Champions League, for example -- are considerably greater. For a discussion on the financial dimensions consult \citeA{GK2020}. Although the framework allows for a weighting scheme, for the sake of simplicity, we consider all thresholds to be equally relevant to each team.} More detailed information on the reward structures per league and season appear in Appendix~\ref{app:End-of-season rewards}.
Individual updating patterns of the reward scheme, e.\,g.\ because the UEFA Europa League place allocated to the national cup winner is transferred and included in the league's season rewards, are explained in Appendix~\ref{app:Indiv-End-of-season rewards}.\par

In Section~\ref{sec:The general framework}, we have outlined how the general framework can be employed for applications with unknown outcome functions and those incorporating the EI itself. Outcomes of football matches do not follow a deterministic rule and can only approximately be described by a probabilistic model. We follow the approach of \citeA{Goller2018}, using an ordered choice model with three outcome probabilities estimated by the Ordered Forest \cite{Lechner2019}, hereafter abbreviated as \textit{ORF}.\footnote{The general framework is not restricted to this specific method and the choice of the underlying outcome model is of second-order (see Appendix~\ref{app:Alternative outcome model}).} To restrict the number of covariates in the ORF model we perform a LASSO-based model selection step. Starting from the second iteration, this set of covariates is extended with the previously estimated EI values (as outlined in Section \ref{sec:Iterative approximation of event importance values}). In addition, we also simulate the exact score of the match, drawn from two independent Poisson distributions, as the goal difference often serves as a tie-breaker in determining the final ranking.\par 

The choice of the Jensen-Shannon divergence as the distance function, specified in equation \eqref{eq:JSD}, follows the argumentation in Section \ref{sec:Distancefunctions} for settings with multiple event outcomes and rewards. We use a scaling factor of $\ln(3)^{-1}$ to constrain the EI to the [0,1] interval and weight the probability distributions $P_i$ by the match outcome probabilities $\{\pi_H, \pi_D, \pi_A\}$ to account for the likelihood of the three outcome scenarios.
\begin{eqnarray}\label{eq:JSD}
    \text{JSD}_{\pi_H,\pi_D,\pi_A}(P_H,P_D,P_A))=\frac{1}{\ln(3)}\left(\mathbf{H}\left(\sum_{i\in \{H,D,A\}} \pi_i P_i\right) - \sum_{i\in \{H,D,A\}} \pi_i \mathbf{H}(P_i) \right) \\
    \textit{where}\ \mathbf{H}(P)=-\sum_{j=1}^mP(x_j)\ln(x_j) \nonumber
\end{eqnarray} 

\subsection{Results}

\subsubsection{Distribution of the estimated values}
\begin{figure}[ht]
    \centering
\includegraphics[width=0.67\textwidth]{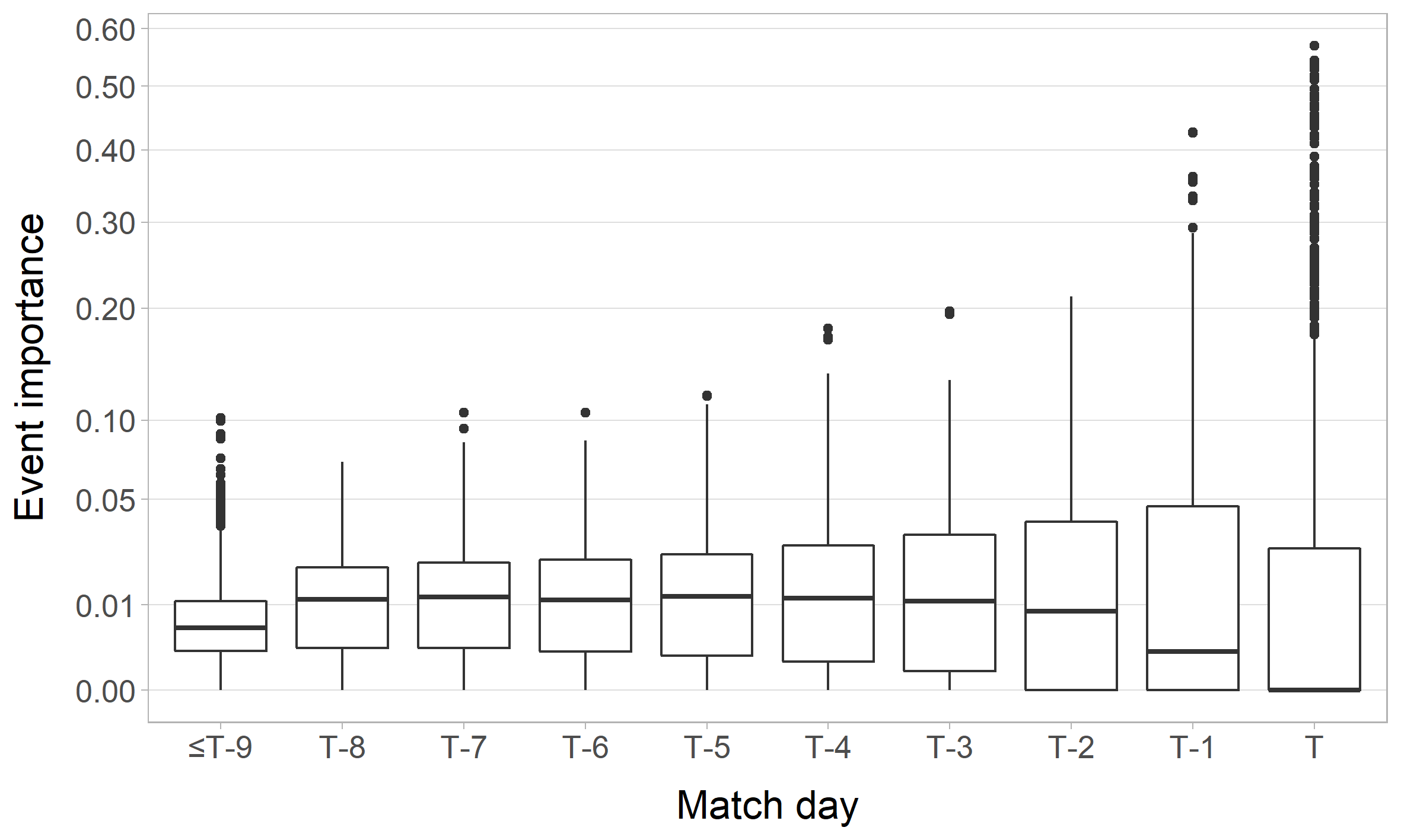}
    \caption{Home and away team's event importance estimates grouped by match day enumerated in relation to the last match day, all seasons, and all leagues. Square-root transformation to y-axis applied}
    \label{fig:back_boxplot_10}
\end{figure}
Figure~\ref{fig:back_boxplot_10} shows the distribution of estimated EI values by match days. For the majority of the season, the estimated EI values are concentrated around a value of about 0.01. In other words, most matches are similarly (un)important for the first parts of the season. Deviations can be observed in pairings between teams that are expected to be close in the final end-of-season standings, as in these matches a positive result implies a negative result for the opponent. This behaviour changes towards the end of the season, with non-relevant matches and frequent outliers of particularly important matches. The uncertainty about the outcome of the rest of the season is reduced with fewer unknown future results, and the results of individual matches can become more decisive for the end-of-season rewards. This results in more pronounced values of the EI towards the end of the season. \par
As an illustrative example, we show the estimated EI values for the last match day in the 2017/18 German Bundesliga 1 season in Appendix~\ref{app:Illustrative example}. 

\subsubsection{Predicting match outcomes}\label{sec:Predicted match outcomes}

To shed light on whether the quantified EI has an impact on outcome prediction, we compare the estimates of a 'baseline' ordered forest model that does not use the EI information with a 'richer' ORF model that includes the estimated EI of both, the home and away team, as additional input.\par

We fit both ORF models on half of the data and predict the outcome probabilities with the two models on the other half. Based on the outcome probabilities we construct the expected points (ExpP) measure by awarding points to the outcomes according to modern football rules -- 3 points for a win, 1 for a draw, and 0 for a loss. This procedure is repeated with swapped training and prediction samples.\par
\begin{figure}[ht]
    \centering
    \includegraphics[width=0.67\textwidth]{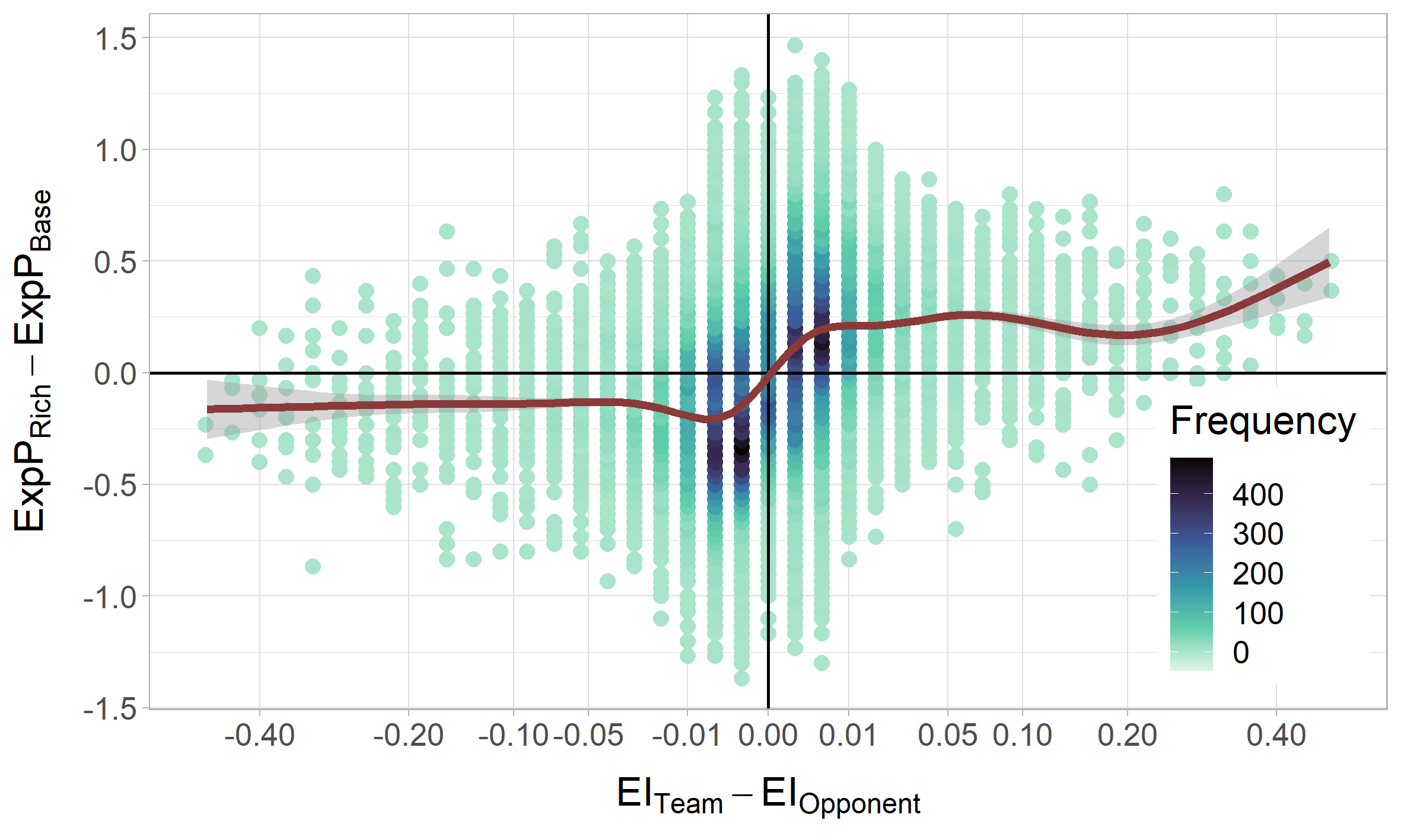}
    \caption{The difference in expected points (ExpP) between the model including EI variables (Rich) and the baseline model (Base) by the difference in event importance (EI) between the team and its opponent. Values are rounded to the nearest grid point. Frequency indicates the number of points on a grid point. The red line denotes a GAM with a 95\% confidence interval. Expected points are averaged over 100 estimates with different sample splits. Square-root transformation to x-axis applied.}
    \label{fig:gg_MI_diff_vs_diff_base_orf_gam}
\end{figure}
Figure~\ref{fig:gg_MI_diff_vs_diff_base_orf_gam} displays the difference in expected points between the rich and the baseline model by the difference in the EI values between the two competing teams. The generalised additive model (GAM) fit on the data confirms that teams with a higher absolute difference in EI are attributed a higher absolute prediction of ExpP with the richer model verifying that the inclusion of the EI variable is relevant for outcome prediction. The variable importance measures of the EI variables in the rich model are shown in Appendix~\ref{app:Variable importance} and provide evidence for the notable contribution of the EI in the outcome model. 

\subsubsection{Prediction power improvement}\label{sec:Prediction power improvement}

In Section~\ref{sec:Predicted match outcomes} we have shown, that the estimated EI values are picked up by an enriched outcome model. This raises the question of whether using EI values in an outcome model improves predictive performance.\par

We compare seven different prediction models to margin-free betting odds of the online betting platform B365.\footnote{The betting odds serve as a benchmark and are collected from the website \url{www.football-data.co.uk}. To ensure comparability with the model predictions, we linearly scale the odds to remove the bookmaker's margin.} The baseline ORF model as described in Section~\ref{sec:Predicted match outcomes} (\textit{ORF}), the richer model including the EI values (\textit{ORF+EI}), and additionally including the difference of EI estimates (\textit{ORF+EI+diff}), an ORF model with a binary importance measure\footnote{A binary variable indicating if the match is still relevant for attaining a better/worse reward or if the match cannot change the reward anymore. Several empirical works use such indicators \cite{Fornwagner2019,Feddersen2021}} (\textit{ORF+naive}), an ORF model (\textit{ORF+add3}) that adds three covariates,\footnote{To investigate if a potential improvement is just induced mechanically by the larger set of covariates. We include travel distance, days since the last match of the home team \& days to the next match of the away team.} an ordered logit model with the baseline variables (\textit{Logit}), and an ordered logit including the EI estimates (\textit{Logit+EI}).\par 
To evaluate the out-of-sample prediction accuracy we randomly split the data into two samples.\footnote{The split is performed on the full-season level to not give the proposed models an unfair advantage over the betting odds.} On one-half of the seasons, the models are fitted, on the other half, the prediction accuracy of the models is measured by the log-likelihood and the Brier score (results for the Brier score can be found in Appendix~\ref{app:alternative measure}). In each repetition, we index the accuracy measures by the results of the benchmark betting-odds model to balance any particular characteristics of the chosen sample.\par
\begin{figure}[ht]
    \centering
    \includegraphics[width=0.67\textwidth]{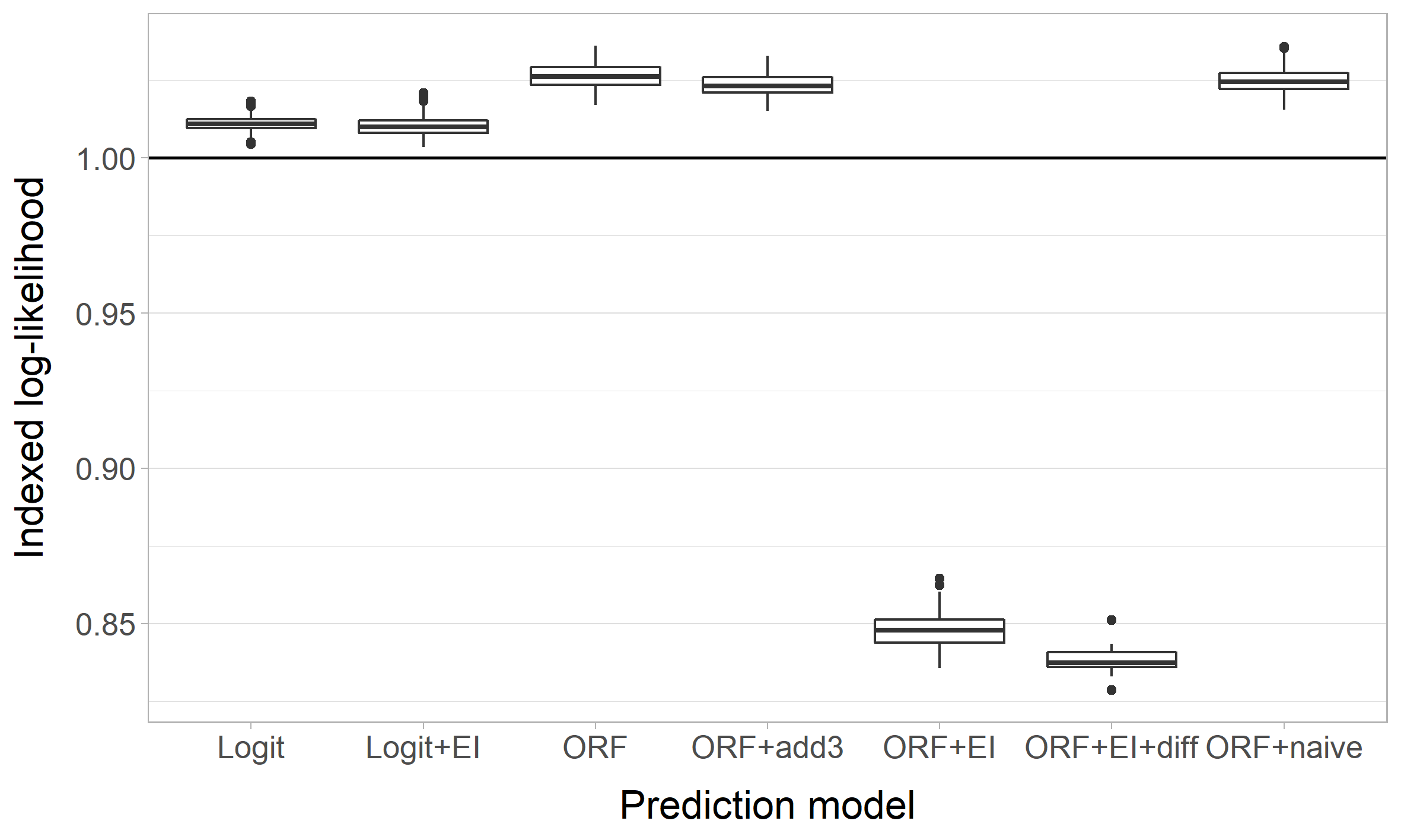}
    \caption{Out-of-sample prediction accuracy of different models over 1000 repetitions. Values are quantified in log-Likelihood and indexed in each repetition by the performance of betting odds.}
    \label{fig:logl_boxplot}
\end{figure}

Figure~\ref{fig:logl_boxplot} shows the out-of-sample prediction accuracy. For the logit model, the addition of the EI results in only a slight improvement, which is probably due to the linearity constraint. Including EI values in the ORF model substantially increases the predictive power, as the additional information contained in the EI variables can be fully utilised, resulting in better performance than the margin-free betting odds. Recording the event importance in a binary variable does not improve the accuracy of the prediction. The model with three added covariates indicates that the increase in prediction power is not just induced by the larger set of covariates.\par
\begin{figure}[ht]
    \centering
    \includegraphics[width=0.67\textwidth]{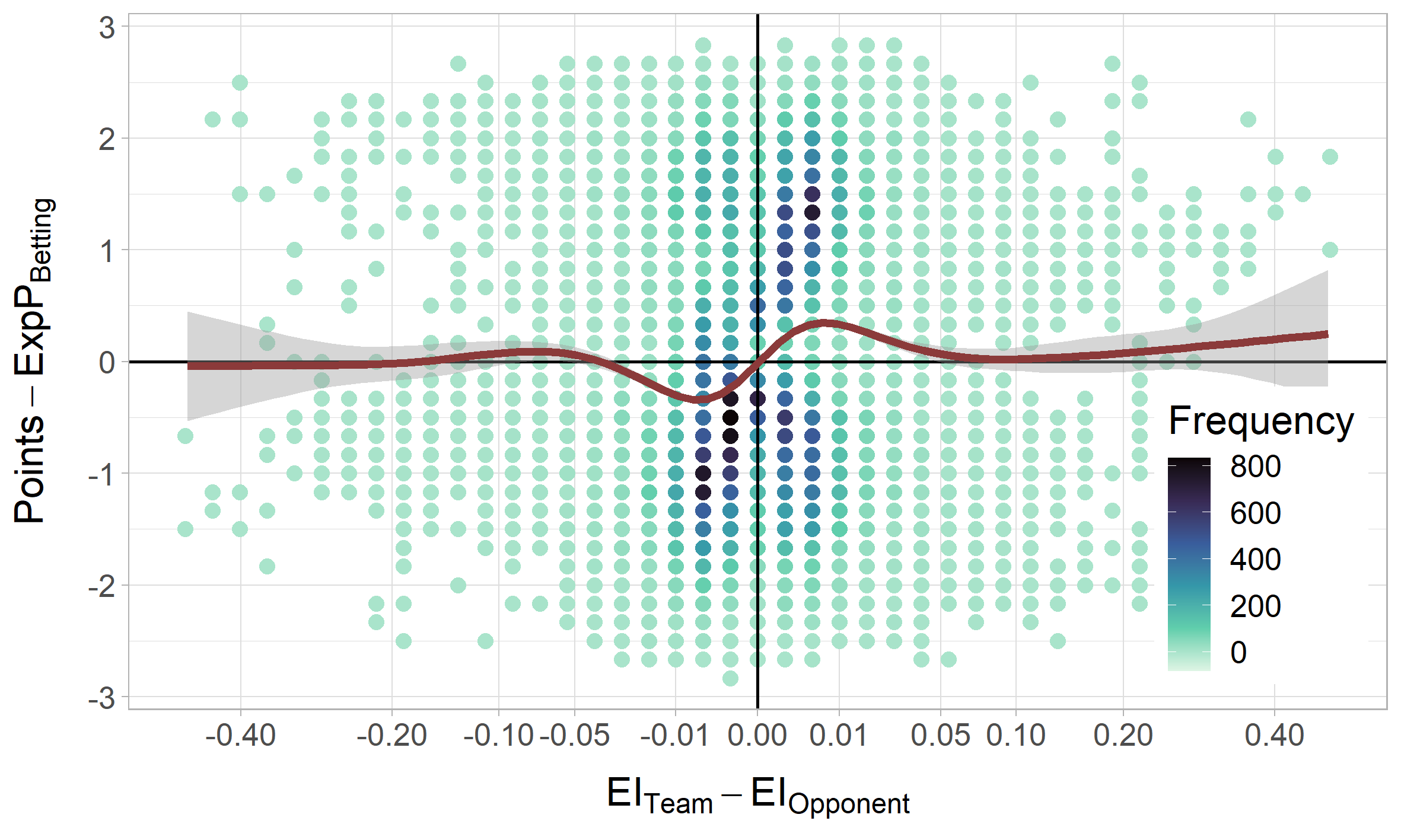}
    \caption{The difference in realised and expected points according to betting odds by the difference in event importance between the team and its opponent. Values are rounded to the nearest grid point. Frequency indicates the number of points on a grid point. The red line denotes GAM with a 95\% confidence interval. Square-root transformation to x-axis applied.}
    \label{fig:gg_EI_diff_gam}
\end{figure}
To break down the improvement by the EI information on the betting odds, we present in Figure~\ref{fig:gg_EI_diff_gam} the difference between the achieved points and the expected points according to the betting odds in relation to the difference in the EI values between the teams and their opponents. A GAM fit on all data points indicates, that in particular across matches where the differences are small, the EI can partly explain the mismatch in the betting odds. For larger EI differences, the EI does not provide additional information to the bookmaker's model. We deduce that the betting odds already cover the unequal incentives when they are particularly pronounced but do not fully account for more subtle disparities in the importance of a match to the competing teams. This is generally in line with and extends the results of \citeA{Feddersen2021}, which show that bookmakers are aware of the impact of different incentives on the outcome of matches on the final match days. We provide additional evidence on the complimentary informational content of the EI measure to the betting odds in Appendix~\ref{app:Improvement on betting odds}.

\subsubsection{Team performance}\label{sec:team_performance}
Besides the usefulness of the EI measure in predictions, we investigate whether the differences in incentives for teams is reflected in-match statistics that record a team's on-field behaviour and performance. For this, we investigate in-match statistics (data source: Opta\footnote{\url{https://www.statsperform.com/opta/}}) for the 2010/11 through 2018/19 German Bundesliga 1 seasons with regards to our EI estimates. The team performance data is collected individually for both teams and pooled for the home and away teams. Outcome variables are totals per match, except for 'Duel win' and 'Tackles win' which are shares. For ease of interpretation, the EI estimates for the home and away teams are each divided into three groups - 'zero' (EI = 0), 'low', and 'high' (above 0.035) EI.\footnote{5\% of the EI values are zero. We therefore choose the \textit{high EI} threshold at the 95 \% quantile of the EI values to obtain balanced groups.}\par
\begin{figure}[ht]
    \begin{subfigure}{.48\textwidth}
      \centering
      \includegraphics[width=\textwidth]{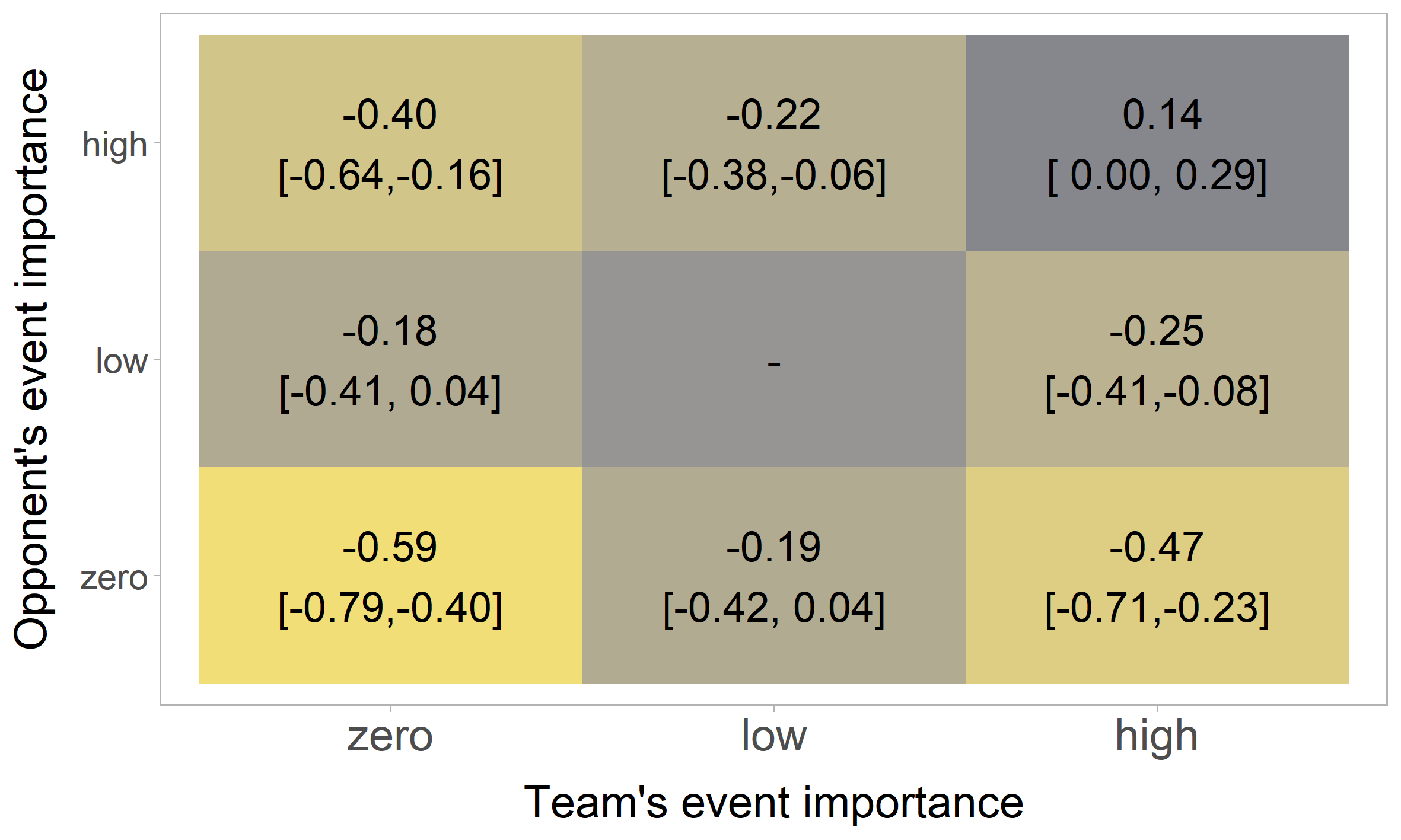}
      \caption{Duels}
      \label{fig:gg_lm_est_duel_pmin}
    \end{subfigure}%
    \hfill
    \begin{subfigure}{.48\textwidth}
      \centering
      \includegraphics[width=\textwidth]{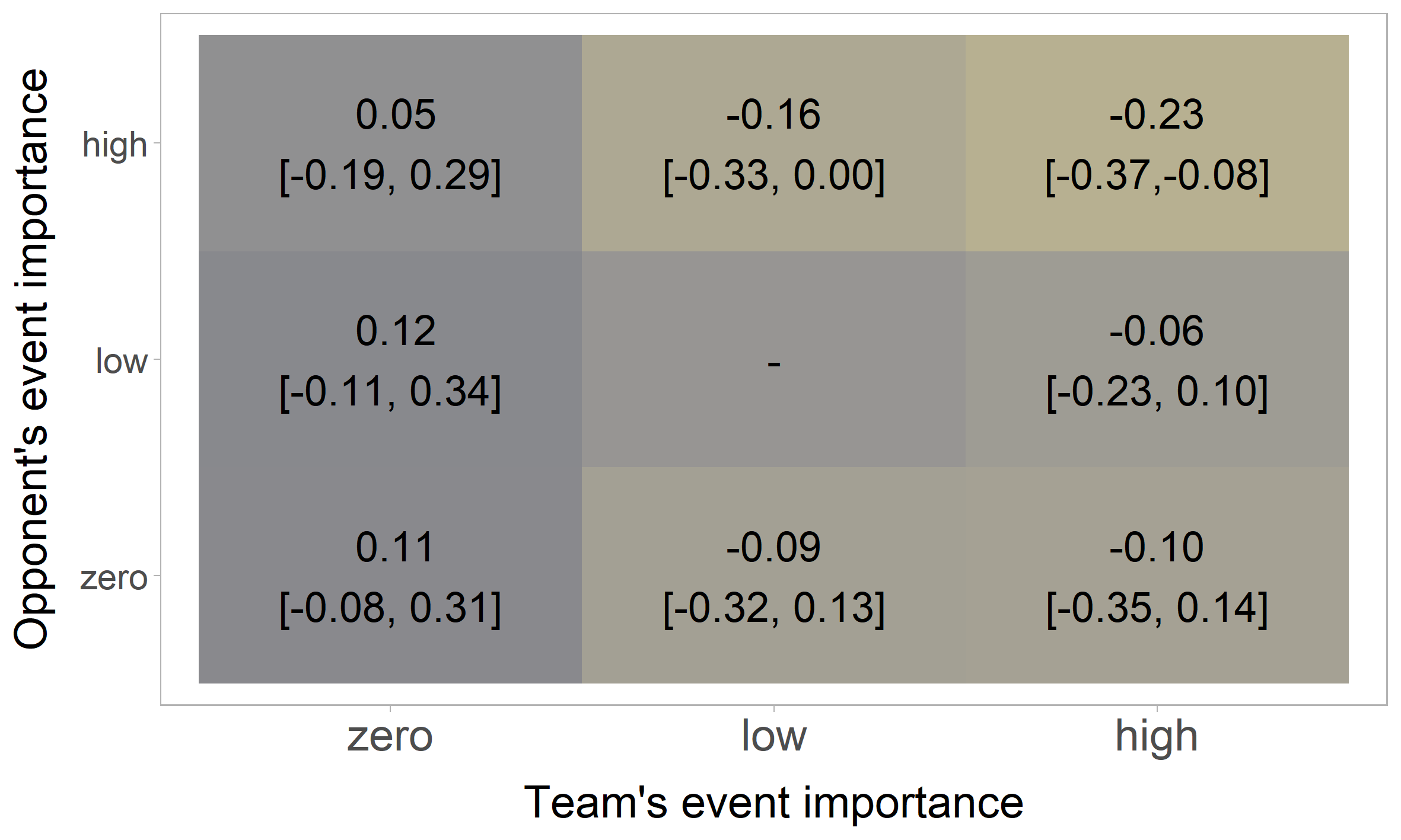}
      \caption{Passes}
      \label{fig:gg_lm_est_pass_pmin}
    \end{subfigure}
  
\vspace{8pt}  
    
    \begin{subfigure}{.48\textwidth}
      \centering
      \includegraphics[width=\textwidth]{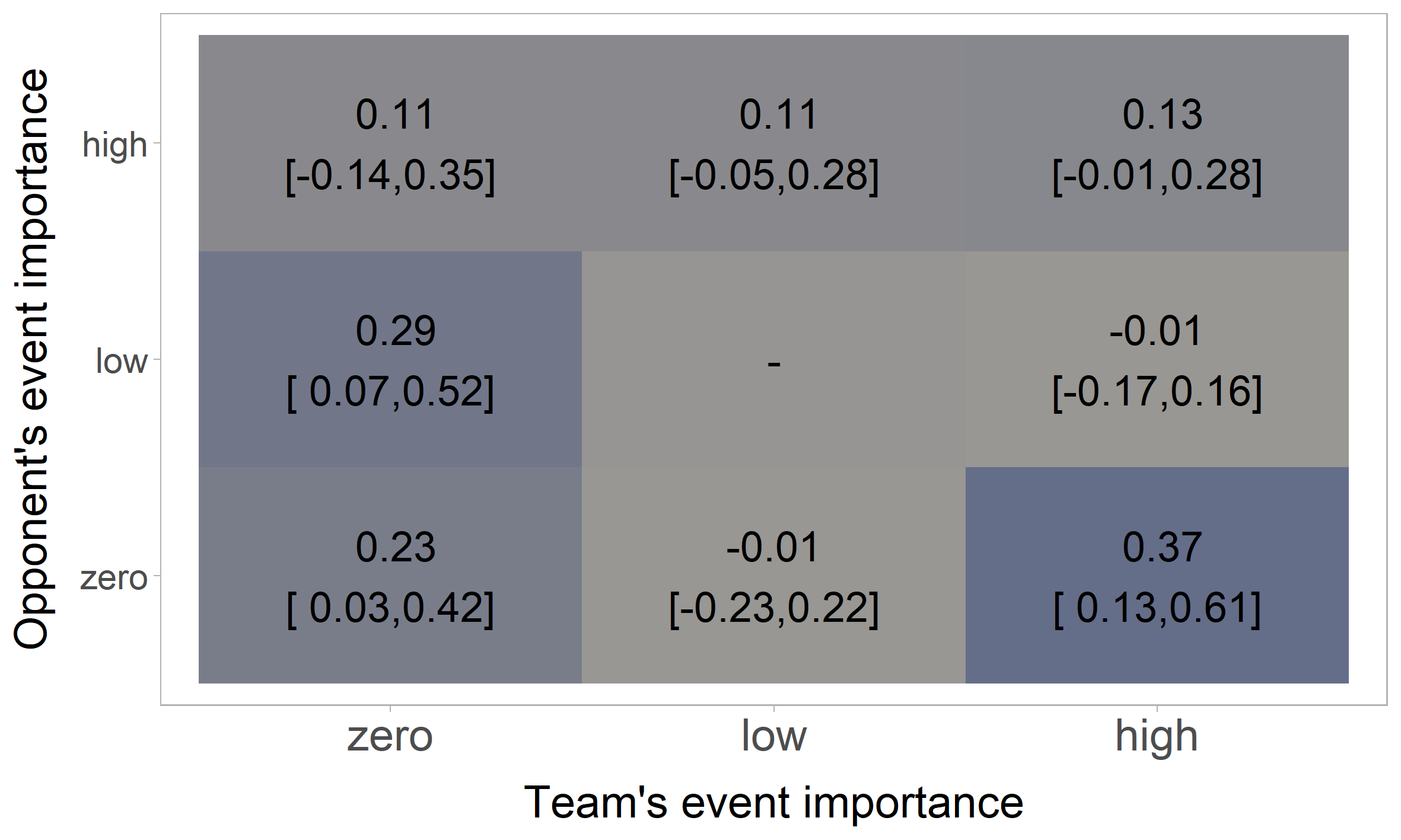}
      \caption{Goals scored}
      \label{fig:gg_lm_est_Goals_for}
    \end{subfigure}%
    \hfill
    \begin{subfigure}{.48\textwidth}
      \centering
      \includegraphics[width=\textwidth]{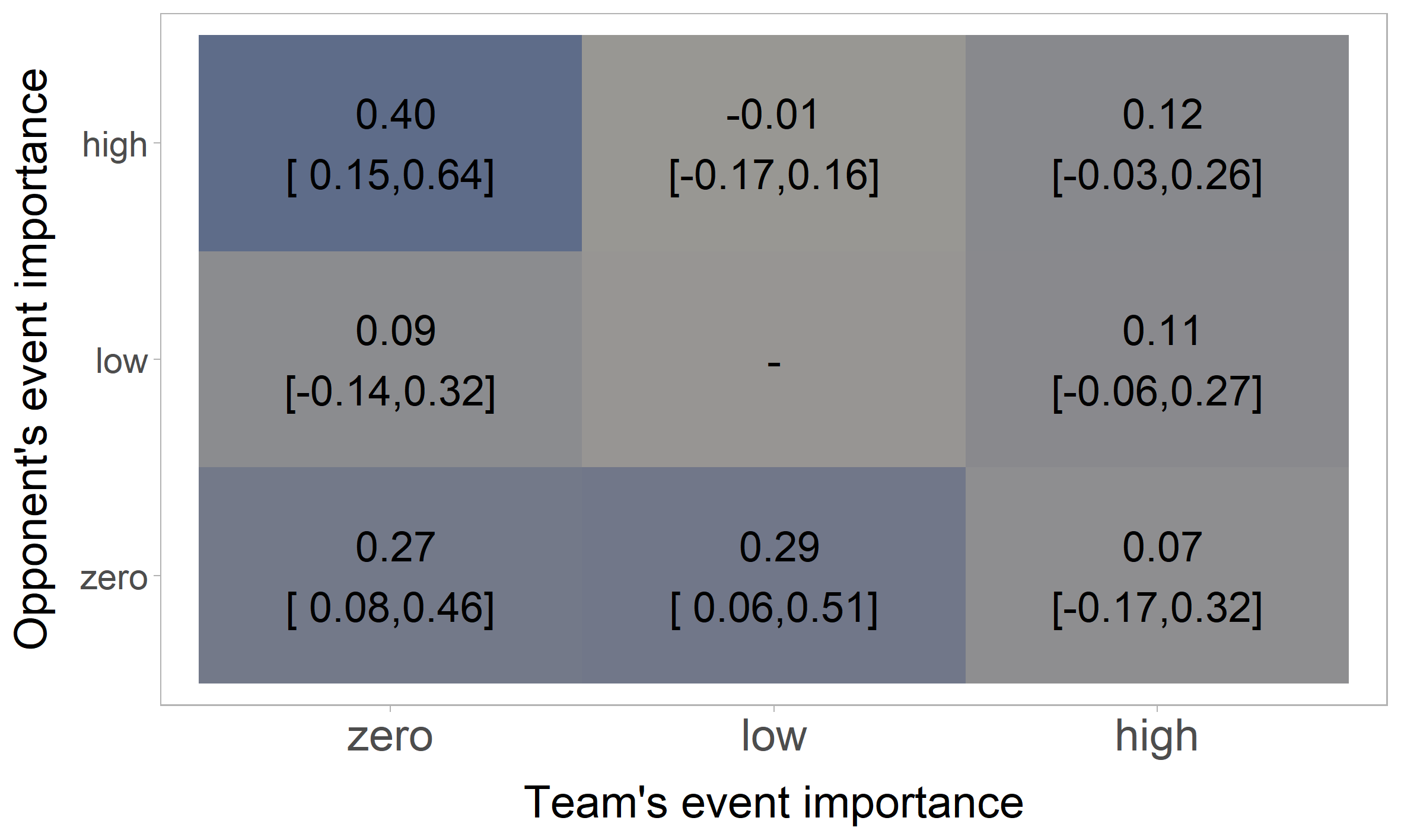}
      \caption{Goals conceded}
      \label{fig:gg_lm_est_Goals_against}
    \end{subfigure}
    \caption{Linear regression estimates of the centred and standardized residuals of different outcomes on the Event importance categories. 95\% confidence intervals are in parentheses. The baseline is low by low category.}
    \label{fig:gg_lm_est_opta}
\end{figure}

Figure~\ref{fig:gg_lm_est_opta} shows the results for four outcomes: duels per game, number of completed passes, number of goals scored, and number of goals conceded.\footnote{In a first step, we run a fixed effect regression for every outcome individually using the combinations 'Team x home/away x season', as well as 'Opponent x home/away x season' fixed effects'. The resulting residuals are centred and standardized by 'Team x home/away x season'. On those scaled residuals we run a regression using again the grid on the EI categories 'zero', 'low', and 'high' for both competing teams.} Complementary results using other outcomes are shown in Appendix~\ref{app:Team performance}. 
The results can be summarised as follows: Teams for which a match is particularly important (i) play more aggressive entering more duels on the pitch, (ii) play more directly towards goal with fewer passes, fewer touches, and more entries into the final third and penalty area, (iii) score more goals. In contrast, teams with zero importance exhibit a more passive style of play and concede more goals.

\subsubsection{Public perception}\label{sec:Public perception}

Sport is entertainment and thrives on public perception. If the calculated EI can represent the (later realised) public interest in a specific match, it could be useful for several purposes -- marketing, ticket pricing, or prime-time broadcasting selection.
With this in mind, we relate the EI to the stadium attendance turnout, as well as social media attention. While attendances are officially reported by clubs, social media attention is captured through club mentions and match hashtags within the 24 hours before kickoff on Twitter.\footnote{Gathered using Twitter API v2. The analysis yields almost identical results when the time period is extended to 48h before the match. Match hashtags are a short abbreviation of the names of the two teams which are regularly used on Twitter, i.\,e.\ \#BVBS04 relates to the match of the teams Borussia Dortmund against Schalke 04. Mentions of club accounts are Twitter tweets containing the account name of a team, e.\,g.\ @LFC - the official Twitter account of the English team Liverpool FC. Due to the inconsistency and lack of use of the aforementioned proxies in the early years and across the leagues, we can only perform this analysis beginning with the 2014/15 season and must exclude the Spanish and Portuguese leagues.} \par
\begin{figure}[ht]
    \begin{subfigure}{.48\textwidth}
      \centering
      \includegraphics[width=\textwidth]{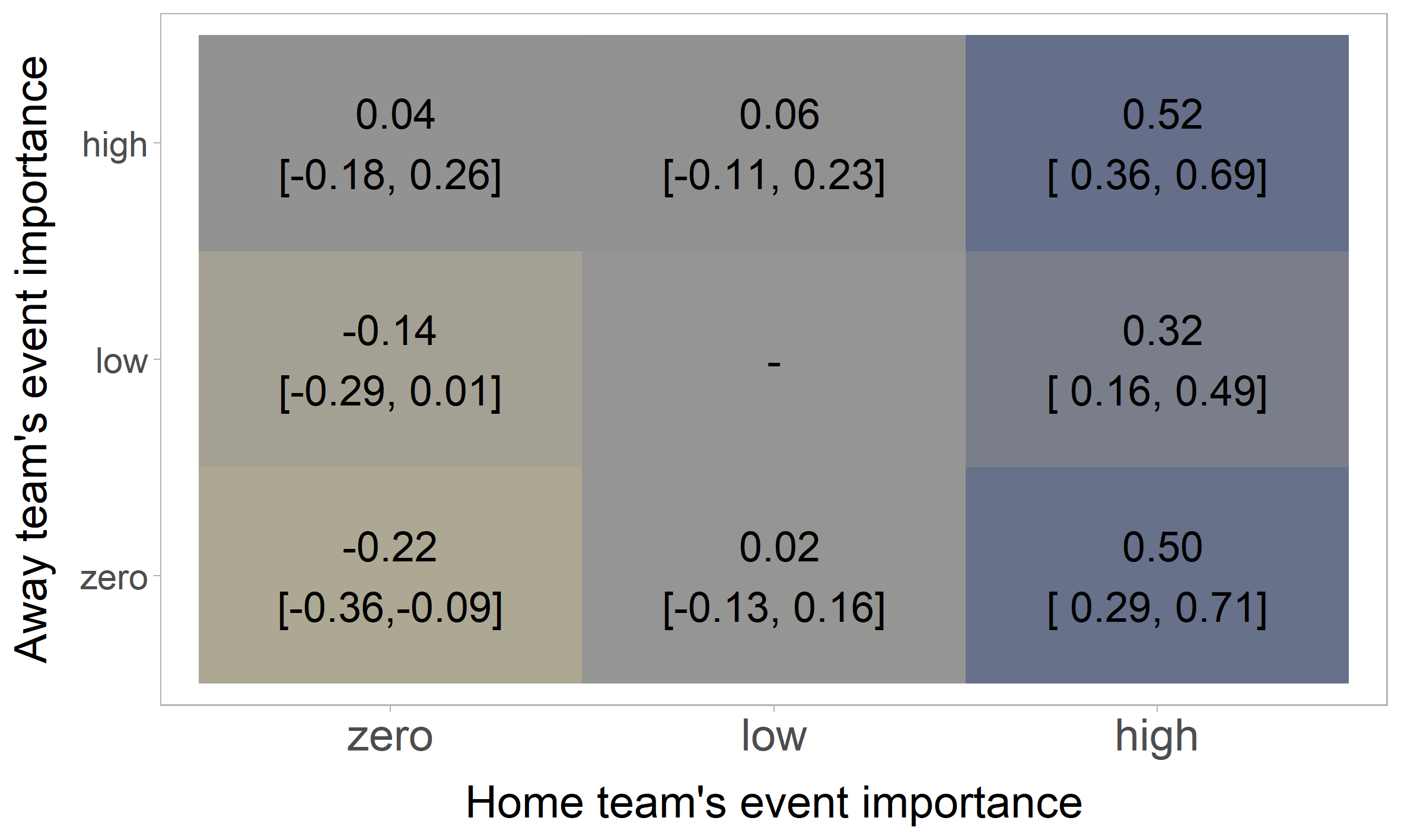}
      \caption{Home team's Twitter account mentions}
      \label{fig:gg_lm_est_H_cumbef_24h}
    \end{subfigure}%
    \hfill
    \begin{subfigure}{.48\textwidth}
      \centering
      \includegraphics[width=\textwidth]{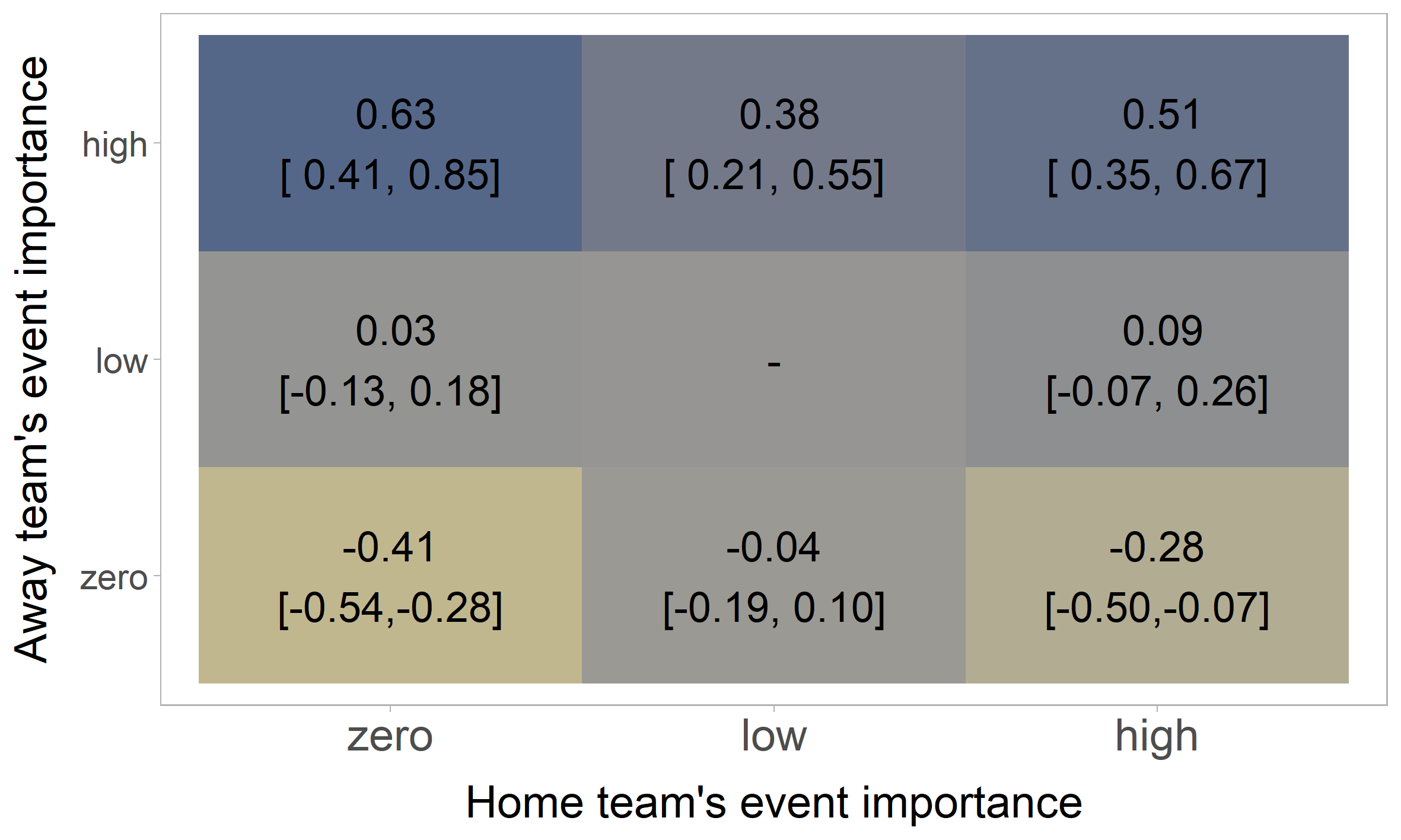}
      \caption{Away team's Twitter account mentions}
      \label{fig:gg_lm_est_A_cumbef_24h}
    \end{subfigure}
    
\vspace{8pt}  
    
    \begin{subfigure}{.48\textwidth}
      \centering
      \includegraphics[width=\textwidth]{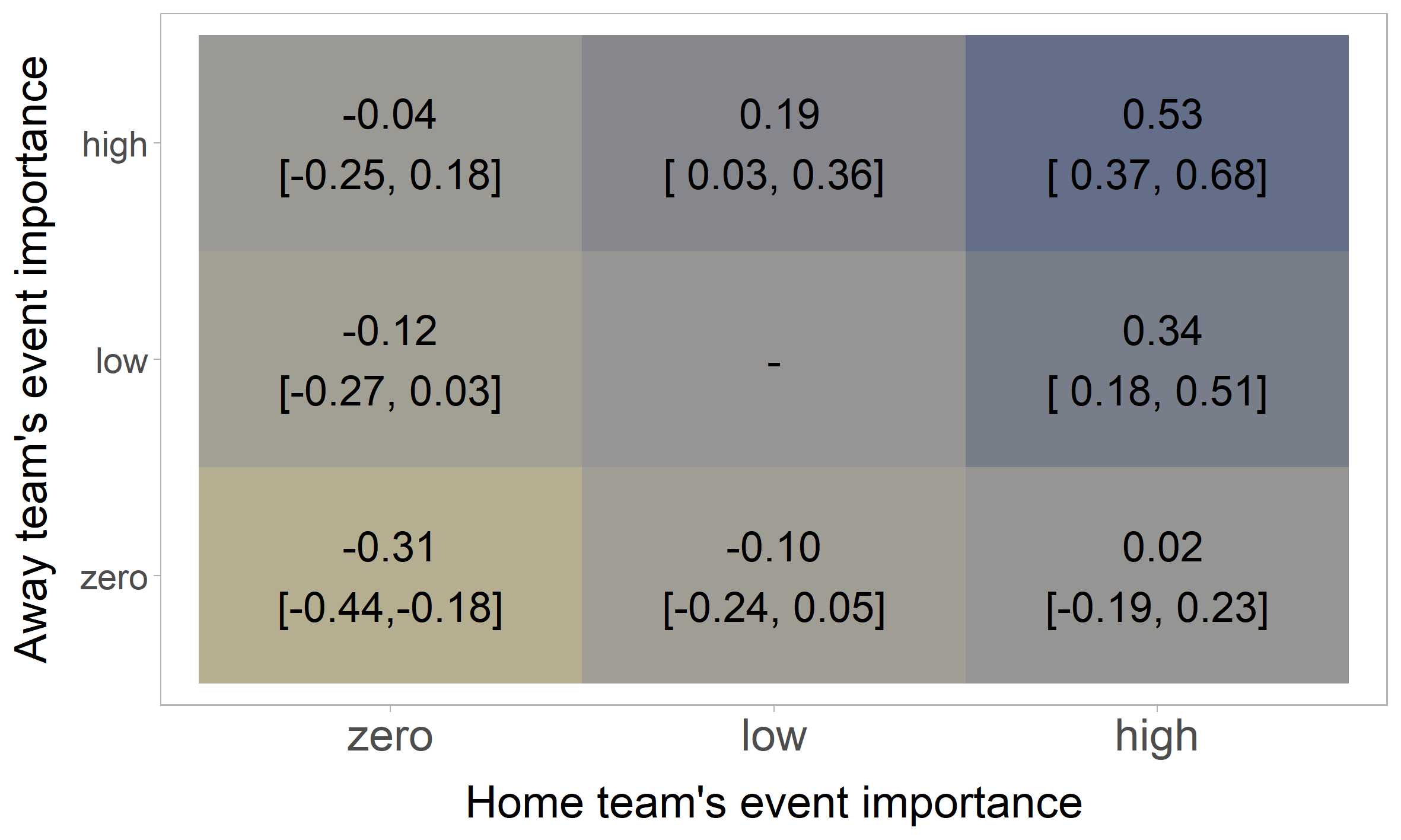}
      \caption{Match-tag Twitter mentions}
      \label{fig:gg_lm_est_count_24h_matchtag}
    \end{subfigure}%
    \hfill
    \begin{subfigure}{.48\textwidth}
      \centering
      \includegraphics[width=\textwidth]{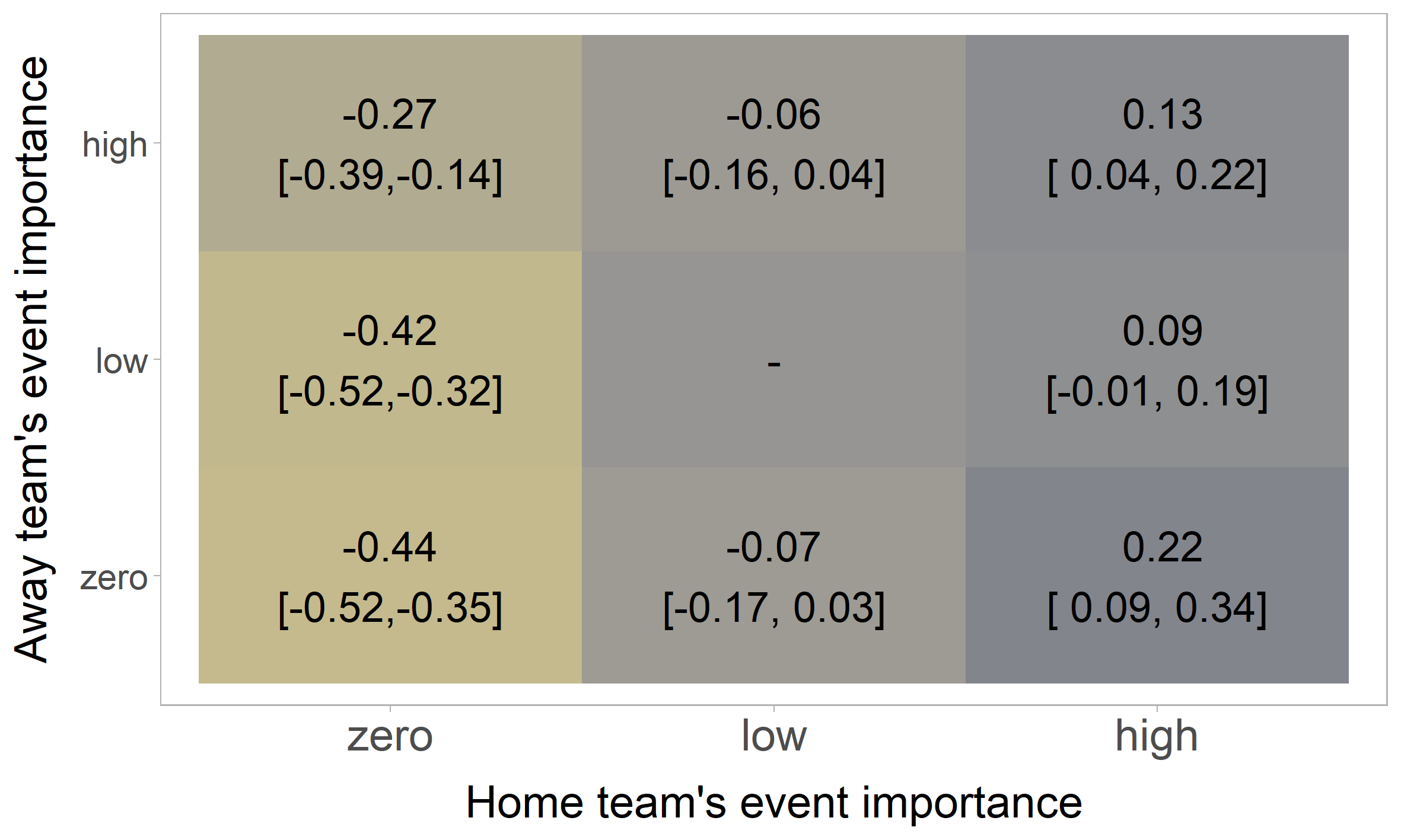}
      \caption{Stadium attendance}
      \label{fig:gg_lm_est_Zuschauer}
    \end{subfigure}
    \caption{Linear regression estimates of the centred and standardized residuals of different outcomes on the event importance (EI) categories. 95\% confidence intervals are in parentheses. The baseline is low by low category.}
    \label{fig:gg_lm_est_public}
\end{figure}

As different clubs have put different emphasis on social media and this has changed over time, we control for the team and season-specific usage of social media.\footnote{The procedure of the analysis is similar to the residual analysis in Section~\ref{sec:team_performance}. In a first step we calculate a linear fixed-effect model containing a FE for every 'team x home/away x season' interaction and for every calendar month and using the Twitter or attendance data, both in logs, as the outcome. In a second step, we centre and standardize the residuals from the linear fixed-effect models by 'league x season' pairs. This is the most natural procedure in our view. By taking nominal outcomes, other FE variants, or standardising over different groups interpretation of results does not change.} In the analysis, we explain the standardised and centred attention measures by linear regression on the EI of the two competing teams. Figure~\ref{fig:gg_lm_est_public} presents the point estimates and 95 \% confidence intervals of the linear regressions.
Stadium attendance is modestly associated with the home team's importance in the match. Here, restrictions on stadium capacity and (pre-sold) season tickets could mitigate the effect. Thus, social media attention might give a more clear picture of realised interest. We find team account mentions are strongly associated with the respective team's EI measure. Similarly, the match-tag mentions increase with both teams' EI. 
This is consistent with and complementary to \citeA{Dobson1992} and \citeA{Lei2013} reporting higher stadium attendance for more important sporting events, and recent findings by \citeA{buraimo2022armchair} that Premier League television audiences are larger for more important matches.

\section{Conclusion}

Public perception and academic research analyse incentives in simple situations where there is a direct link between performance and reward. More complex situations with indirect rewards and therefore unclear implicit incentive structures have received little attention. \par

In this article, we propose a statistical method to quantify the importance of single events in multi-event contests with end-of-contest reward structures. Thanks to its flexibility and generality the procedure covers a multitude of potential applications and can be valuable for various fields, including sales and marketing, human resources, or operations management. Our event importance framework can be adapted to different contest structures seen in society and opens a variety of potential research topics such as behavioural responses involving implicit incentives or operational fairness concerns in contest and reward structures. These include, for example, different valuations due to the order of actions or asymmetric incentives that lead to distorted probabilities of winning in a contest.\par

In an application to European football leagues, we show the association of the quantified importance of a match to in-match behaviour and the performance of the teams. As discrepancies in the EI can lead to altered outcome probabilities, the quantification of the event importance can help to ensure fair tournaments. 
The event importance measure also addresses other stakeholders in the football industry. As we show that the EI measure is consistent with the public interest in terms of social media and stadium attendance it can be useful for dynamic ticket pricing or TV stations that want to broadcast the most attractive match. Lastly, we illustrate the value of the EI measure for predicting match outcomes and point out under which circumstances the bookmakers do not yet account for the event's importance. \par

For the application to the US presidential primaries, we quantify the higher relevance of early election dates induced by the first-winner effect. For small states with a low number of delegates, this can substantially boost their influence on the nomination outcome as otherwise, their votes become irrelevant in many of the primaries. We show that the two investigated hypothetical schedules lead to more equitable distribution in the ratio of event importance values to the number of delegates rewarded by the election. 



\newpage

@article{Apesteguia2016,
author = {Apesteguia, Jose and Palacios-Huerta, Ignacio},
journal = {The American Economic Review},
number = {5},
pages = {2548--2564},
title = {{Psychological Pressure in Competitive Environments : Evidence from a Randomized Natural Experiment}},
volume = {100},
year = {2016}
}
@article{Ariely2009,
abstract = {Workers in a wide variety of jobs are paid based on performance, which is commonly seen as enhancing effort and productivity relative to non-contingent pay schemes. However, psychological research suggests that excessive rewards can, in some cases, result in a decline in performance. To test whether very high monetary rewards can decrease performance, we conducted a set of experiments in the U.S. and in India in which subjects worked on different tasks and received performance-contingent payments that varied in amount from small to very large relative to their typical levels of pay. With some important exceptions, very high reward levels had a detrimental effect on performance. {\textcopyright} 2009 The Review of Economic Studies Limited.},
author = {Ariely, Dan and Gneezy, Uri and Loewenstein, George and Mazar, Nina},
issn = {00346527},
journal = {Review of Economic Studies},
number = {2},
pages = {451--469},
title = {{Large stakes and big mistakes}},
volume = {76},
year = {2009}
}
@article{Arlegi2020,
abstract = {We study the impact of two basic principles of fairness on the structure of elimination-type competitions and perform our analysis by focusing on sports competitions. The first principle states that stronger players should have a larger chance of winning than weaker players, while the second principle provides equally strong players the same chances of being the final winner. We apply these requirements to different kinds of knockout competitions, and characterise the structures satisfying them. In our results, a new competition structure that we call an antler is found to play a referential role.},
author = {Arlegi, Ritxar and Dimitrov, Dinko},
issn = {03772217},
journal = {European Journal of Operational Research},
keywords = {Fairness,Graphs,OR in sports,Seeding rules,Tournament design},
number = {2},
pages = {528--535},
publisher = {Elsevier B.V.},
title = {{Fair elimination-type competitions}},
volume = {287},
year = {2020}
}
@article{Borup2020,
title = {Targeting predictors in random forest regression},
journal = {International Journal of Forecasting},
year = {2022},
issn = {0169-2070},
author = {Daniel Borup and Bent Jesper Christensen and Nicolaj Søndergaard Mühlbach and Mikkel Slot Nielsen},
keywords = {Random forests, Targeted predictors, High-dimensional forecasting, Weak predictors, Variable selection}
}
@article{Bradley2013,
abstract = {The aims of this study were to: (1) quantify match running performance in 5-min periods to determine if players fatigue or modulate high-intensity running according to a pacing strategy, and (2) examine factors impacting high-intensity running such as score line, match importance and the introduction of substitutes. All players were analysed using a computerised tracking system. Maintaining 'high' levels of activity in the first half resulted in a 12 percent reduction (P < 0.01) in the second half for high-intensity running (effect size [ES]: 0.8), while no changes were observed in 'moderate' and 'low' groups (ES: 0.0-0.2). The 'high' group covered less (P < 0.01) high-intensity running in the initial 10-min of the second versus first half (ES: 0.6-0.7), but this was not observed in 'moderate' and 'low' groups (ES: 0.2-0.4). After the most intense periods, players demonstrated an 8 percent drop in high-intensity running (P < 0.05) compared to the match average (ES: 0.2) and this persisted for 5-min before recovering. Players covered similar high-intensity running distances in matches with differing score lines but position-specific trends indica...},
author = {Bradley, Paul S. and Noakes, Timothy D.},
issn = {02640414},
journal = {Journal of Sports Sciences},
keywords = {fatigue,football,match importance,pacing,score,substitutions},
number = {15},
pages = {1627--1638},
pmid = {23808376},
title = {{Match running performance fluctuations in elite soccer: Indicative of fatigue, pacing or situational influences?}},
volume = {31},
year = {2013}
}
@article{Breiman2001,
abstract = {The role of probabilistic methods in discrete mathematics cannot be overestimated. By defining the probability measure on a set of the studied combinatorial objects various numerical characteristics of these objects can be considered as random variables and studied using the methods of probability theory. The advantage of this approach is the well-developed probabilistic analytic techniques that allow us in many cases to obtain results, the proof of which by other methods appears too complicated, if indeed it is at all possible. The application of probabilistic methods is connected with extensive use of the terminology of probability theory. The reader will easily understand however that one speaks in fact about solving enumerative problems of discrete analysis. One of the primary research lines is the study of the limit properties of combinatorial objects manifested at the unlimited increase of the number of elements comprising such objects. It is often possible to represent the distributions of the characteristics of combinatorial objects as conditional distributions of the sums of independent random variables so that they can be studied using asymptotic methods in probability theory, namely limit theorems for sums of independent random variables.},
author = {Breiman, Leo},
isbn = {9783110941975},
journal = {Machine Learning},
keywords = {classification,ensemble,regression},
pages = {5--32},
title = {{Random forests}},
volume = {45},
year = {2001}
}
@article{Cohen-Zada2018,
abstract = {The order of actions in contests may generate different psychological effects which, in turn, may influence contestants' probabilities to win. The Prouhet-Thue-Morse sequence in which the first ‘n' moves is the exact mirror image of the next ‘n' moves should theoretically terminate any advantage to any of the contestants in a sequential pair-wise contest. The tennis tiebreak sequence of serves is the closest to the Prouhet-Thue-Morse sequence that one can find in real tournament settings. In a tiebreak between two players, A and B, the order of the first two serves (AB) is a mirror image of the next two serves (BA), such that the sequence of the first four serves is ABBA. Then, this sequence is repeated until one player wins the tiebreak. This sequence has been used not only in tennis, but also recently in the US TV presidential debates. In this study we analyse 1701 men's and 920 women's tiebreak games from top-tier tournaments between the years 2012 to 2015. Using several different strategies to disentangle the effect of serving first from the effect of selection, we find that, for both genders, serving first does not have any significant effect on the winning probabilities of the two players. Thus, it might be useful for other sports, and contests in general, to consider adopting the ABBA sequence.},
author = {Cohen-Zada, Danny and Krumer, Alex and Shapir, Offer Moshe},
issn = {01672681},
journal = {Journal of Economic Behavior and Organization},
keywords = {Contest,Performance,Sequence,Tennis,Tiebreak},
pages = {106--115},
publisher = {Elsevier B.V.},
title = {{Testing the effect of serve order in tennis tiebreak}},
volume = {146},
year = {2018}
}
@article{Dobson1992,
abstract = {Using match attendance data collected from a postal survey of Football League clubs, separate demand equations are estimated for standing and seated viewing accommodation. Some significant differences between attendance patterns for the two types of accommodation are identified: current form, the championship significance of the match and a geographical distance variable are found to be important determinants of standing attendance, while the club's historical record is of particular importance for seated attendance. The paper also discusses the implications of the results in view of the current moves towards the conversion of stadia to all-seated accommodation. {\textcopyright} 1992, Taylor & Francis Group, LLC. All rights reserved.},
author = {Dobson, S. M. and Goddard, J. A.},
issn = {14664283},
journal = {Applied Economics},
number = {10},
pages = {1155--1163},
title = {{The demand for standing and seated viewing accommodation in the English Football League}},
volume = {24},
year = {1992}
}
@article{Dohmen2008,
abstract = {High rewards or the threat of severe punishment provide strong motivation but also create psychological pressure, which might induce performance decrements. By analyzing the performance of professional football players in penalty kick situations, the paper provides empirical evidence for the existence of detrimental incentive effects. Two pressure variables are considered in particular: (1) the importance of success and (2) the presence of spectators. There are plenty of situations in which pressure arises in the workplace. Knowing how individuals perform under pressure conditions is crucial because it has implications for the design of the workplace and the design of incentive schemes. {\textcopyright} 2006 Elsevier B.V. All rights reserved.},
author = {Dohmen, Thomas J.},
issn = {01672681},
journal = {Journal of Economic Behavior and Organization},
keywords = {Choking under pressure,Paradoxical performance effects of incentives,Social pressure},
number = {3-4},
pages = {636--653},
title = {{Do professionals choke under pressure?}},
volume = {65},
year = {2008}
}
@article{Ehrenberg1990,
abstract = {This paper proposes a tractable model to study the equilibrium diversity of technological progress and shows that equilibrium technological progress may exhibit too little diversity (too much conformity), in particular, foregoing socially beneficial investments in “alternative” technologies that will be used at some point in the future. The presence of future innovations that will replace current innovations imply that social benefits from innovation are not fully internalized. As a consequence, the market favors technologies that generate current gains relative to those that will bear fruit in the future; current innovations in research lines that will be profitable in the future are discouraged because current innovations are typically followed by further innovations before they can be profitably marketed. A social planner would choose a more diverse research portfolio and would induce a higher growth rate than the equilibrium allocation. The diversity of researchers is a partial (imperfect) remedy against the misallocation induced by the market. Researchers with different interests, competences or ideas may choose non-profit maximizing and thus more diverse research portfolios, indirectly contributing to economic growth.},
author = {Ehrenberg, R. and Bognanno, M.},
journal = {Journal of Political Economy},
number = {6},
pages = {1307--1324},
title = {{Do tournaments have incentive effects}},
volume = {98},
year = {1990}
}
@article{Erev1993,
abstract = {Pruitt and Kimmel (1977) regard the problem of external validity as one of the biggest problems of experimental gaming, a problem that seriously limits the relevance of this research tradition to real-life settings and to other areas of social psychology. Following Pruitt and Kimmel′s dictate that “researchers try to generalize their findings,” the present study attempts to generalize recent results by Bornstein, Erev, and Rosen (1990) to real-life settings. Bornstein et al. demonstrated that competition between groups can reduce free riding in an experimental Prisoner′s Dilemma Game. The present study tested the effectiveness of intergroup competition as a solution to free riding in a lifelike orange-picking task. Groups of four subjects picked oranges under three payoff conditions: personal reward, collective reward, and intergroup competition with a reward to the most efficient team. On average, the collective reward rule resulted in a 30 percent loss in production compared to the personal payoff rule. The intergroup competition eliminated this loss of productivity and was more effective the more similar the competing teams were with respect to overall abilities. The implications of these findings are discussed in light of a new approach to game-theoretic equilibrium solutions. {\textcopyright} 1993 by Academic Press, Inc.},
author = {Erev, Ido and Bornstein, Gary and Galili, Rachely},
issn = {10960465},
journal = {Journal of Experimental Social Psychology},
number = {6},
pages = {463--478},
title = {{Constructive intergroup competition as a solution to the free rider problem: A field experiment}},
volume = {29},
year = {1993}
}
@article{Goddard2005,
abstract = {In the previous literature, two approaches have been used to model match outcomes in association football (soccer): first, modelling the goals scored and conceded by each team; and second, modelling win-draw-lose match results directly. There have been no previous attempts to compare the forecasting performance of these two types of model. This paper aims to fill this gap. Bivariate Poisson regression is used to estimate forecasting models for goals scored and conceded. Ordered probit regression is used to estimate forecasting models for match results. Both types of models are estimated using the same 25-year data set on English league football match outcomes. The best forecasting performance is achieved using a 'hybrid' specification, in which goals-based team performance covariates are used to forecast win-draw-lose match results. However, the ifferences between the forecasting performance of models based on goals data and models based on results data appear to be relatively small. {\textcopyright} 2004 International Institute of Forecasters. Published by Elsevier B.V. All rights reserved.},
author = {Goddard, John},
journal = {International Journal of Forecasting},
keywords = {Bivariate Poisson,Football match results,Ordered probit},
number = {2},
pages = {331--340},
title = {{Regression models for forecasting goals and match results in association football}},
volume = {21},
year = {2005}
}
@article{Goller2021,
author = {Goller, Daniel},
doi = {10.1007/s10479-022-04563-0},
journal = {Annals of Operations Research},
title = {{Analysing a built-in advantage in asymmetric darts contests using causal machine learning}},
volume = {forthcom.},
pages = {1--31},
year = {2022}
}
@article{Goller2020,
abstract = {Balancing the allocation of games in sports competitions is an important organizational task that can have serious financial consequences. In this paper, we examine data from 10,142 soccer games played in the top German, Spanish, French, and English soccer leagues between 2007/2008 and 2016/2017. Using a machine learning technique for variable selection and applying a semi-parametric analysis of radius matching on the propensity score, we find that all four leagues have a lower attendance in games that take place on four non-frequently played days than those on three frequently played days. We also find that, in all leagues, there is a significantly lower home advantage for the underdog teams on non-frequent days. Our findings suggest that the current schedule favors underdog teams with fewer home games on non-frequent days. Therefore, to increase the fairness of the competitions, it is necessary to adjust the allocation of the home games on non-frequent days in a way that eliminates any advantage driven by the schedule. These findings have implications for the stakeholders of the leagues, referees' and calendar committees as well as for coaches and players.},
author = {Goller, Daniel and Krumer, Alex},
issn = {03772217},
journal = {European Journal of Operational Research},
keywords = {Performance,Schedule effects,Soccer},
number = {2},
pages = {740--754},
publisher = {Elsevier B.V.},
title = {{Let's meet as usual: Do games played on non-frequent days differ? Evidence from top European soccer leagues}},
volume = {286},
year = {2020}
}
@article{Gonzalez-Diaz2012,
abstract = {Stakes affect aggregate performance in a wide variety of settings. At the individual level, we define the critical ability as an agent's ability to adapt performance to the importance of the situation. We identify individual critical abilities of professional tennis players, relying on point-level data from twelve years of the US Open tournament. We establish persistent heterogeneity in critical abilities. We find a significant statistical relationship between identified critical abilities and overall career success, which validates the identification procedure and suggests that response to pressure is a significant factor for success. {\textcopyright} 2012 Elsevier B.V.},
author = {Gonz{\'{a}}lez-D{\'{i}}az, Julio and Gossner, Olivier and Rogers, Brian W.},
issn = {01672681},
journal = {Journal of Economic Behavior and Organization},
keywords = {Career success,Critical ability,Heterogeneity,Performance,Pressure},
number = {3},
pages = {767--781},
publisher = {Elsevier B.V.},
title = {{Performing best when it matters most: Evidence from professional tennis}},
volume = {84},
year = {2012}
}
@article{Harb-Wu2019,
abstract = {Performing in front of a supportive audience increases motivation. However, it also creates psychological pressure, which may impair performance, especially in precision tasks. In this paper, we exploit a unique setting in which professionals compete in a real-life contest with large monetary rewards in order to assess how they perform in front of a supportive audience. Using the task of shooting in the sprint competitions of professional biathlon events over a period of 16 years, we find that for both genders, biathletes from the top quartile of the ability distribution miss significantly more shots when competing in their home country compared to competing abroad. Our results are in line with the hypothesis that high expectations to perform well in front of a friendly audience prompt individuals to choke when performing skill-based tasks.},
author = {Harb-Wu, Ken and Krumer, Alex},
issn = {01672681},
journal = {Journal of Economic Behavior and Organization},
keywords = {Biathlon,Choking under pressure,Home advantage,Paradoxical performance effects of incentives,Social pressure},
pages = {246--262},
publisher = {Elsevier B.V.},
title = {{Choking under pressure in front of a supportive audience: Evidence from professional biathlon}},
volume = {166},
year = {2019}
}
@book{Hastie2009,
address = {2nd. ed. New York},
author = {Hastie, Trevor and Tibshirani, Robert and Friedman, Jerome},
isbn = {9781479932115},
publisher = {Springer},
title = {{The Elements of Statistical Learning - Data mining, inference, and prediction}},
year = {2009}
}
@article{Jennett1984,
author = {Jennett, Nicholas},
journal = {Scottish Journal of Political Economy},
number = {1},
pages = {176--198},
title = {{Attendances, Uncertainty of Outcome and Policy in Scottish League Football}},
volume = {31},
year = {1984}
}
@article{Klaassen2001,
abstract = {This article tests whether points in tennis are independent and identically distributed (iid). We model the probability of winning a point on service and show that points are neither independent nor identically distributed: winning the previous point has a positive effect on winning the current point, and at “important” points it is more difficult for the server to win the point than at less important points. Furthermore, the weaker a player, the stronger are these effects. Deviations from iid are small, however, and hence the iid hypothesis will still provide a good approximation in many cases. The results are based on a large panel of matches played at Wimbledon 1992–1995, in total almost 90,000 points. Our panel data model takes into account the binary character of the dependent variable, uses random effects to capture the unobserved part of a player's quality, and includes dynamic explanatory variables. {\textcopyright} 2001 American Statistical Association.},
author = {Klaassen, Franc J.G.M. and Magnus, Jan R.},
isbn = {0162145017531},
issn = {1537274X},
journal = {Journal of the American Statistical Association},
keywords = {Binary choice,Dependence,Dynamic panel data,Linear probability model,Nonidentical distribution,Random effects,Tennis},
number = {454},
pages = {500--509},
title = {{Are points in tennis independent and identically distributed? Evidence from a dynamic binary panel data model}},
volume = {96},
year = {2001}
}
@article{Krumer2017,
abstract = {The order of actions in contests may have a significant effect on performance. In this study, we examine the role of the schedule in round-robin tournaments with sequential games between three and four contestants. Our empirical analysis, based on soccer FIFA World Cups and UEFA European Championships, and on two Olympic wrestling events, reveals that there is a substantial advantage to the contestant who competes in the first and third matches. Our findings are in line with the hypothesis that winning probabilities in multi-stage contests with sequential games are endogenously depending on the schedule of contests as predicted by game-theoretical models. We also discuss possible ways to reduce the effect of the schedule.},
author = {Krumer, Alex and Lechner, Michael},
issn = {00142921},
journal = {European Economic Review},
keywords = {First-mover advantage,Performance,Schedule effects,Soccer,Wrestling},
pages = {412--427},
publisher = {Elsevier B.V.},
title = {{First in first win: Evidence on schedule effects in round-robin tournaments in mega-events}},
volume = {100},
year = {2017}
}
@article{Krumer2017a,
abstract = {We study round-robin tournaments with either three or four symmetric players whose values of winning are common knowledge. With three players there are three rounds, each of which includes one pair-wise game such that each player competes in two rounds only. The player who wins two games wins the tournament. We characterize the subgame perfect equilibrium and show that each player's expected payoff and probability of winning is maximized when he competes in the first and the last rounds. With four players there are three rounds, each of which includes two sequential pair-wise games where each player plays against a different opponent in every round. We again characterize the subgame perfect equilibrium and show that a player who plays in the first game of each of the first two rounds has a first-mover advantage as reflected by a significantly higher winning probability as well as by a significantly higher expected payoff than his opponents.},
author = {Krumer, Alex and Megidish, Reut and Sela, Aner},
issn = {01761714},
journal = {Social Choice and Welfare},
number = {3},
pages = {633--658},
publisher = {Springer Berlin Heidelberg},
title = {{First-mover advantage in round-robin tournaments}},
volume = {48},
year = {2017}
}
@article{Lechner2019,
abstract = {In econometrics so-called ordered choice models are popular when interest is in the estimation of the probabilities of particular values of categorical outcome variables with an inherent ordering, conditional on covariates. In this paper we develop a new machine learning estimator based on the random forest algorithm for such models without imposing any distributional assumptions. The proposed Ordered Forest estimator provides a flexible estimation method of the conditional choice probabilities that can naturally deal with nonlinearities in the data, while taking the ordering information explicitly into account. In addition to common machine learning estimators, it enables the estimation of marginal effects as well as conducting inference thereof and thus providing the same output as classical econometric estimators based on ordered logit or probit models. An extensive simulation study examines the finite sample properties of the Ordered Forest and reveals its good predictive performance, particularly in settings with multicollinearity among the predictors and nonlinear functional forms. An empirical application further illustrates the estimation of the marginal effects and their standard errors and demonstrates the advantages of the flexible estimation compared to a parametric benchmark model. A software implementation of the Ordered Forest is provided in GAUSS as well as in the R-package orf available on CRAN.},
archivePrefix = {arXiv},
arxivId = {1907.02436},
author = {Lechner, Michael and Okasa, Gabriel},
eprint = {1907.02436},
issn = {2331-8422},
journal = {arXiv:1907.02436},
title = {{Random Forest Estimation of the Ordered Choice Model}},
year = {2019}
}
@article{Lei2013,
abstract = {Sports teams have incentives to put more effort into games with an immediate effect on standings compared to games that do not, possibly affecting outcome uncertainty. We develop a measure of game outcome uncertainty, game importance (GI), that captures how each game affects a team ' s standing and can be calculated for individual games. Results show that observed variation in GI explains observed variation in attendance, game outcomes, and margin of victory at MLB games over the 1994 through 2010 seasons, suggesting that GI is an influential indicator in fans ' attendance decisions, consistent with the uncertainty of outcome hypothesis.},
author = {Lei, Xinrong and Humphreys, Brad R.},
issn = {15590410},
journal = {Journal of Quantitative Analysis in Sports},
keywords = {Game Attendance,Game Importance, Game Outcome,Major League Baseball,Uncertainty Of Outcome},
number = {1},
pages = {25--36},
title = {{Game importance as a dimension of uncertainty of outcome}},
volume = {9},
year = {2013}
}
@article{Link2016,
abstract = {This research explores the influence of match importance on player activity in professional soccer. Therefore, we used an observational approach and analyzed 1,211 matches of German Bundesliga and 2nd Bundesliga. The importance measurement employed is based on post season consequences of teams involved in a match. This means, if a match result could potentially influence the final rank, and this rank would lead to different consequences for a team, such as qualification for Champions League opposed to qualification for Europe League, then this match is classified as important; otherwise not. Activity was quantified by TOTAL DISTANCE COVERED, SPRINTS, FAST RUNS, DUELS, FOULS and ATTEMPTS. Running parameters were recorded using a semi-automatic optical tracking system, while technical variables were collected by professional data loggers. Based on our importance classification, low important matches occurred at the beginning of round 29. A two-way ANOVA indicates significantly increased FAST RUNS (+4 percent, d = 0.3), DUELS (+16 percent, d = 1.0) and FOULS (+36 percent, d = 1.2) in important matches compared to low important ones. For FAST RUNS and FOULS, this effect only exists in Bundesliga. A comparison of the two leagues show that TOTAL DISTANCE COVERED (+3 percent, d = 0.9), SPRINTS (+25 percent, d = 1.4) and FAST RUNS (+15 percent, d = 1.4) are higher compared to 2nd Bundesliga, whilst FOULS is less in Bundesliga (-7 percent, d = 0.3). No difference in player activity was found between matches at the beginning of a season (round 1-6) and at the end of a season (round 29-34). We conclude that match importance influences player activity in German professional soccer. The most reasonable explanation is a conscious or unconscious pacing strategy, motivated by preserving abilities or preventing injury. Since this tendency mainly exists in Bundesliga, this may suggest that more skilled players show a higher awareness for the need of pacing.},
author = {Link, Daniel and {De Lorenzo}, Michael F.},
issn = {19326203},
journal = {PLoS ONE},
number = {6},
pages = {1--10},
pmid = {27281051},
title = {{Seasonal pacing - Match importance affects activity in professional soccer}},
volume = {11},
year = {2016}
}
@article{Moreira2013,
author = {Moreira, Alexandre and Freitas, Camila and Nakamura, Fabio and Drago, Gustavo and Drago, Murilo and Aoki, Marcelo},
journal = {The Journal of Strength and Conditioning Research},
number = {1},
pages = {202--207},
title = {{Effect of Match Importance on Salivary Cortisol and Immunoglobulin A Responses in Elite Young Volleyball Players}},
volume = {27},
year = {2013}
}
@article{Paserman2010,
author = {Paserman, Daniele M.},
journal = {mimeo},
title = {{Gender Differences in Performance in Competitive Environments? Evidence from Professional Tennis Players}},
year = {2010}
}
@article{Stefani2007,
abstract = {Comprehensive rankings of football teams have become an important, and occasionally controversial, feature of many football codes. The rationale behind these systems needs to be understood. The historical evolution of the current eight codes is briefly traced from just two: association football and rugby football, played under various forms in the mid 19th century. Based on current rules, the eight codes fall into four groups: first, Australian Rules football and Gaelic football; second, American college football, American professional football and Canadian football; third, rugby union and rugby league; and fourth, soccer. Comprehensive rating systems exist for three codes. For American college football, the Bowl Championship Series or BCS system places top US college football teams into a national championship game and other important "bowl games". That system combines two normally incompatible components, an objective adjustive computer component and a subjective human-poll component. The composite has been controversial in four of the nine years of service, when the computer component differed from the human component resulting in major changes that favored the human component each time. For rugby union, the International Rugby Board or IRB system employs a predictor/corrector adjustment in which defeating a weak team provides less gain than defeating a strong team while losing to a weak team elicits a much larger negative adjustment than losing to a strong team, arguably a fair and efficient methods for rating competitors. For soccer, FIFA have improved the previous rating systems with a new and simpler system which takes into account strength of opponents and game importance; however, all losses are treated as equal regardless of the opponents, and home advantage is ignored. An Elo based system, employing many of features of the IRB system, appears to have advantages over the FIFA system.},
author = {Stefani, Ray and Pollard, Richard},
issn = {2194-6388},
journal = {Journal of Quantitative Analysis in Sports},
keywords = {football, American football, rugby union, FIFA, so},
number = {3},
title = {{Football Rating Systems for Top-Level Competition: A Critical Survey}},
volume = {3},
year = {2007}
}
@article{Wager2018,
abstract = {Many scientific and engineering challenges—ranging from personalized medicine to customized marketing recommendations—require an understanding of treatment effect heterogeneity. In this article, we develop a nonparametric causal forest for estimating heterogeneous treatment effects that extends Breiman's widely used random forest algorithm. In the potential outcomes framework with unconfoundedness, we show that causal forests are pointwise consistent for the true treatment effect and have an asymptotically Gaussian and centered sampling distribution. We also discuss a practical method for constructing asymptotic confidence intervals for the true treatment effect that are centered at the causal forest estimates. Our theoretical results rely on a generic Gaussian theory for a large family of random forest algorithms. To our knowledge, this is the first set of results that allows any type of random forest, including classification and regression forests, to be used for provably valid statistical inference. In experiments, we find causal forests to be substantially more powerful than classical methods based on nearest-neighbor matching, especially in the presence of irrelevant covariates.},
archivePrefix = {arXiv},
arxivId = {1510.04342},
author = {Wager, Stefan and Athey, Susan},
eprint = {1510.04342},
issn = {1537274X},
journal = {Journal of the American Statistical Association},
keywords = {Adaptive nearest neighbors matching,Asymptotic normality,Potential outcomes,Unconfoundedness},
number = {523},
pages = {1228--1242},
publisher = {Taylor & Francis},
title = {{Estimation and Inference of Heterogeneous Treatment Effects using Random Forests}},
volume = {113},
year = {2018}
}
@article{Wright2014,
abstract = {This paper surveys the academic OR/analytics literature describing research into the laws and rules of sports and sporting competitions. The literature is divided into post hoc analyses and proposals for future changes, and is also divided into laws/rules of sports themselves and rules/organisation of tournaments or competitions. The survey outlines a large number of studies covering 21 sports in many parts of the world. The analytical approaches most commonly used are found to be various forms of regression analysis and simulation. Issues highlighted by this survey include the different views of what constitutes fairness and the frequency with which changes produce unintended consequences. {\textcopyright} 2013 Elsevier B.V. All rights reserved.},
author = {Wright, Mike},
issn = {03772217},
journal = {European Journal of Operational Research},
keywords = {Analytics,OR in Sport,Review,Rules,Survey,Tournaments},
number = {1},
pages = {1--8},
publisher = {Elsevier B.V.},
title = {{OR analysis of sporting rules - A survey}},
volume = {232},
year = {2014}
}
@article{Lahvicka2015,
abstract = {This article presents a new method of calculating match importance. Match importance is defined as strength of relationship between the match result and a given season outcome. Probabilities of all necessary match result-season outcome combinations are estimated by Monte Carlo simulation. Using actual results of 12 seasons of English Premier League and betting odds, it is shown that both match result and season outcome predictions are realistic. The method provides results that are close to Jennett's approach; however, it does not require ex post information and can be used for any type of season outcome.},
author = {Lahvi{\v{c}}ka, Jiř{\'{i}}},
journal = {Journal of Sports Economics},
keywords = {Monte Carlo,match importance,seasonal uncertainty},
number = {4},
pages = {390--409},
title = {{Using Monte Carlo Simulation to Calculate Match Importance: The Case of English Premier League}},
volume = {16},
year = {2015}
}
@article{Soebbing2013,
abstract = {A growing body of literature indicates that sports teams face incentives to lose games at the end of the season. This incentive arises from a league's entry draft policy. We use data from betting markets to confirm the existence of tanking, or the perception of tanking, in the National Basketball Association (NBA). Results from a Seemingly Unrelated Regression (SUR) model of point spreads and point differences in NBA games indicate that betting markets believe that tanking takes place in the NBA, even though the evidence that tanking actually exists is mixed. Other NBA policy changes also affect betting market outcomes. {\textcopyright} 2011 Western Economic Association International.},
author = {Soebbing, Brian P. and Humphreys, Brad R.},
isbn = {7804923406},
issn = {10743529},
journal = {Contemporary Economic Policy},
number = {2},
pages = {301--313},
title = {{Do gamblers think that teams tank? Evidence from the NBA}},
volume = {31},
year = {2013}
}
@article{Rohde2017,
abstract = {Europe's professional football clubs engage in intraseasonal races to win the league, qualify for Union of European Football Associations (UEFA) competitions, and avoid relegation. In these competitions, playing talent is a scarce good as players invest effort in the form of fitness and the risk of injuries. Thus, managers face incentives to adjust effort levels by means of changing the value of starting squads. This paper analyzes whether managers save efforts in the absence of financial incentives or ahead of more important games. The theoretical model extends previous papers from the contest theory and shirking literature to a match-day panel regression analysis of strategic behavior. We build on two unbalanced panels covering 10 consecutive seasons in the German Bundesliga (n=6,120) and UEFA Champions League (n=1,920). The insights generated in this article are important for national league administrators as well as the UEFA in aligning their competitions towards each other and increasing incentives towards participating teams.},
author = {Rohde, Marc and Breuer, Christoph},
issn = {1930076X},
journal = {International Journal of Sport Finance},
keywords = {Contest theory,Financial incentives,League design,Managerial effort,Soccer,Sport economics},
number = {2},
pages = {160--182},
title = {{Financial incentives and strategic behavior in european professional football: A match day analysis of starting squads in the German Bundesliga and UEFA Competitions}},
volume = {12},
year = {2017}
}
@article{Kendall2017,
abstract = {Wright (2014) recently presented a survey of sporting rules from an Operational Research (OR) perspective. He surveyed 21 sports, which consider the rules of sports and tournaments and whether changes have led to unintended consequences. The paper concludes: “Overall, it would seem that this is just a taster and there may be plenty more such studies to come”. In this paper we present one such study. This is an interdisciplinary paper, which cuts across economics, sport and Operational Research. We recognize that the paper could have been published in any of these disciplines but for the sake of continuity with the paper that motivated this study, we wanted to publish this paper in an OR journal. We look at specific examples where the rules of sports have led to unforeseen and/or unwanted consequences. We hope that the paper will be especially useful to sports administrators, helping them to review what has not previously worked and also encouraging them to engage with the scientific community when considering making changes. We believe that this is the first time that such a comprehensive review of sporting rules, which have led to unexpected consequences, has been published in the scientific literature.},
author = {Kendall, Graham and Lenten, Liam J.A.},
issn = {03772217},
journal = {European Journal of Operational Research},
keywords = {Incentives,Sport,Strategies},
number = {2},
pages = {377--394},
publisher = {Elsevier B.V.},
title = {{When sports rules go awry}},
volume = {257},
year = {2017}
}
@article{Paul2003,
abstract = {In recent years the National Hockey League (NHL) has put policies in place to boost attendance. Specifically, these changes have been to curb violence, increase scoring, and move to an unbalanced schedule featuring more games against regional rivals. This research looks at variations in game-to-game attendance in the NHL, focusing on these policy changes. It is found that violence, specifically fighting, tends to attract fans in large numbers across the United States and Canada. Surprisingly, increases in scoring, ceteris paribus, tend to depress attendance. The change in scheduling by the NHL, however, has been a success, with divisional rivals increasing attendance in U.S. cities and additional contests against other Canadian teams increasing attendance in Canada.},
author = {Paul, Rodney J.},
journal = {American Journal of Economics and Sociology},
number = {2},
pages = {345--364},
title = {{Variations in NHL attendance: The impact of violence, scoring, and regional rivalries}},
volume = {62},
year = {2003}
}
@article{Goddard2004,
abstract = {An ordered probit regression model estimated using 10 years' data is used to forecast English league football match results. As well as past match results data, the significance of the match for end-of-season league outcomes, the involvement of the teams in cup competition and the geographical distance between the two teams' home towns all contribute to the forecasting model's performance. The model is used to test the weak-form efficiency of prices in the fixed-odds betting market. A strategy of selecting end-of-season bets with a favourable expected return according to the model appears capable of generating a positive return. Copyright {\textcopyright} 2004 John Wiley & Sons, Ltd.},
author = {Goddard, John and Asimakopoulos, Ioannis},
issn = {02776693},
journal = {Journal of Forecasting},
keywords = {Efficiency, ordered probit,Sports betting},
number = {1},
pages = {51--66},
title = {{Forecasting Football Results and the Efficiency of Fixed-odds Betting}},
volume = {23},
year = {2004}
}
@article{Feddersen2021,
abstract = {Research Question: : The study asks whether bookmakers alter the closing betting odds on European football matches due to potential variation in team incentives to exert effort to win games based on position in the standings. Effort is a multidimensional concept and teams have many ways to alter effort supplied in a match. For example, the players included or left out of the starting 11, the tactics used in a specific match, substitutions made during a match, and other factors under team control, can be interpreted as changes in effort supplied. Research Methods: OLS models explaining observed variation in betting odds are estimated for roughly 25,000 regular season football games in 8 top European domestic leagues from 2000–2001 through 2018–2019. Results: Regression results indicate bookmakers alter betting odds on matches involving clubs in the middle of the domestic league standings that have no chance of qualifying for the Champions/Europa Leagues or being relegated. In other words, bookmakers and bettors believe these teams will not put forth maximum effort to compete in and win these matches. We find similar results for teams already assured of relegation and teams that already clinched the league championship. Implications: Overall, the research contributes to literature analyzing the impact of incentives in sporting contests. As the world of professional sport and sports gambling become increasingly interconnected and integrated, the results of the study present opportunities not only for collaboration between the parties but also an increased understanding on the role league design has on individual match outcomes.},
author = {Feddersen, Arne and Humphreys, Brad R. and Soebbing, Brian P.},
issn = {1746031X},
journal = {European Sport Management Quarterly},
keywords = {Tournament theory,football,incentives,sports betting},
number = {forthcom.},
pages = {1--17},
publisher = {Taylor & Francis},
title = {{Contest incentives, team effort, and betting market outcomes in European football}},

year = {2021}
}
@article{Fornwagner2019,
abstract = {This paper analyzes data from a tournament, namely the National Hockey League regular scheduled season of games, which provides incentives to increase effort in order to reach the playoffs and incentives to decrease effort once a team has been eliminated from playoff considerations because of the entry draft. Our results show that teams react to these dual incentives - they win more games when there is still a chance to reach the playoffs and lose more after being eliminated from playoff considerations. One can argue that losing more games after having no more chance to reach the playoffs could be the result of lower motivation or disappointment. This is the first study to show that this is not the only explanation for a higher amount of lost games. Instead, we find that there is a concrete strategy behind losing.},
author = {Fornwagner, Helena},
issn = {01674870},
journal = {Journal of Economic Psychology},
keywords = {3720,Dual incentives,National Hockey League,Natural field experiment,Sport data,Strategic losing,Tournaments},
pages = {1--12},
title = {{Incentives to lose revisited: The NHL and its tournament incentives}},
volume = {75},
year = {2019}
}
@article{Scarf2009,
abstract = {Operational Research may be used to compare different designs for a sporting contest or tournament. This paper considers a methodology for this purpose. We propose a number of tournament metrics that can be used to measure the success of a sporting contest or tournament, and describe how these metrics may be evaluated for a particular tournament design. Knowledge of these measures can then be used to compare competing designs, such as round-robin, pure knockout and hybrids of these designs. We show, for example, how the design of the tournament influences the outcome uncertainty of the tournament and the number of unimportant matches within the tournament. In this way, where new designs are proposed, the implications of these designs may be explored within a modelling paradigm. In football (soccer), the UEFA Champions League has adopted a number of designs over its 50 year history; the design of the tournament has been modified principally in response to the changing demands of national league football and television - the paper uses this particular tournament to illustrate the methodology. {\textcopyright} 2008 Elsevier B.V. All rights reserved.},
author = {Scarf, Philip and Yusof, Muhammad Mat and Bilbao, Mark},
issn = {03772217},
journal = {European Journal of Operational Research},
keywords = {Competitive balance,OR in sport,Simulation,Sports tournaments},
number = {1},
pages = {190--198},
publisher = {Elsevier B.V.},
title = {{A numerical study of designs for sporting contests}},
volume = {198},
year = {2009}
}
@article{Scarf2008,
abstract = {A quantitative measure of "match importance" is useful in a number of decision problems, for example: as a metric in tournament design; for selecting matches for broadcasting; for scheduling matches in a tournament; and for assigning referees. To date measures of match importance used in such analyses have been relatively na{\"{i}}ve. We discuss a general measure that considers the effect of a particular match on the end of tournament position, given the results of all other matches, some played, some predicted. We use logistic regression to predict matches and Monte Carlo simulation to compute the match importance measure, and apply these to soccer matches in the English Football Association Premier League. {\textcopyright} 2006 Elsevier Ltd. All rights reserved.},
author = {Scarf, Philip A. and Shi, Xin},
issn = {03050548},
journal = {Computers and Operations Research},
keywords = {Championship significance,Contest,Outcome uncertainty,Sport},
number = {7},
pages = {2406--2418},
title = {{The importance of a match in a tournament}},
volume = {35},
year = {2008}
}
@article{Corona2017,
abstract = {Identifying the decisive matches in international football tournaments is of great relevance for a variety of decision makers such as organizers, team coaches and/or media managers. This paper addresses this issue by analyzing the role of the statistical approach used to estimate the outcome of the game on the identification of decisive matches on international tournaments for national football teams. We extend the measure of decisiveness proposed by Geenens (2014) in order to allow us to predict or evaluate the decisive matches before, during and after a particular game on the tournament. Using information from the 2014 FIFA World Cup, our results suggest that Poisson and kernel regressions significantly outperform the forecasts of ordered probit models. Moreover, we find that although the identification of the most decisive matches is independent of the model considered, the identification of other key matches is model dependent. We also apply this methodology to identify the favorite teams and to predict the most decisive matches in 2015 Copa America before the start of the competition. Furthermore, we compare our forecast approach with respect to the original measure during the knockout stage.},
author = {Corona, Francisco and {De Dios Tena Horrillo}, Juan and Wiper, Michael Peter},
issn = {15590410},
journal = {Journal of Quantitative Analysis in Sports},
keywords = {Kernel regression,Poisson model,entropy,game importance,ordered probit model},
number = {1},
pages = {11--23},
title = {{On the importance of the probabilistic model in identifying the most decisive games in a tournament}},
volume = {13},
year = {2017}
}
@article{Schilling1994,
author = {Schilling, Mark F .},
journal = {Mathematics Magazine},
number = {4},
pages = {282--288},
title = {{The Importance of a Game}},
volume = {67},
year = {1994}
}
@article{Geenens2014,
abstract = {In sport tournaments in which teams are matched two at a time, it is useful for a variety of reasons to be able to quantify how important a particular game is. The need for such quantitative information has been addressed in the literature by several more or less simple measures of game importance. In this paper, we point out some of the drawbacks of those measures and we propose a different approach, which rather targets how decisive a game is with respect to the final victory. We give a definition of this idea of game decisiveness in terms of the uncertainty about the eventual winner prevailing in the tournament at the time of the game. As this uncertainty is strongly related to the notion of entropy of a probability distribution, our decisiveness measure is based on entropy-related concepts. We study the suggested decisiveness measure on two real tournaments, the 1988 NBA Championship Series and the UEFA 2012 European Football Championship (Euro 2012), and we show how well it agrees with what intuition suggests. Finally, we also use our decisiveness measure to objectively analyse the recent UEFA decision to expand the European Football Championship from 16 to 24 nations in the future, in terms of the overall attractiveness of the competition. {\textcopyright} 2013 Elsevier B.V. All rights reserved.},
author = {Geenens, Gery},
issn = {03772217},
journal = {European Journal of Operational Research},
keywords = {Applied probability,Entropy,Game importance,OR in sports,Uncertainty modeling},
number = {1},
pages = {156--168},
publisher = {Elsevier B.V.},
title = {{On the decisiveness of a game in a tournament}},
volume = {232},
year = {2014}
}
@article{Goossens2012,
abstract = {Recently, most clubs in the highest Belgian football division have become convinced that the format of their league should be changed. Moreover, the TV station that broadcasts the league is pleading for a more attractive competition. However, the clubs have not been able to agree on a new league format, mainly because they have conflicting interests. In this paper, we compare the current league format, and three other formats that have been considered by the Royal Belgian Football Association. We simulate the course of each of these league formats, based on historical match results. We assume that the attractiveness of a format is determined by the importance of its games; our importance measure for a game is based on the number of teams for which this game can be decisive to reach a given goal. Furthermore, we provide an overview of how each league format aligns with the expectations and interests of each type of club. {\textcopyright} 2010 Springer Science+Business Media, LLC.},
author = {Goossens, Dries R. and Beli{\"{e}}n, Jeroen and Spieksma, Frits C.R.},
issn = {02545330},
journal = {Annals of Operations Research},
keywords = {Football,Match importance,Optimization,Simulation,Tournament design},
number = {1},
pages = {223--240},
title = {{Comparing league formats with respect to match importance in Belgian football}},
volume = {194},
year = {2012}
}
@article{VanDijk2001,
abstract = {In the reported experiment different payment schemes are examined on their incentive effects. Payments based on individual, team and relative performance are compared. Subjects conducted computerized tasks that required substantial effort. The results show that individual and team payment induced the same effort levels. In team production free-riding occurred, but it was compensated by many subjects providing more effort than in case of individual pay. Effort was higher, but more variable in tournaments, while in case of varying abilities workers with relatively low ability worked very hard and drove up effort of the others. Finally, attitudes towards work and other workers differed strongly between conditions. {\textcopyright} 2001 Elsevier Science B.V. All rights reserved.},
author = {{Van Dijk}, Frans and Sonnemans, Joep and {Van Winden}, Frans},
isbn = {3102052541},
issn = {00142921},
journal = {European Economic Review},
keywords = {Experiment,Payment schemes},
number = {2},
pages = {187--214},
title = {{Incentive systems in a real effort experiment}},
volume = {45},
year = {2001}
}

@article{Szymanski2003,
author = {Szymanski, Stefan},
isbn = {9780230274273},
journal = {Journal of Economic Literature},
pages = {1137--1187},
title = {{The economic design of sporting contests}},
volume = {XLI},
year = {2003}
}

@article{Sheremeta2018,
abstract = {Group contests are ubiquitous. Some examples include warfare between countries, competition between political parties, team-incentives within firms, and rent-seeking. In order to succeed, members of the same group have incentives to cooperate with each other by expending effort. However, since effort is costly, each member also has an incentive to abstain from expending any effort and instead free ride on the efforts of other members. Contest theory predicts that the intensity of competition between groups and the amount of free riding within groups depend on the group size, sharing rule, group impact function, contest success function, and heterogeneity of players. We review experimental studies testing these theoretical predictions. Most studies find significant over-expenditure of effort relative to the theory and significant heterogeneity of behavior within and between groups. Also, most studies find support for the comparative statics predictions of the theory (with the exception of the “group size paradox”). Finally, studies show that there are effective mechanisms that can promote within-group cooperation and conflict resolution mechanisms that can de-escalate and potentially eliminate between-group conflict.},
author = {Sheremeta, Roman M.},
issn = {14676419},
journal = {Journal of Economic Surveys},
keywords = {Contests,Experiments,Groups},
number = {3},
pages = {683--704},
title = {{Behavior in Group Contests: a Review of Experimental Research}},
volume = {32},
year = {2018}
}
@article{Rosen1985,
abstract = {The role of rewards for maintaining performance incentives in multi-stage, sequential games of survival is studied. The sequential structure is a statistical design-of-experiments for selecting and ranking contestants. It},
author = {Rosen, Sherwin},
journal = {National Bureau of Economic Research Working Paper Series},
number = {1668},
title = {{Prizes and Incentives in Elimination Tournaments}},
year = {1985}
}
@article{Prendergast1999,
author = {Prendergast, Canice},
issn = {00220515},
journal = {Journal of Economic Literature},
number = {1},
pages = {7--63},
title = {{The provision of incentives in firms}},
volume = {37},
year = {1999}
}
@article{Lazear2000,
author = {Lazear, Edward},
journal = {The American Economic Review, Papers and Proceedings},
number = {2},
pages = {410--414},
title = {{The Power of Incentives}},
volume = {90},
year = {2000}
}
@article{KleinTeeselink2020,
abstract = {This paper examines how within-match variation in incentives affects the performance of darts players. The game of darts offers an attractive naturally occurring research setting, because performance can be observed at the individual level and without obscuring effects of risk considerations and behavior of others. We analyze four data sets covering a total of 29,381 darts matches of professional, amateur, and youth players. We find that amateur and youth players display a sizable performance decrease at decisive moments. Professional players appear less susceptible of such choking under pressure. Our results speak to a growing literature on the limits of increasing incentives as a recipe for better performance.},
author = {{Klein Teeselink}, Bouke and {Potter van Loon}, Rogier J.D. and van den Assem, Martijn J. and van Dolder, Dennie},
issn = {01672681},
journal = {Journal of Economic Behavior and Organization},
keywords = {Choking under pressure,Darts,Incentives,Performance pressure},
number = {August},
pages = {38--52},
title = {{Incentives, performance and choking in darts}},
volume = {169},
year = {2020}
}
@article{Iqbal2019,
abstract = {The discouragement effect of being the lagging player in multi-stage contests is a well-documented phenomenon. In this study, we utilize data from 2447 Davis Cup matches in team tennis tournaments to test the effect of being behind or ahead on individuals' performance with and without intermediate prizes. Using several different strategies to disentangle the effect of being ahead in the interim score from the effect of selection, we find the usual discouragement effect. However, the discouragement effect disappears after the introduction of intermediate prizes in the form of ranking points. The lagging favorite had close to a 20-percentage point greater probability of winning compared to matches without such a prize. We show that this result is not driven by the selection of better players into tournaments with intermediate prizes. As predicted by previous theoretical studies, our empirical findings suggest that intermediate prizes may mitigate or even eliminate the ahead–behind effects that arise in multi-stage contests.},
author = {Iqbal, Hamzah and Krumer, Alex},
issn = {00142921},
journal = {European Economic Review},
keywords = {Collective decision-making,Discouragement,Multi-stage contests,Tennis},
pages = {364--381},
publisher = {Elsevier B.V.},
title = {{Discouragement effect and intermediate prizes in multi-stage contests: Evidence from Davis Cup}},
volume = {118},
year = {2019}
}
@incollection{Goller2018,
author = {Goller, Daniel and Knaus, Michael C and Lechner, Michael and Okasa, Gabriel},
booktitle = {A Modern Guide to Sports Economics},
chapter = {22},
editor = {Koning, Ruud H. and Kesenne, Stefan},
pages = {335--355},
publisher = {Edward Elgar Publishing},
title = {{Predicting Match Outcomes in Football by an Ordered Forest Estimator}},
year = {2021}
}
@article{Ginsburgh2003,
author = {Ginsburgh, V. and van Ours, J.},
journal = {The American Economic Review},
number = {1},
pages = {289--296},
title = {{Expert Opinion and Compensation : Evidence from a Musical Competition}},
volume = {93},
year = {2003}
}
@article{Lin1991,
  title={Divergence measures based on the Shannon entropy},
  author={Lin, Jianhua},
  journal={IEEE Transactions on Information theory},
  volume={37},
  number={1},
  pages={145--151},
  year={1991},
  publisher={IEEE}
}

@book{Laffont2002,
author = {Laffont, Jean-Jacques and Martimort, David},
mendeley-groups = {MI_project},
publisher = {Princeton University Press},
title = {{The Theory of Incentives: The Principal-Agent Model}},
year = {2002}
}

@article{Cohen-Zada2017,
abstract = {We exploit a unique setting in which two professionals compete in a real-life tennis contest with high monetary rewards in order to assess how men and women respond to competitive pressure. Comparing their performance in low-stakes versus high-stakes situations, we find that men consistently choke under competitive pressure, but with regard to women the results are mixed. Furthermore, even if women show a drop in performance in the more crucial stages of the match, it is in any event about 50 percent smaller than that of men. These findings are robust to different specifications and estimation strategies.},
author = {Cohen-Zada, Danny and Krumer, Alex and Rosenboim, Mosi and Shapir, Offer Moshe},
issn = {01674870},
journal = {Journal of Economic Psychology},
keywords = {Choking,Competitive pressure,Gender,Performance,Tennis},
pages = {176--190},
publisher = {Elsevier B.V.},
title = {{Choking under pressure and gender: Evidence from professional tennis}},
volume = {61},
year = {2017}
}

@article{GK2020,
title = {Let's meet as usual: Do games played on non-frequent days differ? Evidence from top European soccer leagues},
journal = {European Journal of Operational Research},
volume = {286},
number = {2},
pages = {740-754},
year = {2020},
issn = {0377-2217},
author = {Daniel Goller and Alex Krumer},
keywords = {Performance, Schedule effects, Soccer},
abstract = {Balancing the allocation of games in sports competitions is an important organizational task that can have serious financial consequences. In this paper, we examine data from 10,142 soccer games played in the top German, Spanish, French, and English soccer leagues between 2007/2008 and 2016/2017. Using a machine learning technique for variable selection and applying a semi-parametric analysis of radius matching on the propensity score, we find that all four leagues have a lower attendance in games that take place on four non-frequently played days than those on three frequently played days. We also find that, in all leagues, there is a significantly lower home advantage for the underdog teams on non-frequent days. Our findings suggest that the current schedule favors underdog teams with fewer home games on non-frequent days. Therefore, to increase the fairness of the competitions, it is necessary to adjust the allocation of the home games on non-frequent days in a way that eliminates any advantage driven by the schedule. These findings have implications for the stakeholders of the leagues, referees’ and calendar committees as well as for coaches and players.}
}

@article{Alchian1972,
archivePrefix = {arXiv},
arxivId = {https://knowledgehub.unsse.org/es/knowledge-hub/relaciones-entre-los-ods-el-plan-para-una-decada-cooperativa-y-el-balance-social-en-cooperativas-2/},
author = {Alchian, Armen and Demsetz, Harold},
eprint = {/knowledgehub.unsse.org/es/knowledge-hub/relaciones-entre-los-ods-el-plan-para-una-decada-cooperativa-y-el-balance-social-en-cooperativas-2/},
journal = {The American Economic Review},
number = {5},
pages = {777--795},
primaryClass = {https:},
title = {{Production, Information Costs, and Economic Organization}},
volume = {62},
year = {1972}
}

@article{kahn2000sports,
  title={The sports business as a labor market laboratory},
  author={Kahn, Lawrence M},
  journal={Journal of economic perspectives},
  volume={14},
  number={3},
  pages={75--94},
  year={2000}
}

@article{bar2020ask,
  title={Ask not what economics can do for sports-Ask what sports can do for economics},
  author={Bar-Eli, Michael and Krumer, Alex and Morgulev, Elia},
  journal={Journal of Behavioral and Experimental Economics},
  volume={89},
  pages={101597},
  year={2020},
  publisher={Elsevier}
}

@article{lechner2020orf,
  title={orf: Ordered Random Forests},
  author={Lechner, Michael and Okasa, Gabriel},
  journal={R package version 0.1.3},
  year={2020}
}

@article{danziger2011extraneous,
  title={Extraneous factors in judicial decisions},
  author={Danziger, Shai and Levav, Jonathan and Avnaim-Pesso, Liora},
  journal={Proceedings of the National Academy of Sciences},
  volume={108},
  number={17},
  pages={6889--6892},
  year={2011},
  publisher={National Acad Sciences}
}

@article{de2005save,
  title={Save the last dance for me: Unwanted serial position effects in jury evaluations},
  author={de Bruin, W{\"a}ndi Bruine},
  journal={Acta psychologica},
  volume={118},
  number={3},
  pages={245--260},
  year={2005},
  publisher={Elsevier}
}

@article{klumpp2006primaries,
  title={Primaries and the New Hampshire effect},
  author={Klumpp, Tilman and Polborn, Mattias K},
  journal={Journal of Public Economics},
  volume={90},
  number={6-7},
  pages={1073--1114},
  year={2006},
  publisher={Elsevier}
}

@article{preston2003cheating,
  title={Cheating in contests},
  author={Preston, Ian and Szymanski, Stefan},
  journal={Oxford review of economic policy},
  volume={19},
  number={4},
  pages={612--624},
  year={2003},
  publisher={Oxford University Press}
}

@article{taylor2002losing,
  title={Losing to win: Tournament incentives in the National Basketball Association},
  author={Taylor, Beck A and Trogdon, Justin G},
  journal={Journal of Labor Economics},
  volume={20},
  number={1},
  pages={23--41},
  year={2002},
  publisher={The University of Chicago Press}
}

@article{knight2010momentum,
  title={Momentum and social learning in presidential primaries},
  author={Knight, Brian and Schiff, Nathan},
  journal={Journal of political economy},
  volume={118},
  number={6},
  pages={1110--1150},
  year={2010},
  publisher={University of Chicago Press Chicago, IL}
}

@book{mayer2003front,
  title={The front-loading problem in presidential nominations},
  author={Mayer, William G and Busch, Andrew E},
  year={2003},
  publisher={Brookings Institution Press}
}

@article{ridout2008importance,
  title={The importance of being early: Presidential primary front-loading and the impact of the proposed western regional primary},
  author={Ridout, Travis N and Rottinghaus, Brandon},
  journal={PS: Political Science \& Politics},
  volume={41},
  number={1},
  pages={123--128},
  year={2008},
  publisher={Cambridge University Press}
}

@article{harris1987racing,
  title={Racing with uncertainty},
  author={Harris, Christopher and Vickers, John},
  journal={The Review of Economic Studies},
  volume={54},
  number={1},
  pages={1--21},
  year={1987},
  publisher={Wiley-Blackwell}
}

@article{apesteguia2010psychological,
  title={Psychological pressure in competitive environments: Evidence from a randomized natural experiment},
  author={Apesteguia, Jose and Palacios-Huerta, Ignacio},
  journal={American Economic Review},
  volume={100},
  number={5},
  pages={2548--64},
  year={2010}
}

@article{buraimo2022armchair,
  title={Armchair fans: Modelling audience size for televised football matches},
  author={Buraimo, Babatunde and Forrest, David and McHale, Ian G and Tena, J d D},
  journal={European Journal of Operational Research},
  volume={298},
  number={2},
  pages={644--655},
  year={2022},
  publisher={Elsevier}
}

@Article{Nielsen2021,
AUTHOR = {Nielsen, Frank},
TITLE = {On a Generalization of the Jensen–Shannon Divergence and the Jensen–Shannon Centroid},
JOURNAL = {Entropy},
VOLUME = {22},
YEAR = {2020},
NUMBER = {2},
ARTICLE-NUMBER = {221},
ISSN = {1099-4300},
ABSTRACT = {The Jensen&ndash;Shannon divergence is a renown bounded symmetrization of the Kullback&ndash;Leibler divergence which does not require probability densities to have matching supports. In this paper, we introduce a vector-skew generalization of the scalar    &alpha;   -Jensen&ndash;Bregman divergences and derive thereof the vector-skew    &alpha;   -Jensen&ndash;Shannon divergences. We prove that the vector-skew    &alpha;   -Jensen&ndash;Shannon divergences are f-divergences and study the properties of these novel divergences. Finally, we report an iterative algorithm to numerically compute the Jensen&ndash;Shannon-type centroids for a set of probability densities belonging to a mixture family: This includes the case of the Jensen&ndash;Shannon centroid of a set of categorical distributions or normalized histograms.}
}

@incollection{mcfadden_conditional_1974,
	address = {New York},
	title = {Conditional logit analysis of qualitative choice behavior},
	language = {en},
	booktitle = {Frontiers in {Econometrics}},
	publisher = {Academic press},
	author = {McFadden, Daniel},
	editor = {Zarembka, Paul},
	year = {1974},
	pages = {105--142}
}

@data{Dataverse,
author = {Goller, Daniel and Heiniger, Sandro},
publisher = {Harvard Dataverse},
title = {{Replication code and results for: 'A general framework to quantify the event importance in multi-event contests'}},
UNF = {UNF:6:LktvG47zGdoPNd+HgJeKLA==},
year = {2022},
version = {V1},
doi = {https://doi.org/10.7910/DVN/F3QA9N}
}

\vfill

\bibliographystyle{apacite}
\bibliography{main}

\newpage

\appendix
\section{Details to Algorithm~\ref{algo: general event importance}}\label{app:Details to algorithm}
The Algorithm~\ref{algo: general event importance} works as follows: Line~\ref{line:loop possible states} loops over all possible outcomes $\mathbcal{Y}_{e_{t,i}}$ for the particular event $e_{t,i}$. This loop calculates the reward distribution at the end of the contest once for each possible outcome. Note that all variables in this loop are depending on the value of the loop iterator $\mathbcal{y}_{e_{t, i}}$. To improve readability, we denote this by the superscript * instead of the more intuitive subscript $\mathbcal{y}_{e_{t, i}}$ which prevents an additional subscript level. All variables are subject to a probability distribution which is formed by the continuous evaluations of the outcome model out(). \par
At the beginning of each iteration of the loop, the outcome for event $e_{t,j}^*$ is set to the iterator outcome $\mathbcal{y}_{e_{t,i}}$. In case there are other simultaneous events $e_{t,j}$ at time $t$, the loop in line~\ref{line:loop same time events} evaluates their outcome probabilities according to the outcome model $\text{out}(x_{e_{t,j}})$. Next, the outcome-dependent elements of the future schedule and covariates are generated according to the outcome probabilities of $y_{t^-}^*$, as well as $\mathcal{T}_{t^-}$ and $x_{t^-}$. \par
Line~\ref{line:future events} loops over the future times $t'$ until the end of the contest $\mathcal{T_{t^+}}^*$. Note that this set can be adapted or extended at run time by the gen() steps in case of an outcome-dependent schedule. In each iteration step, first, the outcome probabilities $y_{e_{t'}}^*$ for all events at this particular time $t'$ are determined according to the outcome model, and then the contest schedule and the covariates are generated, in case they depend on the past outcomes. When the end of the contest $\mathcal{T}^*$ has been reached, line~\ref{line:final ranking reward} determines the reward distribution $r_{k,\mathbcal{y}_{e_{t, i}}}$ for contestant $k$ based on the outcomes of all single events in the contest. \par
The final step, after the outer loop over all possible outcomes has finished, calculates the distance between the reward distributions which quantifies the event importance $\text{EI}_{e_{t, i},k}$.

\newpage
\section{2020 Democratic Party presidential primaries}\label{App:frontloading}
\begin{table}[H] 
	\footnotesize
	\centering 
	\caption{Data and results summary on 2020 Democratic Party presidential primaries.} 
	\label{tab:Primaries summary} 
	\def\arraystretch{0.7}
	\begin{tabular}{llrrrr} 
		\toprule
		& & & \multicolumn{3}{c}{ avg. EI}\\[0.4ex]\cline{4-6}\\[-1ex]
		State/Territory & Election Date & Delegates & Actual & Rank increase & Random \\ 
		\midrule 
Alabama & March 3, 2020 & 52 & 0.046 & 0.071 & 0.048 \\
Alaska & April 10, 2020 & 15 & 0.012 & 0.027 & 0.018 \\
American Samoa & March 3, 2020 & 6 & 0.009 & 0.060 & 0.010 \\
Arizona & March 17, 2020 & 67 & 0.054 & 0.081 & 0.060 \\
Arkansas & March 3, 2020 & 31 & 0.028 & 0.045 & 0.031 \\
California & March 3, 2020 & 415 & 0.462 & 0.386 & 0.388 \\
Colorado & March 3, 2020 & 67 & 0.061 & 0.079 & 0.060 \\
Connecticut & August 11, 2020 & 60 & 0.044 & 0.076 & 0.058 \\
Delaware & July 7, 2020 & 21 & 0.017 & 0.037 & 0.023 \\
Democrats Abroad & March 10, 2020 & 13 & 0.011 & 0.023 & 0.014 \\
District of Columbia & June 2, 2020 & 20 & 0.015 & 0.036 & 0.019 \\
Florida & March 17, 2020 & 219 & 0.182 & 0.189 & 0.194 \\
Georgia & June 9, 2020 & 105 & 0.083 & 0.105 & 0.096 \\
Guam & June 6, 2020 & 7 & 0.006 & 0.060 & 0.010 \\
Hawaii & May 22, 2020 & 24 & 0.019 & 0.043 & 0.029 \\
Idaho & March 10, 2020 & 20 & 0.017 & 0.036 & 0.020 \\
Illinois & March 17, 2020 & 155 & 0.127 & 0.140 & 0.139 \\
Indiana & June 2, 2020 & 82 & 0.060 & 0.082 & 0.074 \\
Iowa & February 3, 2020 & 41 & 0.227 & 0.056 & 0.039 \\
Kansas & May 2, 2020 & 39 & 0.031 & 0.056 & 0.039 \\
Kentucky & June 23, 2020 & 54 & 0.040 & 0.070 & 0.054 \\
Louisiana & July 11, 2020 & 54 & 0.038 & 0.069 & 0.045 \\
Maine & March 3, 2020 & 24 & 0.022 & 0.042 & 0.022 \\
Maryland & June 2, 2020 & 96 & 0.072 & 0.097 & 0.089 \\
Massachusetts & March 3, 2020 & 91 & 0.082 & 0.092 & 0.079 \\
Michigan & March 10, 2020 & 125 & 0.107 & 0.118 & 0.112 \\
Minnesota & March 3, 2020 & 75 & 0.067 & 0.076 & 0.072 \\
Mississippi & March 10, 2020 & 36 & 0.030 & 0.053 & 0.035 \\
Missouri & March 10, 2020 & 68 & 0.057 & 0.069 & 0.062 \\
Montana & June 2, 2020 & 19 & 0.014 & 0.033 & 0.020 \\
Nebraska & May 12, 2020 & 29 & 0.023 & 0.042 & 0.031 \\
Nevada & February 22, 2020 & 36 & 0.144 & 0.053 & 0.036 \\
New Hampshire & February 11, 2020 & 24 & 0.110 & 0.043 & 0.025 \\
New Jersey & July 7, 2020 & 126 & 0.094 & 0.119 & 0.105 \\
New Mexico & June 2, 2020 & 34 & 0.025 & 0.049 & 0.030 \\
New York & June 23, 2020 & 274 & 0.206 & 0.233 & 0.246 \\
North Carolina & March 3, 2020 & 110 & 0.100 & 0.108 & 0.101 \\
North Dakota & March 10, 2020 & 14 & 0.012 & 0.025 & 0.016 \\
Northern Marianas & March 14, 2020 & 6 & 0.007 & 0.077 & 0.008 \\
Ohio & April 28, 2020 & 136 & 0.110 & 0.128 & 0.123 \\
Oklahoma & March 3, 2020 & 37 & 0.033 & 0.055 & 0.035 \\
Oregon & May 19, 2020 & 61 & 0.048 & 0.075 & 0.055 \\
Pennsylvania & June 2, 2020 & 186 & 0.140 & 0.173 & 0.160 \\
Puerto Rico & July 12, 2020 & 51 & 0.031 & 0.069 & 0.052 \\
Rhode Island & June 2, 2020 & 26 & 0.019 & 0.047 & 0.026 \\
South Carolina & February 29, 2020 & 54 & 0.188 & 0.072 & 0.053 \\
South Dakota & June 2, 2020 & 16 & 0.012 & 0.029 & 0.017 \\
Tennessee & March 3, 2020 & 64 & 0.058 & 0.079 & 0.062 \\
Texas & March 3, 2020 & 228 & 0.215 & 0.202 & 0.207 \\
Utah & March 3, 2020 & 29 & 0.027 & 0.044 & 0.030 \\
Vermont & March 3, 2020 & 16 & 0.015 & 0.028 & 0.015 \\
Virgin Islands & June 6, 2020 & 7 & 0.006 & 0.054 & 0.010 \\
Virginia & March 3, 2020 & 99 & 0.089 & 0.100 & 0.092 \\
Washington & March 10, 2020 & 89 & 0.076 & 0.090 & 0.080 \\
West Virginia & June 9, 2020 & 28 & 0.020 & 0.049 & 0.029 \\
Wisconsin & April 7, 2020 & 84 & 0.069 & 0.085 & 0.072 \\
Wyoming & April 17, 2020 & 14 & 0.012 & 0.024 & 0.016 \\[0.4ex]
		\bottomrule
	\end{tabular} 
\end{table}

\newpage
\section{European football leagues}

\subsection{League structures}\label{app:Appendix_European_football leagues}
We focus on the seven major European football leagues: The Dutch 'Eredivisie', the English 'Premier League', the French 'Ligue 1', the German 'Bundesliga 1', the Italian 'Serie A', the Portuguese 'Primeira Liga', and the Spanish 'Primera Division'. These leagues are leaders in terms of success in international competitions, financial capabilities, competitiveness and stadium attendance.\par 
The investigated European football leagues are organised in a double round-robin tournament, in which clubs face each other twice during the season, once at each home ground. Thus, a season consists of $2(n-1)$ match days, where $n$ is the number of competing teams in the league - that is 16, 18, or 20. Table~\ref{tab:Data summary} summarises the number of teams by the league, the number of observed distinct teams, and the total number of matches in the covered time period.\par
\begin{table}[ht]
	\footnotesize
	\centering 
	\caption{Data summary on leagues.} 
	\label{tab:Data summary} 
	\def\arraystretch{1.2}
	\begin{tabular}{lccc} 
		\toprule
		League & Teams per season & Total matches & Observed teams \\ 
		\midrule 
		Bundesliga 1 & 18 & 3'060 & 28 \\
		Eredivisie & 18 & 3'060 & 26 \\
		La Liga & 20 & 3'800 & 35 \\
		Ligue 1 & 20 & 3'800 & 37 \\
		Premier League & 20 & 3'800 & 36 \\
		Primeira Liga  & 16 or 18 & 2'730 & 30 \\
		Serie A & 20 & 3'800 & 34 \\[0.4ex]
		\bottomrule
	\end{tabular} 
\end{table}
To level out the encounters over the whole season, the season is split into two halves in which each pairing is played once. With the exception of the English Premier League, all of the covered leagues feature a mid-season break around December/January -- a season usually starts in August/September and finishes in May. On a match day, or round, $\frac{n}{2}$ matches are played with every team competing in one pairing.\par

Winning a match is rewarded with 3 points, draws are valued with 1 point, and losses with 0 points.. At the end of the season, when all matches have been played, the points of all match days are added up and the teams are ranked in descending order in the total sum of points. In the event of a tie in the ranking, the number of goals scored/achieved or the result of the match pairings between the teams concerned shall be decisive, depending on the league-specific tie-breaker rules.

\subsection{List of covariates}\label{app: List of covariates}
Table~\ref{tab:Covariates list} lists all covariates used in the dataset and whether they have been selected by the model selection procedure. \par
{
    \footnotesize
    \def\arraystretch{0.6}

    	\begin{xltabular}{14cm}{Xlcc}
    	   \caption{List of all covariates in the data.\label{tab:Covariates list} } \\
            \toprule  & Description  & Type & Selected\\  \midrule 
            \endfirsthead
            
            \multicolumn{4}{c}{\tablename\ \thetable{} -- continued from previous page} \\
            \toprule & Description  & Type & Selected\\  \midrule 
            \endhead
            
            \midrule \multicolumn{4}{|r|}{{Continued on next page}} \\
            \endfoot
        
            \hline
            \endlastfoot
        
    		\multicolumn{4}{l}{\textbf{One variable per game}}\\
    		\midrule
    		& Season & numeric & - \\
            & Bundesliga & binary & - \\
            & Eredivisie & binary & - \\
            & La Liga & binary & - \\
            & Ligue 1 & binary & - \\
            & Premier League & binary & yes \\
            & Primeira Liga & binary & - \\
            & Serie A & binary & - \\
    		\midrule
            & Kick-off [hour] & numeric & - \\
            & Day of the week & numeric & yes \\
            & Weekend match (Friday to Sunday) & binary & - \\
            & Match during a public holiday & binary & - \\
            & Match during Christmas holidays & binary & - \\
            & Match before an international competition break & binary & - \\
            & Match after an international competition break & binary & - \\
            & Match before a national team break & binary & - \\
            & Match after a national team break & binary & - \\
            & Match during Africa Cup & binary & - \\
            & Match after Asian Nations Cup & binary & - \\
    		\midrule
            & Stadium capacity & binary & - \\
            & Travel distance between home and away [km] & numeric & - \\
            & Travel distance between home and away [min] & numeric & - \\
    		\midrule
            & Season before World Cup & binary & -\\
            & Season after World Cup & binary & yes \\
            & Season before European Championships & binary & -\\
            & Season after European Championships & binary & -\\
    		\midrule
            & Home total market value (MV) - Away total MV & numeric & yes \\
            & Home average MV - Away average MV & numeric & yes \\
            & Home standardized MV - Away standardized MV & numeric & yes \\
            & Home total MV / Away total MV & numeric & - \\
            & Home standardized MV / Away standardized MV & numeric & - \\
     		\midrule
            & Home average height - Away average height & numeric & - \\
            & Home avg. height Top-11 - Away avg. height Top-11 & numeric & yes \\
    		\midrule
            \multicolumn{4}{l}{\textbf{Variables once for each team}}\\
    		\midrule
            & Team plays Champions League this season & binary & - \\
            & Team plays Europa League this season & binary & - \\
            & Team plays Champions or Europa League this season & binary & home \\
    		\midrule
            & Size of squad & numeric & home \\
            & Number of new players this season & numeric & - \\
            & Days since last match in any competition & numeric & - \\
            & Days until next match in any competition & numeric & - \\
    		\midrule
            & Points in last league match & numeric & home \\
            & Average points in last 2 league matches & numeric & home \\
            & Average points in last 3 league matches & numeric & away \\
            & Average points in last 4 league matches & numeric & - \\
            & Average points in all previous league matches & numeric & home \\
    		\midrule
            & Total market value (MV) of all players & numeric & home \\
            & Average MV & numeric & - \\
            & Median MV & numeric & away \\
            & Standard deviation of MV & numeric & - \\
            & Skewness of MV & numeric & - \\
            & Total MV standardized by league and season & numeric & - \\
            & Herfindahl-Hirschman-Index of MV & numeric & - \\
            & Normalized Herfindahl-Hirschman-Index of MV & numeric & - \\
            & Top 1-3 MV / Top 9-11 MV & numeric & away \\
            & Top 1-3 MV / Top 12-14 MV & numeric & - \\
            & Top 1-3 MV / Top 12-14 MV & numeric & - \\
            & Standard deviation of MV / Average MV & numeric & - \\
    		\midrule
            & Share all players right-footed & numeric & - \\
            & Share all players left-footed & numeric & - \\
            & Share all players both feet & numeric & away \\
            & Share Top-3 players right-footed & numeric & away \\
            & Share Top-3 players left-footed & numeric & - \\
            & Share Top-3 players both feet & numeric & - \\
            & Share Top-11 players right-footed & numeric & - \\
            & Share Top-11 players left-footed & numeric & - \\
            & Share Top-11 players both feet & numeric & home \\
    		\midrule
            & Average height all players & numeric & - \\
            & Min height all players & numeric & - \\
            & Max height all players & numeric & - \\
            & Standard deviation of height all players & numeric & - \\
            & Average height Top-11 & numeric & - \\
            & Min height Top-11 & numeric & - \\
            & Max height Top-11 & numeric & - \\
            & Standard deviation of height Top-11 & numeric & - \\
    		\midrule
            & Min age & numeric & home \& away \\
            & Max age & numeric & - \\
            & Average age & numeric & - \\
            & Median age & numeric & home \& away \\
            & Standard deviation age & numeric & - \\
            & Standard deviation age / Average age & numeric & - \\
            & Average age of Top-11 & numeric & home \\
            & Average age of 1-11 / Average age of 12-21 & numeric & home \\
            & Average age of players above 20y & numeric & away \\
            \bottomrule 
    \end{xltabular}
    
}

\subsection{Specific framework algorithm} \label{app:Specific framework algorithm}

\subsubsection{Application-specific algorithm}
\begin{algorithm}[H]
    \DontPrintSemicolon
    \caption{AllEventImportances}\label{algo:specific event importance}
    \SetKwComment{Comment}{$\triangleright$}{}
    \KwData{$\mathcal{T}, x_{\text{VarSel}}, y_{\text{VarSel}}, x, y$}
    \KwResult{Event importance for all leagues and seasons}
    \Begin{
        $x \gets \text{VarSel}(x, x_{\text{VarSel}}, y_{\text{VarSel}})$\;
        iter $\gets$ 1\;
        \Repeat{$\text{iter}>3$}{
            \uIf{\text{iter}=1}{
                $x_{\text{train}} \gets x$\;
            }
            \Else{
                $x_{\text{train}} \gets \{x,\text{EI}\}$\;
            }
            $\text{ORF} \gets \text{trainORF}(x_{\text{train}},y) \qquad \triangleright$ {\small Alternatively an ordered logit/probit}\;
            \ForAll{LS in \{Leagues $\times$ Seasons\}}{
                \ForAll{$t$ in $\mathcal{T}^{\text{LS}}$, \textit{backwards}}{
                    \ForAll{$e_{t,i}$ in $e_t$}{
                        \ForAll{$\mathbcal{y}$ in \{H,D,A\}}{
                            \uIf{$(t\neq t_{\max}) \land (y_{e_{t,i}}=\mathbcal{y}) \land (\lvert e_t \rvert=1)$}{
                                $r_{k,\mathbcal{y}} \gets r_{k,e_{t+1}}$
                            }
                            \Else{
                                $y_{e_{t,i}}^* \gets \mathbcal{y}$\;
                                \ForAll{$e_{t,j}$ in $e_t$ with $j\neq i$}{
                                    $y_{e_{t,j}}^* \gets       \text{drawOutcome}(\text{ORF}(x_{e_{t,j}}))$\;
                                }
                                $x_{t^+}^* \gets \text{gen}(\mathcal{T}_{t^-},x_{t^-},y_{t^-}^*)$\;
                                \ForAll{$t'$ in $\mathcal{T}_{t^+}^{\text{LS}}$}{
                                    $y_{e_{t'}}^* \gets \text{drawOutcome}(\text{ORF}(x_{e_{t'}}^*))$\;
                                    $x_{t'^+}^* \gets \text{gen}(\mathcal{T}^{\text{LS}},x_{t^{'-}}^*,y_{t^{'-}}^*)$\;
                                }
                                $r_{\mathbcal{y}} \gets \text{rew}\bigcup_{\mathcal{T}^{\text{LS}}} y_e^*)$\; 
                            }
                        }
                        $r_{e_{t,i}} \gets \{z_{H},z_{D},z_{A}\}$ \;
                        \ForAll{$k$ in \{Home team, Away team\}}{
                            $\text{EI}_{e_{t,i},k} \gets \text{JSD}(r_{k,e_{t,i}},\text{ORF}(x_{e_{t,i}}))$\;
                        }
                    }
                    $\text{EI}_t \gets \bigcup_{e_t} \text{EI}_{e_{t,i},k}$\;
                }
                $\text{EI}_{\mathcal{T}^{\text{LS}}} \gets \bigcup_{\mathcal{T}^{\text{LS}}} \text{EI}_{e_t}$\;
            }
             $\text{EI} \gets \bigcup_{LS} \text{EI}_{\mathcal{T}^{\text{LS}}}$\;
            iter $\gets$ iter$+1$\;
          }
      \Return{$\text{EI}$}
    }
\end{algorithm}
\vspace{6 pt}

Algorithm~\ref{algo:specific event importance} describes the determination of all the event importance values in the specific framework of the presented application.\par

We estimate the EI in an iterative approach as described in Section~\ref{sec:Iterative approximation of event importance values} as the EI values are an important feature in the outcome model. The comparison of estimated EI values after different numbers of iterations in Appendix~\ref{app:Compare EI by iteration} indicates that three iteration steps are already enough to obtain a sufficient convergence of the EI estimation. The approximation of the final ranking distribution with a Monte Carlo simulation as described in Section~\ref{sec:Approximation of the probability distributions} is performed with \textit{$N_\text{MC}=7500$} runs.

\subsubsection{End-of-season rewards} \label{app:End-of-season rewards}

The ultimate goal of each team is to become champions of the respective season. However, only few teams are usually capable of participating in the race for the first place. Yet, the end-of-season ranking is relevant for various matters. First, the winner is crowned champion of the respective season. 
Second, the participants of the international competitions for the next season will be determined. There are multiple international competitions in European club football. The UEFA Champions League is the highest-rated competition, where the best-ranked teams in the leagues participate. The next best-positioned teams participate in the UEFA Europa League. The distribution of the starting places for the upcoming season granted by UEFA by the league associations is determined by hard thresholds in the ranking. The number of allocated spots is dependent on the previous success of the association's teams and is defined in the UEFA Access List and determined by the UEFA 5-years ranking. The official access list for all years can be found in the UEFA document library: \url{https://www.uefa.com/insideuefa/documentlibrary/}.  \par
Third, relegation to lower leagues is decided in the fixed breakdown of the available spots. Depending on the league, relegation can be direct or decided by a play-off match-up between a candidate for relegation and a candidate for promotion, while direct relegation always applies to the lowest-placed candidates, if both variants exist. Fourth, financial rewards, i.\,e.\ money from broadcasting rights, are determined by the final league table. However, this gradation is less relevant for the team, the coaches, and the players. Becoming champions, qualifying for next year's European Cup or not being relegated to a lower league is what we assume is more in the focus of the involved entities. While relegation denotes a massive cut in financial benefits, reputation, and attractiveness for a club, participation in international tournaments, or becoming champion of the league tremendously boosts them. For a discussion on the financial dimensions consult \citeA{GK2020}.

Table~\ref{tab:Reward structures} shows the different number of international and relegation ranks observed in the investigated leagues during the considered time period. Which reward structure has been applied in which season across the leagues is shown in Table~\ref{tab:Reward table}. In the Dutch league, the last qualification spot for the Europa League is determined in a play-off format including four teams. These ranking positions therefore are aggregated to one reward.

\begin{landscape}
    \begin{table}
    	\footnotesize
    	\centering 
    	\caption{Code for every reward structure in the data.} 
    	\label{tab:Reward structures} 
    	\def\arraystretch{1.2}
    	\begin{tabularx}{10cm}{Xcccc} 
    		\toprule
    		 & Champions  & Europa & Relegation & Direct\\ 
    		Code &  league & league & play-off & relegation\\ 
    		\midrule 
    		 4/3/DDD & 4 & 3 & 0 & 3\\
    		 4/3/PDD & 4 & 3 & 1 & 2\\
    		 3/3/DDD & 3 & 3 & 0 & 3\\
    		 3/3/DD & 3 & 3 & 0 & 2\\
    		 3/3/PDD & 3 & 3 & 1 & 2\\
    		 3/3/PD & 3 & 3 & 1 & 1\\
    		 2/4/DD & 2 & 4 & 0 & 2\\
    		 2/3/DDD & 2 & 3 & 0 & 3\\
    		 2/3/DD & 2 & 3 & 0 & 2\\
    		 2/3/PPD & 2 & 3 & 2 & 1\\
    		 2/2/DD & 2 & 2 & 0 & 2\\
    		 2/2/PPD & 2 & 2 & 2 & 1\\
    		\bottomrule
    	\end{tabularx}

    \end{table}

    \begin{table}
    	\footnotesize
    	\centering 
    	\caption{Implemented rewards structure for each league and season.} 
    	\label{tab:Reward table} 
    	\def\arraystretch{1.2}
    	\begin{tabularx}{23cm}{Xllllllllll} 
    		\toprule
    		League & 2009/10 & 2010/11 & 2011/12 & 2012/13 & 2013/14 & 2014/15 & 2015/16 & 2016/017 & 2017/18 & 2018/19\\ 
    		\midrule 
    		 Bundesliga & 3/3/PDD & 3/3/PDD & 4/3/PDD & 4/3/PDD & 4/3/PDD & 4/3/PDD & 4/3/PDD & 4/3/PDD & 4/3/PDD & 4/3/PDD \\
    		 Eredivisie & 2/3/PPD & 2/2/PPD & 2/3/PPD & 2/3/PPD & 2/3/PPD & 2/2/PPD & 2/2/PPD & 2/2/PPD & 2/2/PPD & 2/2/PPD \\
    		 La Liga & 4/3/DDD & 4/3/DDD & 4/3/DDD & 4/3/DDD & 4/3/DDD & 4/3/DDD & 4/3/DDD & 4/3/DDD & 4/3/DDD & 4/3/DDD \\
    		 Ligue 1 & 3/3/DDD & 3/3/DDD & 3/3/DDD & 3/3/DDD & 3/3/DDD & 3/3/DDD & 3/3/DDD & 3/3/PDD & 3/3/PDD & 3/3/PDD \\
    		 Premier League & 4/3/DDD & 4/3/DDD & 4/3/DDD & 4/3/DDD & 4/3/DDD & 4/3/DDD & 4/3/DDD & 4/3/DDD & 4/3/DDD & 4/3/DDD \\
    		 Primeira Liga & 2/3/DD & 2/4/DD & 3/3/DD & 3/3/DD & 3/3/PD & 3/3/DD & 3/3/DD & 3/3/DD & 2/3/DD & 2/3/DDD \\ 
    		 Serie A & 4/3/DDD & 4/3/DDD & 3/3/DDD & 3/3/DDD & 3/3/DDD & 3/3/DDD & 3/3/DDD & 3/3/DDD & 4/3/DDD & 4/3/DDD \\
    		 \bottomrule
    	\end{tabularx} 
    \caption*{\footnotesize \raggedright ~~~~ Note: Codes are explained in Table~\ref{tab:Reward structures}. The 2006/07 through 2008/09 seasons are excluded as they are only used for model selection.}
    \end{table}
\end{landscape}

\subsubsection{Individually different end-of-season rewards} \label{app:Indiv-End-of-season rewards}

Among the strengths of the proposed algorithm is that individually different and asymmetric reward schemes can be incorporated. 
In the application of European football leagues, differential reward schemes are not uncommon.
All national league associations in the investigated European football leagues award a starting place for the UEFA Europa League to the winners of national cup tournaments. In the case of two national cup competitions, one spot is reserved for each winner. If a cup winner would also qualify for an international tournament by finishing high enough in the league (or winning the second cup title), the allocation of European starting places is regulated by the league associations. Typically, the place reserved for the cup winner is transferred to an additional place determined by the league table. This has consequences for this particular team, as its reward scheme distributed via the league table has changed, as well as for all other teams such that a lower position in the league table is already sufficient for them to qualify for a European starting place. \par
We consider the possibility of this special case from the moment the final pairing of the cup tournament is fixed. In every simulation run, we evaluate at the end of the season, whether all teams that are still in the hunt for the cup win at the moment of the initial event meet the required configuration in the ranking. If this condition is met, the threshold in the rankings for the Europa League is lowered by one rank. Depending on how likely it is that the qualifying place is transferred to the league ranking based on the initial situation, the probability distribution of the final rewards is the weighted average of the two possibilities. If there are two cup tournaments in one country, this procedure is evaluated for both tournaments, taking into consideration that a team possibly plays in both finals.\par
Other reasons for individually different reward schemes for clubs are e.\,g.\ the exclusion of teams from international competitions, winning the UEFA Champions/Europa League, or legal issues of teams. All of them are taken into account as individually different reward schemes but are not addressed in detail.

\subsubsection{Details of the outcome model}

We estimate the outcomes of football matches following the approach of \citeA{Goller2018}, which has proven successful in the outcome prediction of football matches. Our initial data set contains 133 variables. We perform a LASSO-based model selection step on the initial set of variables using the subset of the 2006/07 to 2008/09 seasons across all leagues. The variable selection with a linear LASSO model uses the points won by the home team as the outcome variable, 10 folds cross-validation, and the optimal $\lambda$ at minimum MSE. The 2006/07 to 2008/09 seasons are excluded from the remainder of the analysis. After this data-driven variable selection, 26 ($\sim 20\%$) variables remain in the model - the full list of variables, as well as those selected, can be found in Appendix~\ref{app: List of covariates}. This model selection procedure is consistent with \citeA{Borup2020}, who find that for predictive models with many covariates, variable selection benefits prediction accuracy and usually 10-30\% of the original set of covariates turns out to be optimal.\par

The ORF predicts probabilities of ordered outcomes in a flexible way, building on Random Forests developed in \citeA{Breiman2001}. This machine learning method is tailored for the predictions of outcomes that appear in a natural order. The ORF model is estimated using the R-package \texttt{orf} with 1000 trees, without honesty option, minimal node size is 5, 5 randomly selected variables considered at each split and a sub-sampling rate of 2/3. The general framework is not restricted to this specific method and the choice of the underlying outcome model is of second-order (see Appendix~\ref{app:Alternative outcome model}). 
Other well-suited methods to predict match outcomes could be employed as well, like ordered logit, ordered probit, Poisson distribution-based models, or others. The exact score of the match is drawn from two independent Poisson distributions, for which the $\lambda$ parameters of the Poisson distributions of the scored goals are estimated on the full data set. The obtained values are 1.55 for the home team and 1.16 for the away team.

\subsection{Additional results}

\subsubsection{An illustrative example of EI estimates}\label{app:Illustrative example}
To illustrate the EI estimates, we show in Table~\ref{table:Ranking Bundesliga} the ranking of the 2017/18 Bundesliga 1 season before the last match day and in Table~\ref{table:Matchday Bundesliga} the fixtures of the final match day together with the estimated EI values. Several teams have already settled in their reward area, e.\,g.\ teams ranked $10^{\text{th}}$ to $14^{\text{th}}$, for which the EI estimates are zero. The teams ranked $4^{\text{th}}$ to $8^{\text{th}}$ are in strong competition for the qualification for the European club tournaments, while Freiburg, Wolfsburg, and Hamburg fight against relegation to Bundesliga 2. In consequence, those teams have a particularly high EI estimate. Currently ranked $16^{\text{th}}$ Wolfsburg has the highest EI estimate of all teams, as a win in their last match can lift them out of the relegation zone and losing could result in direct relegation. For Dortmund and M.gladbach, it is highly unlikely that they climb either up or down to a different reward, as this requires not only a particular outcome of their own match but also of other matches (including a sufficiently large shift in the goal difference on top) -- this is reflected in EI estimates very close to 0.\par

\begin{table}[ht]
    \caption{The final match day of the 2017/18 German Bundesliga 1 season\label{table:Final match day Bundesliga}}
    \parbox[t]{.4\linewidth}{
    \subcaption{Ranking prior to match day 34\label{table:Ranking Bundesliga}}
    \centering
    \begin{tabular}{r l c r c} 
     \hline\hline\\ [-1.5ex] 
     \rotatebox{90}{Rank} & \rotatebox{90}{Team} & \rotatebox{90}{Played} & \rotatebox{90}{GD} & \rotatebox{90}{Points} \\ [0.5ex] 
     \hline
     1 & Bayern M. & 33 & 67 & 84 \\
     \hline
     2 & Schalke 04 & 33 & 15 & 60 \\
     3 & Dortmund & 33 & 19 & 55 \\
     4 & Hoffenheim & 33 & 16 & 52 \\
     \hline
     5 & Leverkusen & 33 & 13 & 52 \\
     6 & RB Leipzig & 33& 0 & 50 \\
     \hline
     7 & Frankfurt & 33 & 1 & 49 \\
     8 & Stuttgart & 33 &  -3 & 48 \\
     9 & M.gladbach & 33 &-4 & 47 \\
     10 & Hertha BSC & 33 & 1 & 43 \\
     11 & Augsburg & 33 & -1 & 41 \\
     12 & Bremen & 33 & -4 & 39 \\
     13 & Hannover & 33 & -9 & 39 \\
     14 & Mainz 05 & 33 & -13 & 36\\
     15 & Freiburg & 33 & -26 & 33 \\
     \hline
     16 & Wolfsburg & 33 & -15 & 30 \\
     \hline
     17 & Hamburg & 33 & -25 & 28 \\
     18 & Köln & 33 & -32 & 22 \\ [0.5ex]
     \hline\hline
    \end{tabular}
    }\hfill
    \parbox[t]{.55\linewidth}{
    \centering
    \subcaption{Results of match day 34 and EI estimates.\label{table:Matchday Bundesliga}}
    \begin{tabular}{r@{}l l l} 
    \\
     \hline\hline\\ [-1.5ex] 
     & & \multicolumn{2}{c}{EI} \\\cmidrule(lr){3-4}
     \multicolumn{2}{c}{Match result} & \multicolumn{1}{c}{H} & \multicolumn{1}{c}{A}\\ [1ex] 
     \hline 
     Bayern M.\; 1:&4 \;Stuttgart & 0.00 & 0.18 \\ [0.5ex] 
     Hoffenheim\; 3:&1 \;Dortmund & 0.21 & 0.00 \\ [0.5ex] 
     Hertha BSC\; 2:&6 \;Leipzig & 0.00 & 0.10 \\ [0.5ex] 
     Freiburg\; 2:&0 \;Augsburg & 0.16 & 0.00 \\ [0.5ex] 
     Schalke 04\; 1:&0 \;Frankfurt & 0.00 & 0.11 \\ [0.5ex] 
     Leverkusen\; 3:&2 \;Hannover & 0.08 & 0.00 \\ [0.5ex] 
     Hamburg\; 2:&1 \;M.gladbach & 0.09 & 0.02 \\ [0.5ex] 
     Mainz 05\; 1:&2 \;Bremen & 0.00 & 0.00 \\ [0.5ex] 
     Wolfsburg\; 4:&1 \;Köln & 0.24 & 0.00 \\ [0.75ex]
     \hline\hline
    \end{tabular}
    }
    \caption*{\raggedright\footnotesize Note: In (a): GD refers to the difference between scored and conceded goals. Lines split the different reward areas in decreasing ordering (Championship, Champions League, Europa League, none, Relegation Playoffs, Relegation). In (b): EI provides the estimated event importance for the home (H) and away (A) team.}
\end{table}

The results of the fixtures in Table~\ref{table:Matchday Bundesliga} suggest, that incentives, as measured by the EI, might have an impact on the performance of the teams. Teams with a large EI estimate succeed over teams with a low or zero EI value, even though their opponents are much stronger on paper, i.\,e.\ Stuttgart beating the champions Bayern M. or Hamburg winning against M.gladbach. 

\subsubsection{Compare EI by iteration}\label{app:Compare EI by iteration}
\begin{figure}[ht]
    \centering
    \begin{subfigure}{.48\textwidth}
      \includegraphics[width=\textwidth]{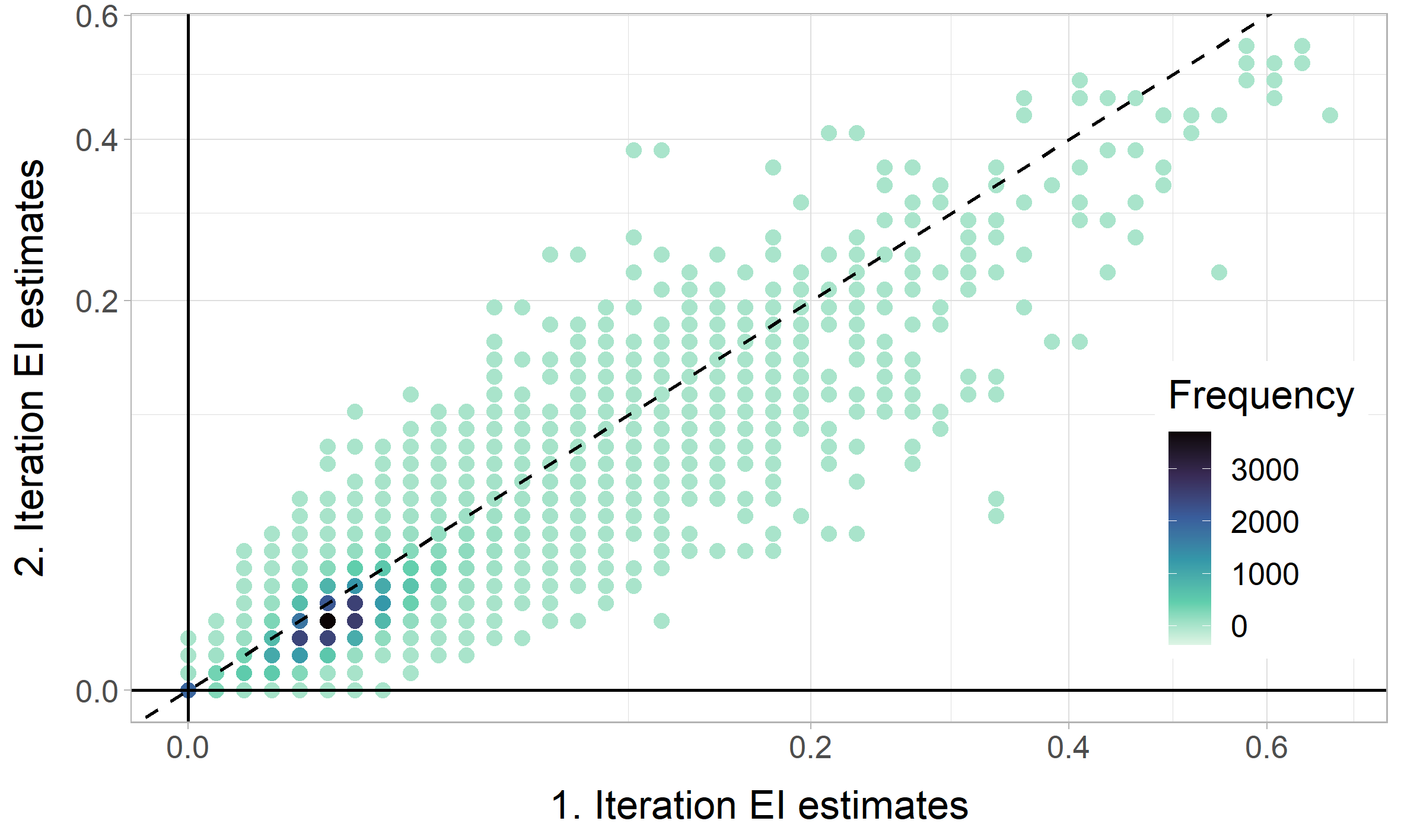}
      \caption{Iterations 1 and 2}
      \label{fig:gg_compare_iterations1}
    \end{subfigure}%
    \hfill
    \begin{subfigure}{.48\textwidth}
      \includegraphics[width=\textwidth]{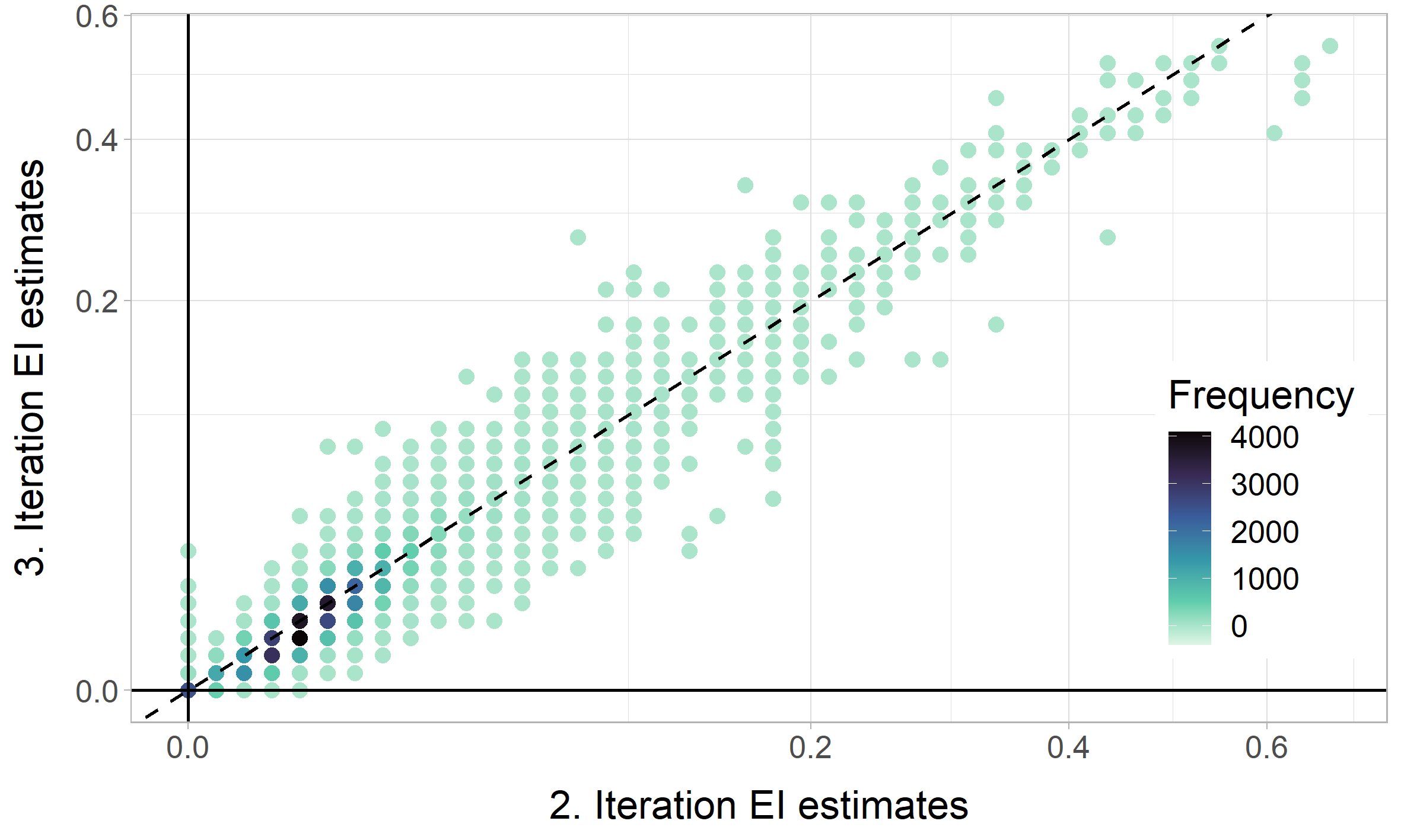}
      \caption{Iterations 2 and 3}
      \label{fig:gg_compare_iterations2}
    \end{subfigure}
    
    \begin{subfigure}{.48\textwidth}
      \includegraphics[width=\textwidth]{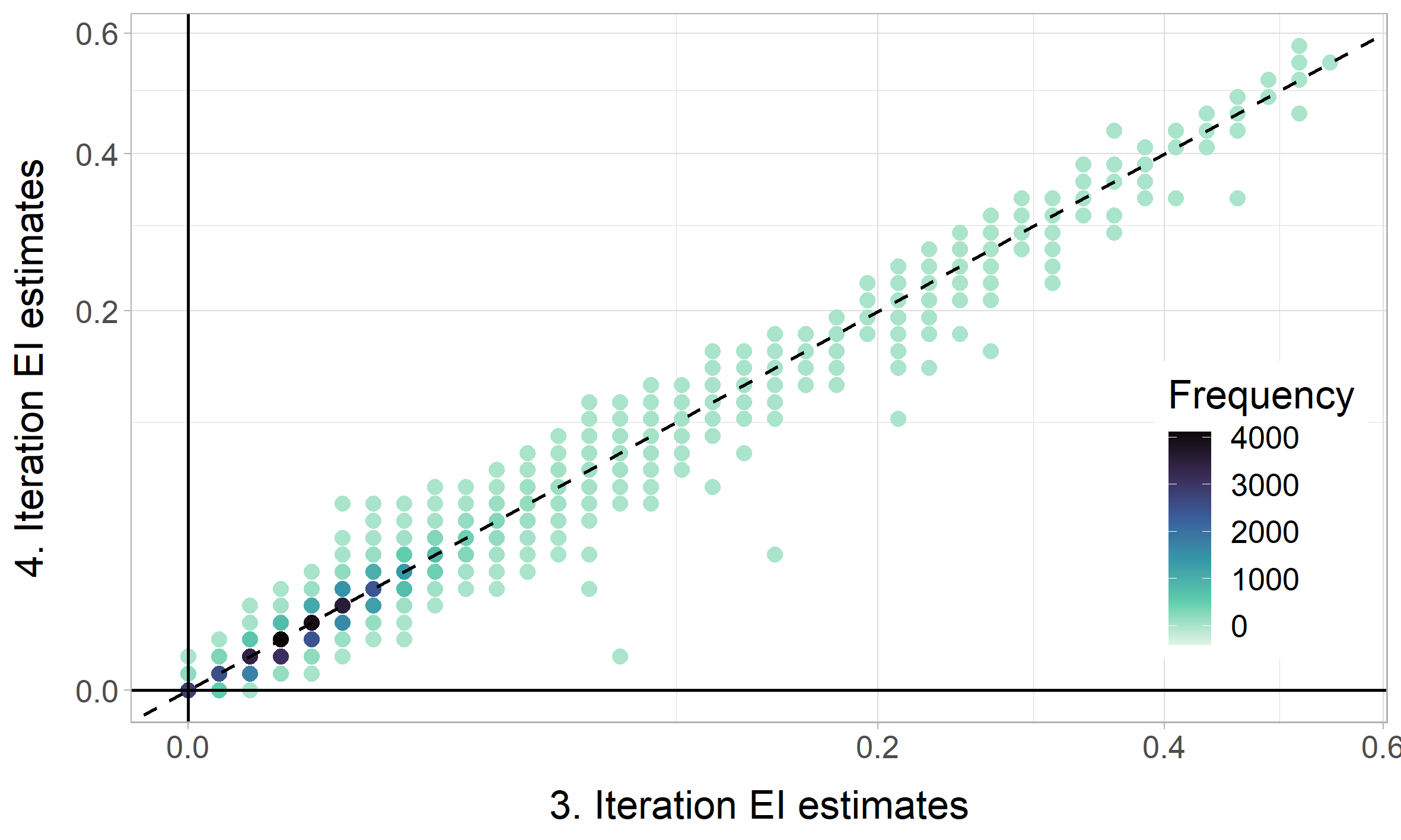}
      \caption{Iterations 3 and 4}
      \label{fig:gg_compare_iterations3}
    \end{subfigure}
    \caption{Comparison of EI estimates for different iterations}
    \label{fig:gg_compare_iterations}
\end{figure}

Figure~\ref{fig:gg_compare_iterations} shows the convergence of the estimated EI values over successive iterations. The second iteration of Algorithm~\ref{algo:specific event importance} incorporates the EI estimates of the first iteration into the outcome model which has a notable effect on the results. Integrating the refined estimates adds only a little information to the outcome model which is affirmed by the very high correlation of 0.96 for the estimated values of iterations 2 and 3. The fourth iteration increases the correlation to the previous estimates to 0.98, which cannot be substantially increased anymore.\par
Because of the random component in Monte Carlo simulation, two consecutive iterations will never return the exact same values. More iterations cannot alleviate this source of randomness and exhibit similar correlations. Convergence can be improved further by increasing the number of runs in the Monte Carlo simulation.

\FloatBarrier

\subsubsection{Variable importance in the outcome model}\label{app:Variable importance}
Table~\ref{tab:Variable importance} displays the permutation-based variable importance of the rich ORF model including the EI estimates as input variables. The EI variables are among the most important variables which supports the evidence of the relevance of the EI in the prediction of the match outcomes. If the difference between the home and away EI is also included in the model, this third variable also has considerable variable importance, which is, however, lower than that of the home or away EI.
\begin{table}[ht]
    \footnotesize
	\centering 
	\caption{Variable importance in the ORF model including the event importance estimates.}
	\label{tab:Variable importance} 
	\def\arraystretch{1.2}
	\begin{tabular}{lr} 
		\toprule
		 Variable & Importance\\ 
		\midrule 
		 Away EI & 0.052\\
		 Home average MV - Away average MV & 0.036\\
		 Home EI & 0.036\\
		 Home standardized MV - Away standardized MV & 0.033\\
		 Home total market value (MV) - Away total MV & 0.028\\
		 Home total market value (MV) of all players & 0.007\\
		 Away median MV & 0.005 \\
		 Home average points in all previous league matches & 0.005\\
		 Home plays Champions or Europa League & 0.002 \\
		 Home average age of Top-11 & 0.001 \\
		 All other variables & $\leq 0.001$ \\
		\bottomrule
	\end{tabular}
	\caption*{\raggedright \footnotesize Note: Permutation-based variable importance in rich ORF model (Base covariates + Home EI + Away EI). Only the top 10 variables are shown.}
\end{table}

\FloatBarrier

\subsubsection{Alternative outcome model}\label{app:Alternative outcome model}
\begin{figure}[ht]
    \centering
    \includegraphics[width=0.67\textwidth]{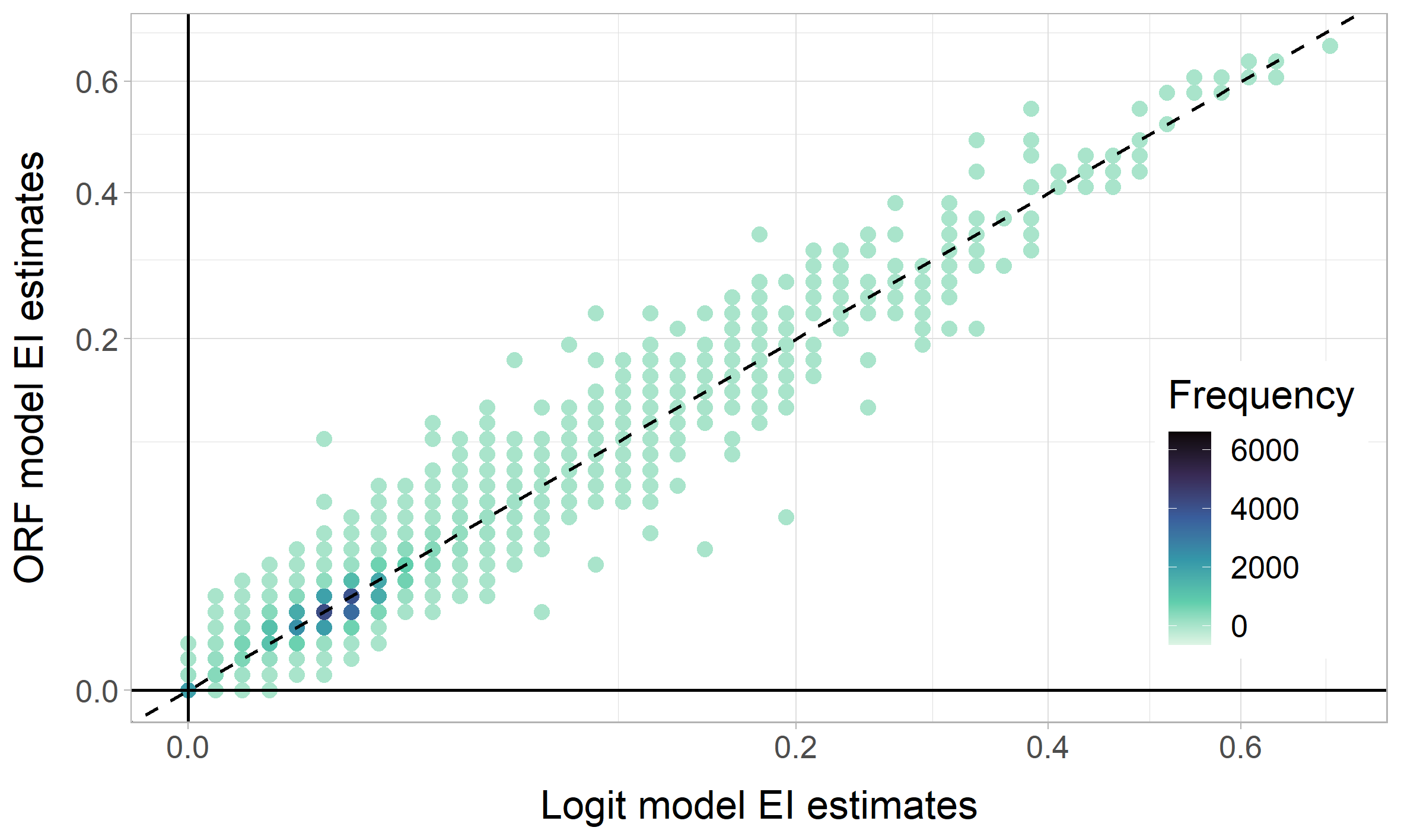}
    \caption{Event importance (EI) estimates for the ORF and logit outcome model. Values are rounded to the nearest grid point. Frequency indicates the number of points on a grid point. Square-root
transformation to x and y-axis applied.}
    \label{fig:gg_compare_orf_logit}
\end{figure}

As described in Section~\ref{sec:Sp_Ap_Fw} we use an ordered choice model with outcome probabilities estimated with an Ordered Forest (ORF). In Figure~\ref{fig:gg_compare_orf_logit} we compare the estimates between an ORF and an ordered logit outcome model. The estimated EI values are highly correlated (0.97) which suggests that the framework does not rely on a specific outcome model.  

\FloatBarrier

\subsubsection{Improvement on betting odds}\label{app:Improvement on betting odds}
\begin{figure}[ht]
    \centering
    \begin{subfigure}{.48\textwidth}
      \includegraphics[width=\textwidth]{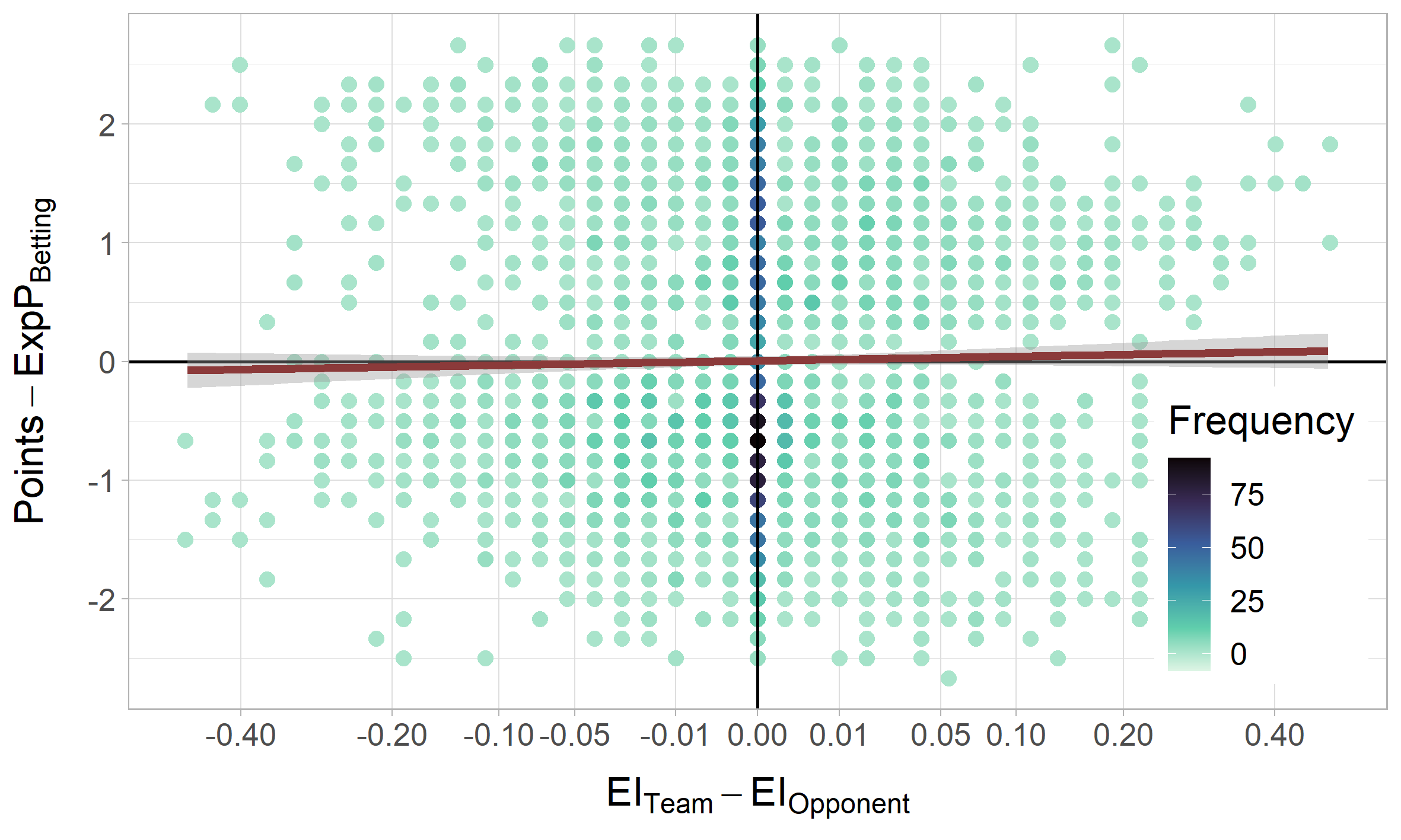}
      \caption{At least one contestant with EI=0}
      \label{fig:gg_EI_diff_gam_subsample1}
    \end{subfigure}%
    \hfill
    \begin{subfigure}{.48\textwidth}
      \includegraphics[width=\textwidth]{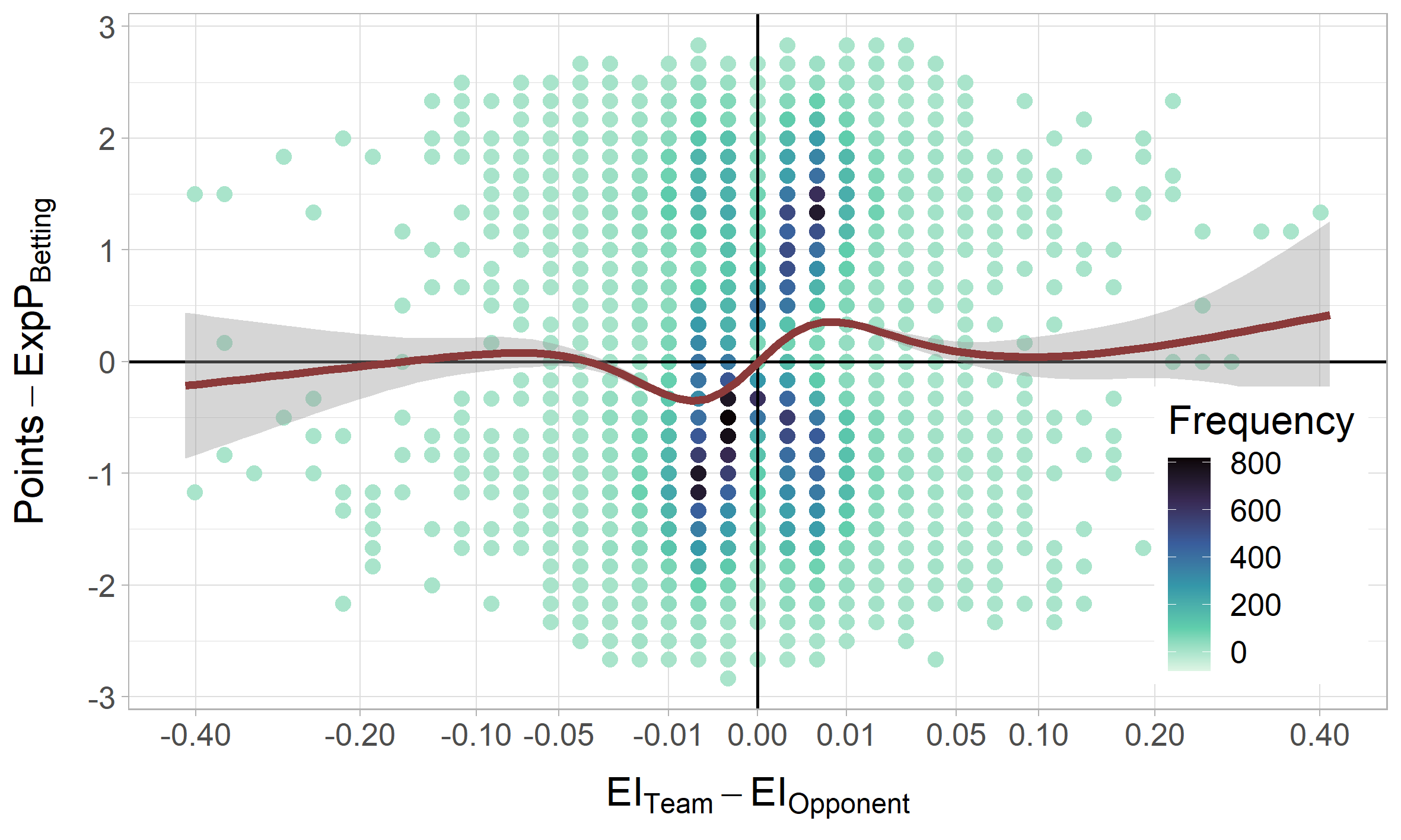}
      \caption{Both contestants with EI>0}
      \label{fig:gg_EI_diff_gam_subsample2}
    \end{subfigure}
        \caption{The difference in realised and expected points by the event importance for the team and its opponent, in samples with at least one contestant with null EI (a) and no contestant with null EI (b). Values are rounded to the nearest grid point. Frequency indicates the number of points on a grid point. Square-root transformation to x-axis applied.}
    \label{fig:gg_EI_diff_gam_subsample}
\end{figure}

In Section~\ref{sec:Prediction power improvement} we provide evidence that bookmakers are successful in integrating imbalances in incentives induced by large differences in the EI measure but fail to reflect the subtle differences in the EI values in the betting odds. To support this hypothesis we split the data into cases where at least one of the teams exhibits a zero EI (Figure~\ref{fig:gg_EI_diff_gam_subsample1}) and all other matches where both competing teams have a positive estimated EI (Figure~\ref{fig:gg_EI_diff_gam_subsample2}). The easily grasped case -- a team is settled in its final reward area -- is well accounted for in the betting odds as no difference between the expected and realised points can be explained by the EI measure. In matches between teams that both have a positive EI, their EI difference has still valuable information content which is not entirely integrated into the betting odds. \par

\begin{figure}[ht]
    \centering
    \includegraphics[width=0.67\textwidth]{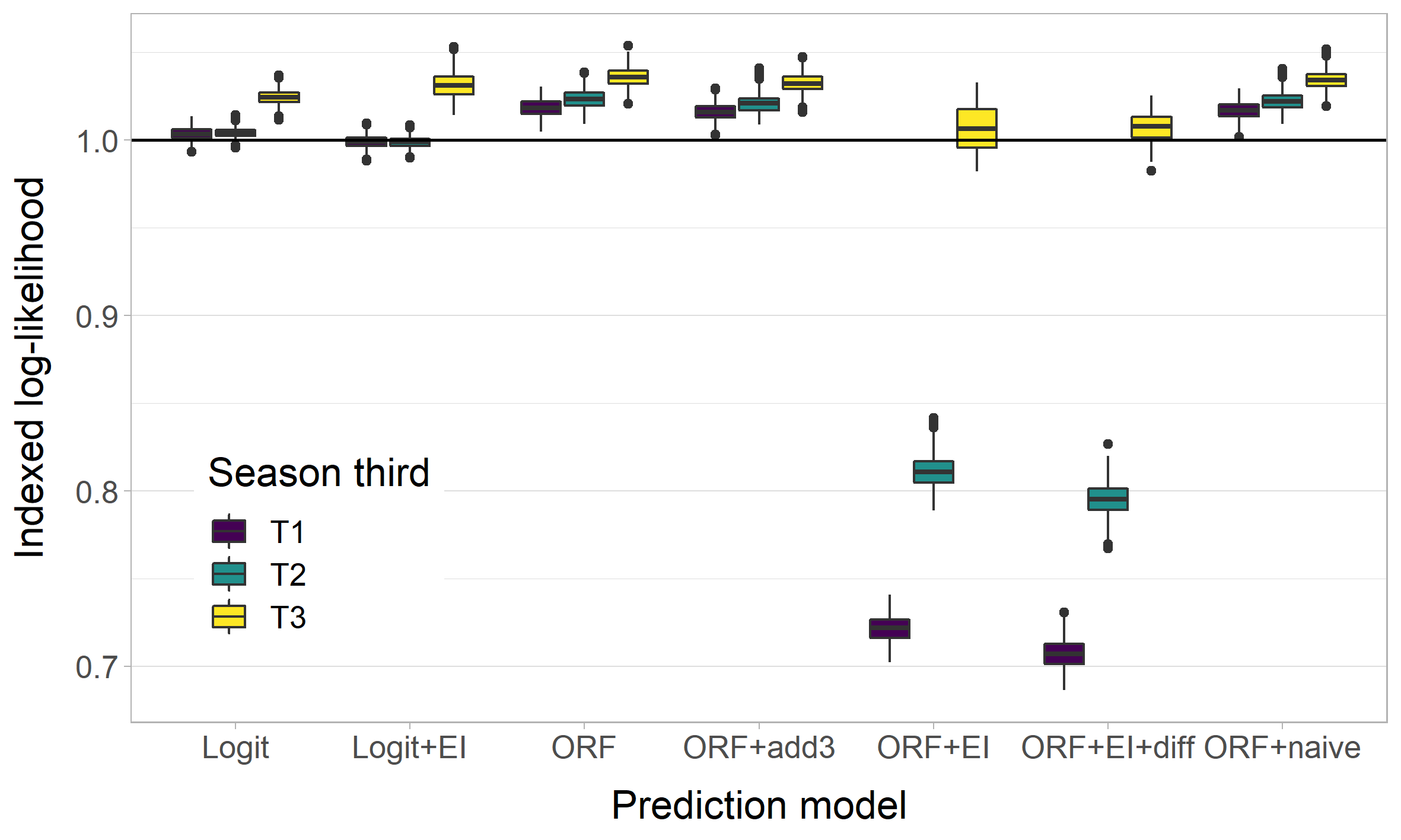}
    \caption{Indexed log-likelihood of out-of-sample predictions of different models split in the first (T1), second (T2), and last (T3) third of the season over 1000 replications.}
    \label{fig:logl_boxplot_thirds}
\end{figure}

To complete this analysis, we split the prediction sample into three groups indicating whether the matches are played in the first, second, or third chronological third of a season in Figure~\ref{fig:logl_boxplot_thirds}. This split points out, that the improvement relative to the betting odds arises from the first and second third of the season where differences in the EI are still very small and hence are not reflected in the betting odds.\footnote{In general, the matches in the last tier of the season are less accurately predicted by all models. This can be explained by the decreasing capacity of the available covariates to comprehensively characterise the teams with the duration of the season and by the fewer available matches with highly disparate EI values in the training data, which are frequent in the last third of the season.}

\FloatBarrier

\subsubsection{Alternative predictive power measure}\label{app:alternative measure}
\begin{figure}[ht]
    \centering
    \includegraphics[width=0.67\textwidth]{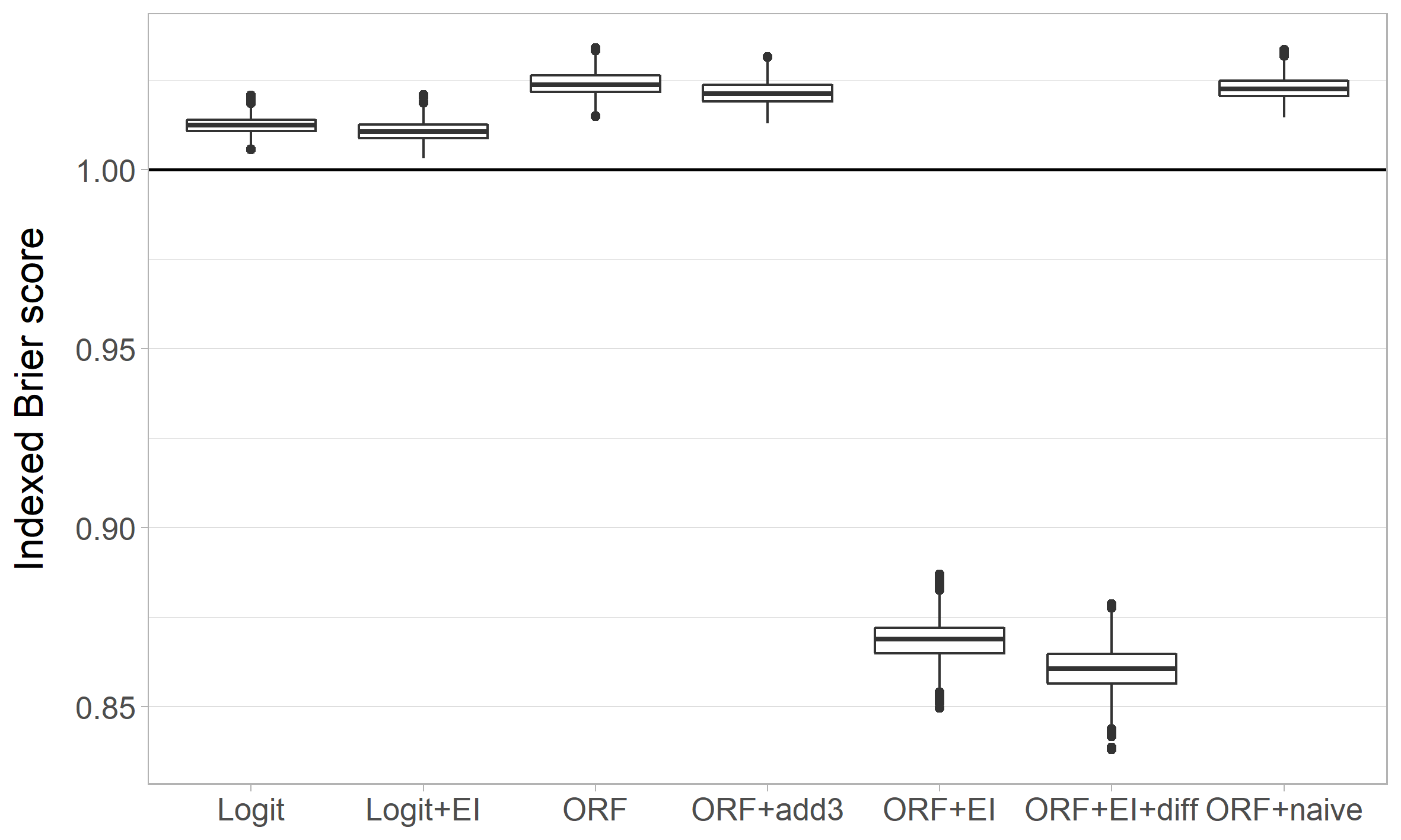}
    \caption{Brier score out-of-sample prediction accuracy of different models, indexed by the performance of betting odds.}
    \label{fig:brier_boxplot}
\end{figure}

\begin{figure}[ht]
    \centering
    \includegraphics[width=0.67\textwidth]{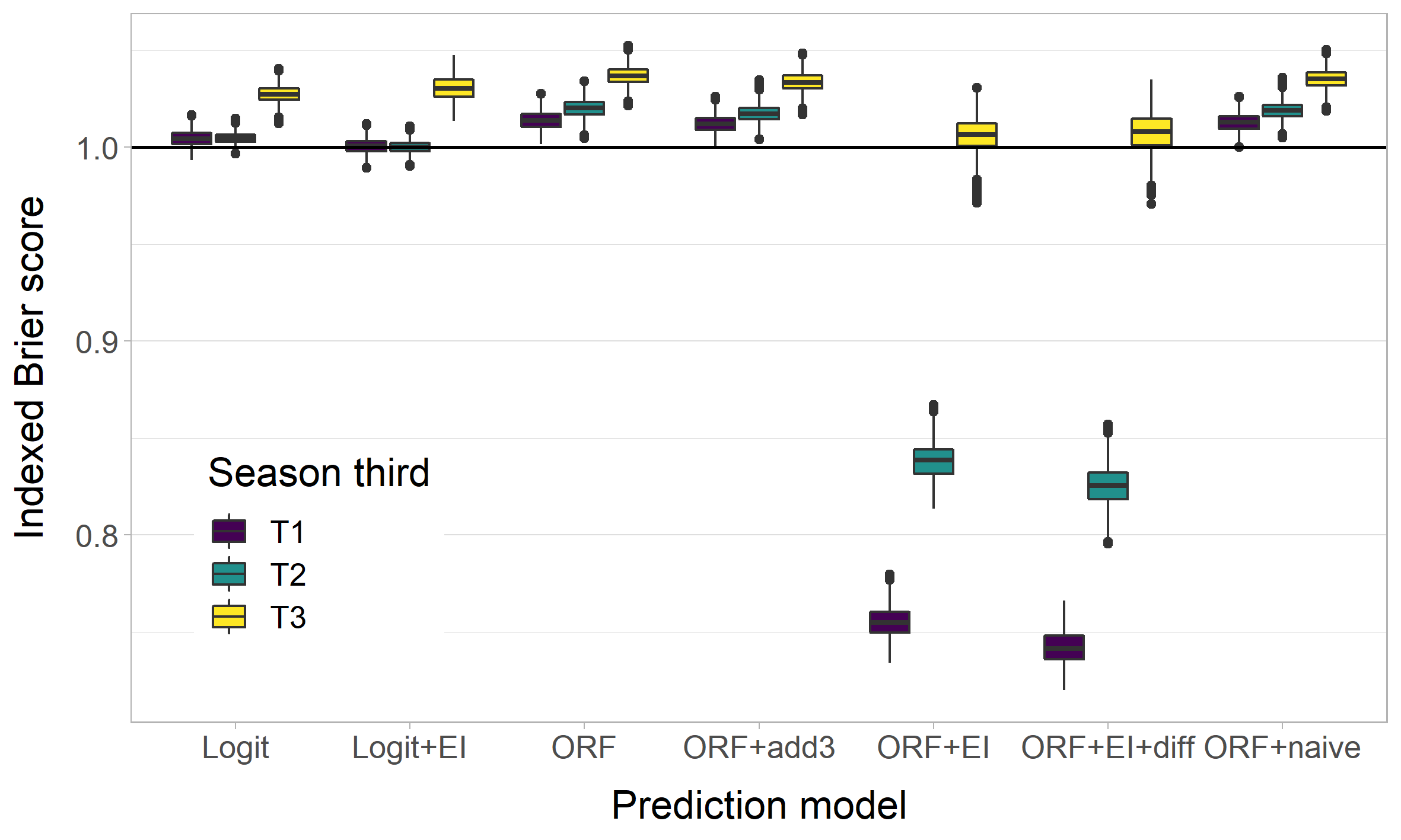}
    \caption{Brier score of the out-of-sample prediction accuracy of different models split in the first (T1), second (T2), and last (T3) third of the season}
    \label{fig:brier_boxplot_thirds}
\end{figure}

To highlight the independence of the out-of-sample prediction power analysis on the measure, we provide the results of the identical analysis procedure measuring the prediction accuracy with the Brier score. Figures~\ref{fig:brier_boxplot} \&~\ref{fig:brier_boxplot_thirds} are the equivalent plots to Figures~\ref{fig:logl_boxplot} \&~\ref{fig:logl_boxplot_thirds} using the Brier score instead of the log-likelihood and exhibit the equivalent patterns independent of the measure.

\FloatBarrier

\subsubsection{Team performance: Additional outcomes}\label{app:Team performance}
\begin{figure}[ht]
    \begin{subfigure}{.48\textwidth}
      \centering
      \includegraphics[width=\textwidth]{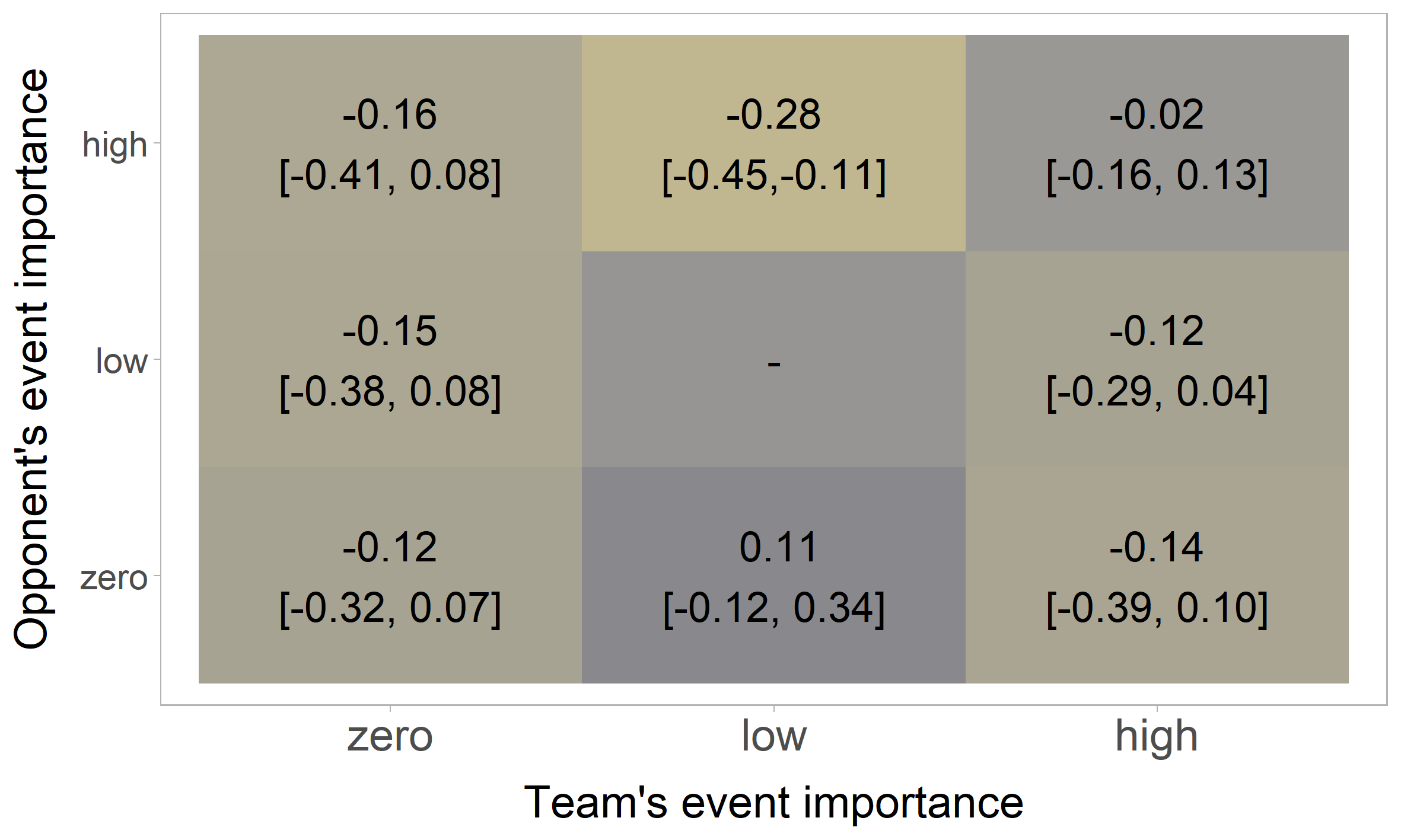}
      \caption{Tackles}
      \label{fig:gg_lm_est_tackle_pmin}
    \end{subfigure}%
    \hfill
    \begin{subfigure}{.48\textwidth}
      \centering
      \includegraphics[width=\textwidth]{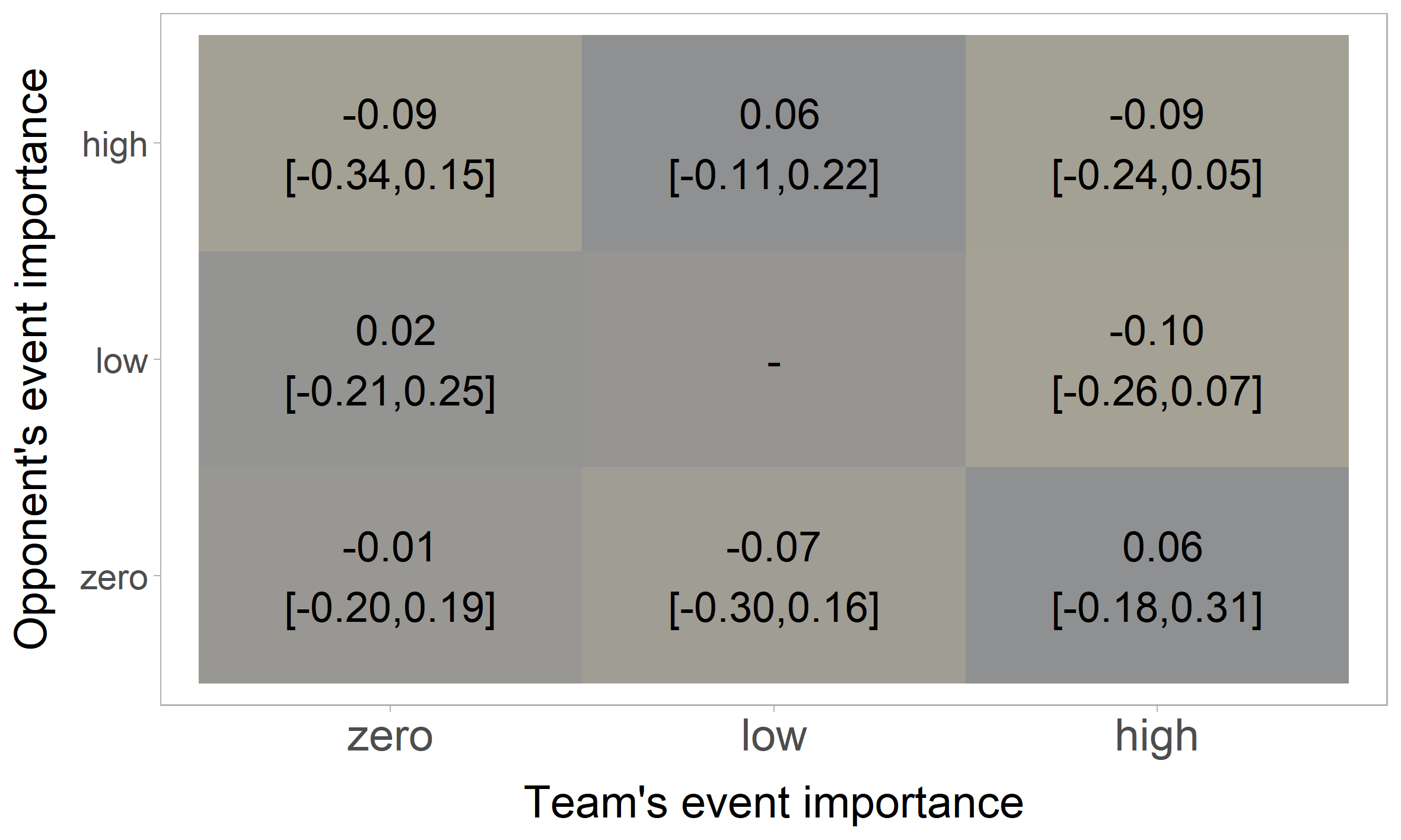}
      \caption{Tackles win share}
      \label{fig:gg_lm_est_tackle_win_share}
    \end{subfigure}
\end{figure}  
\begin{figure}\ContinuedFloat
    \begin{subfigure}{.48\textwidth}
      \centering
      \includegraphics[width=\textwidth]{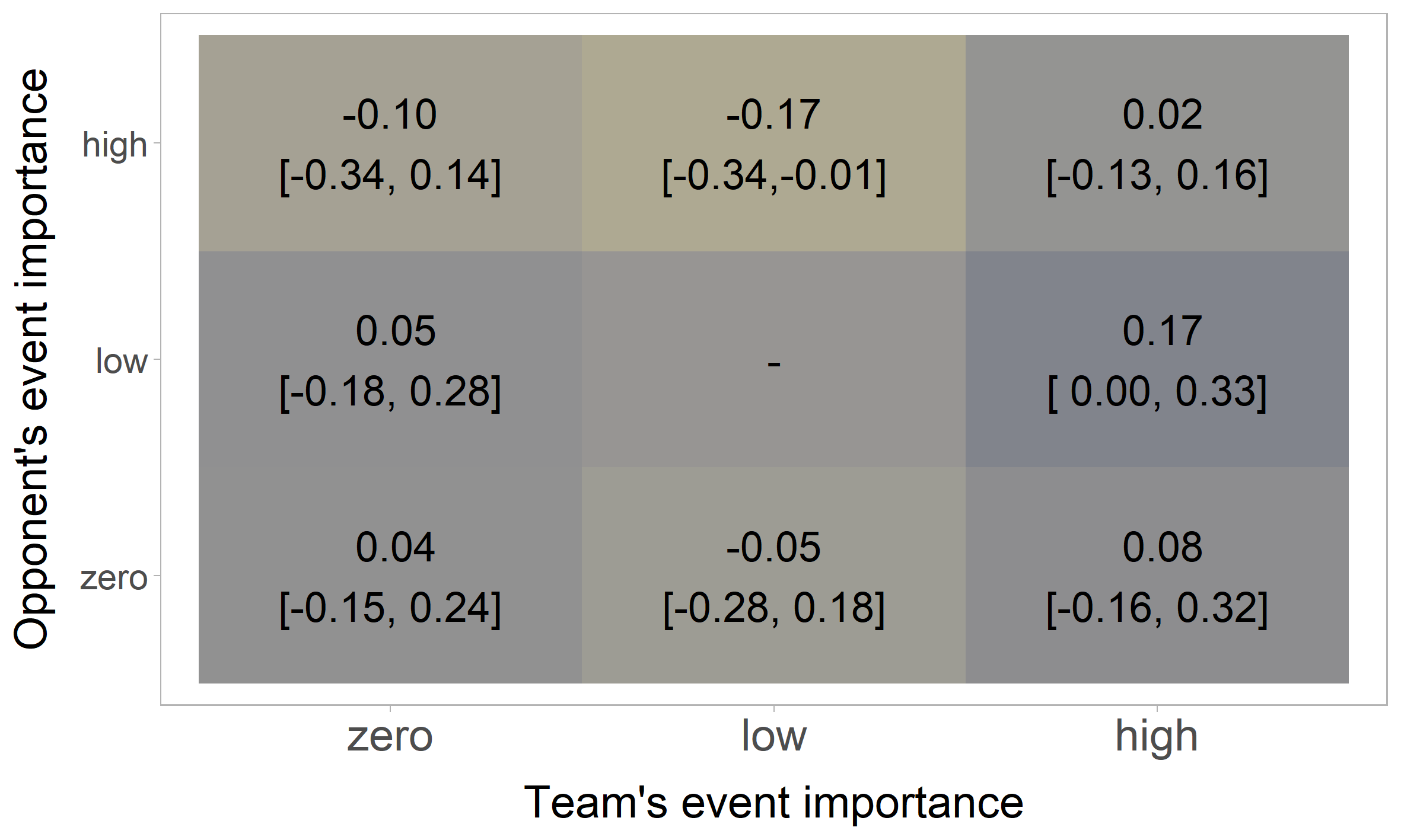}
      \caption{Duels win share}
      \label{fig:gg_lm_est_duel_win_share}
    \end{subfigure}%
    \hfill
    \begin{subfigure}{.48\textwidth}
      \centering
      \includegraphics[width=\textwidth]{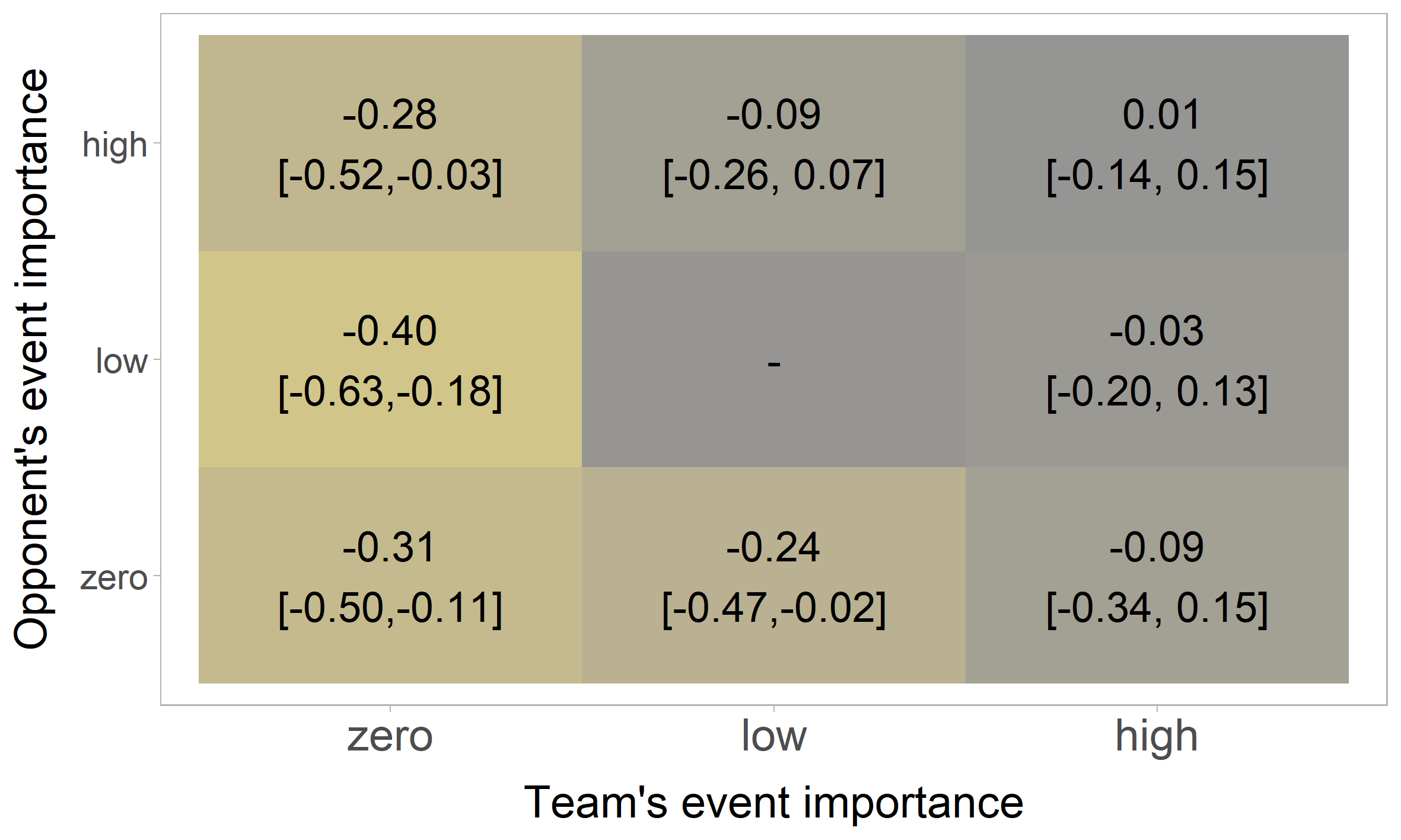}
      \caption{Fouls}
      \label{fig:gg_lm_est_fouls_pmin}
    \end{subfigure}
\end{figure}
\begin{figure}\ContinuedFloat
    \begin{subfigure}{.48\textwidth}
      \centering
      \includegraphics[width=\textwidth]{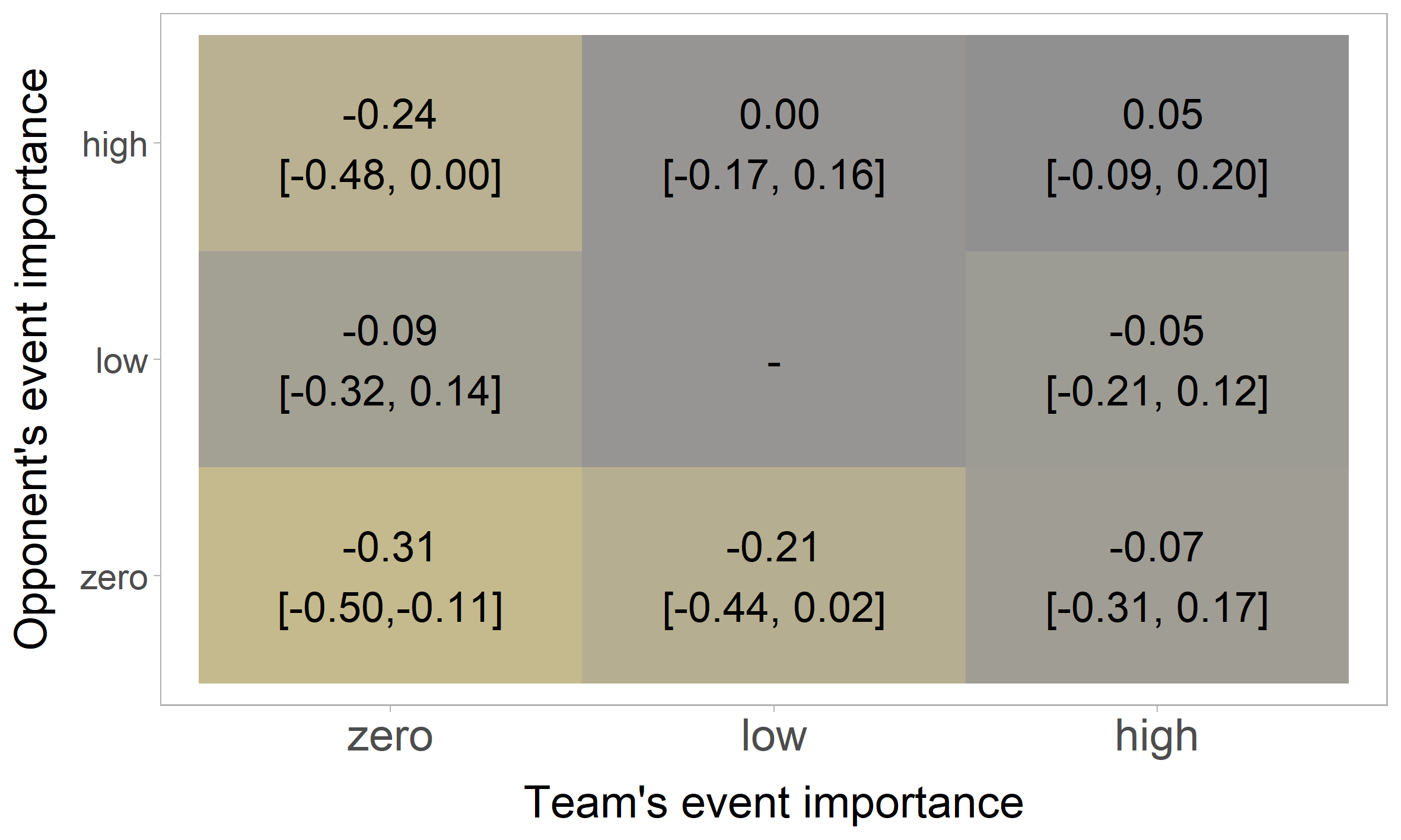}
      \caption{Yellow cards}
      \label{fig:gg_lm_est_yellow_card_pmin}
    \end{subfigure}%
    \hfill
    \begin{subfigure}{.48\textwidth}
      \centering
      \includegraphics[width=\textwidth]{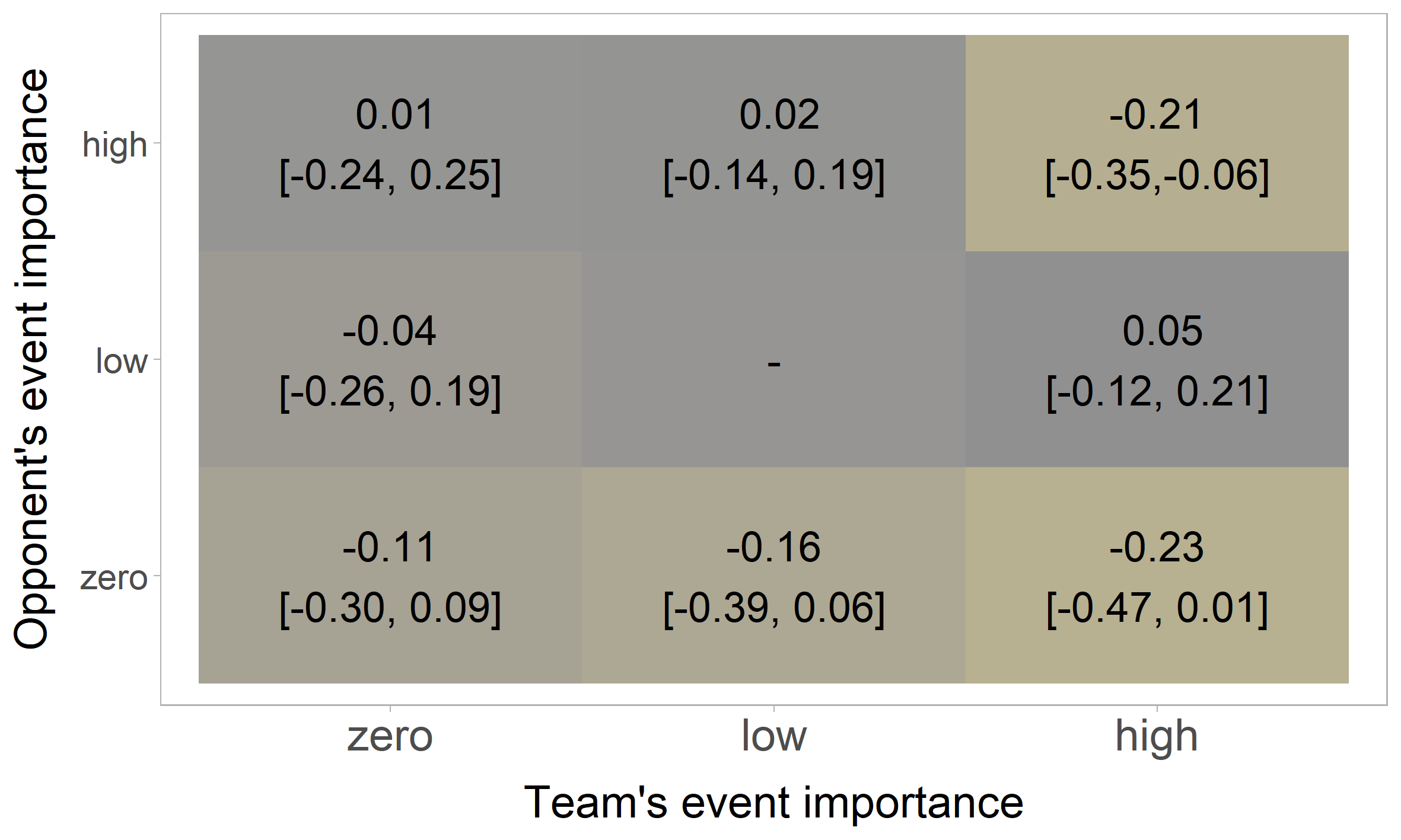}
      \caption{Sent off}
      \label{fig:gg_lm_est_sent_off_pmin}
    \end{subfigure}
\end{figure}
\begin{figure}\ContinuedFloat
    \begin{subfigure}{.48\textwidth}
      \centering
      \includegraphics[width=\textwidth]{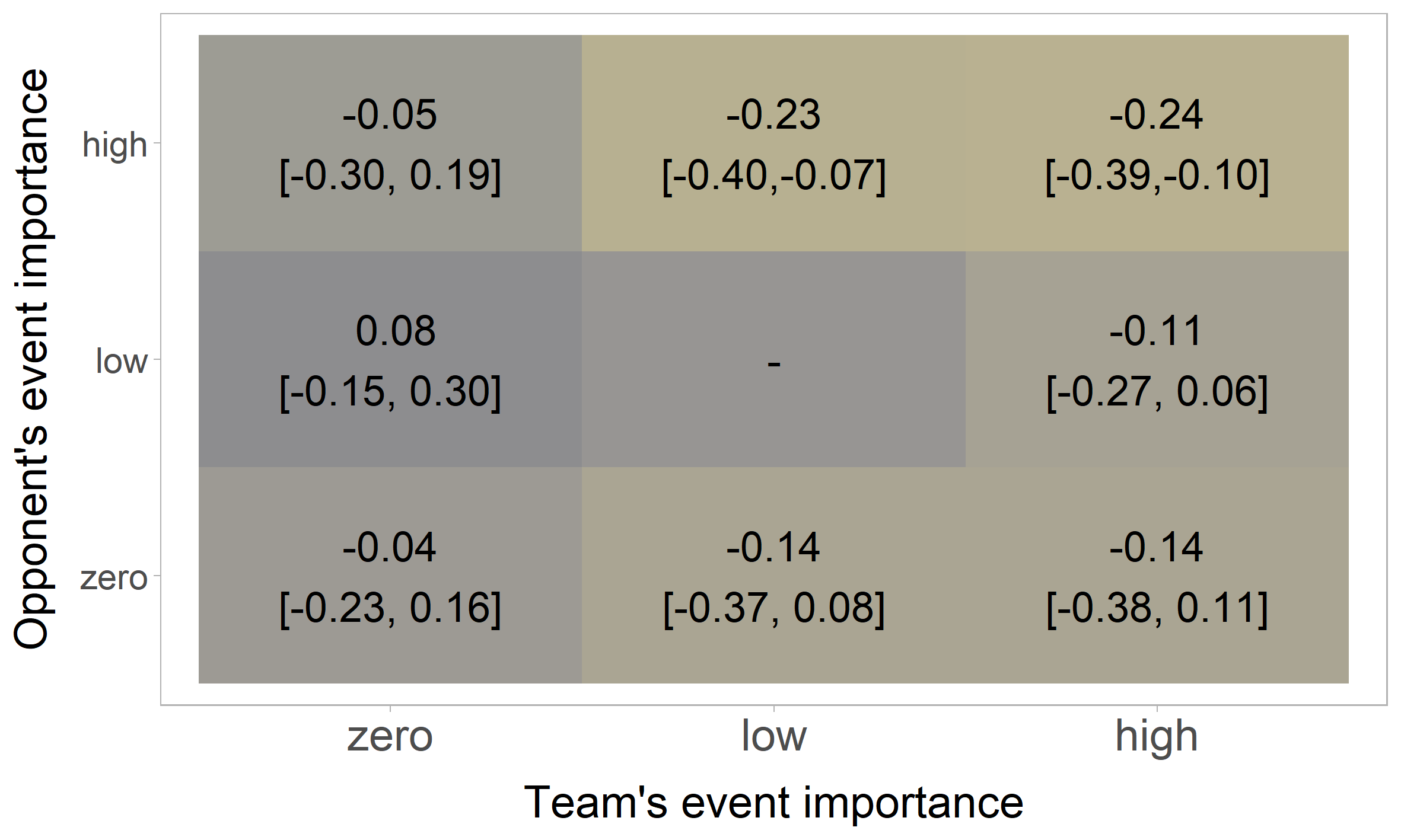}
      \caption{Touches}
      \label{fig:gg_lm_est_touches_pmin}
    \end{subfigure}
    \hfill
    \begin{subfigure}{.48\textwidth}
      \centering
      \includegraphics[width=\textwidth]{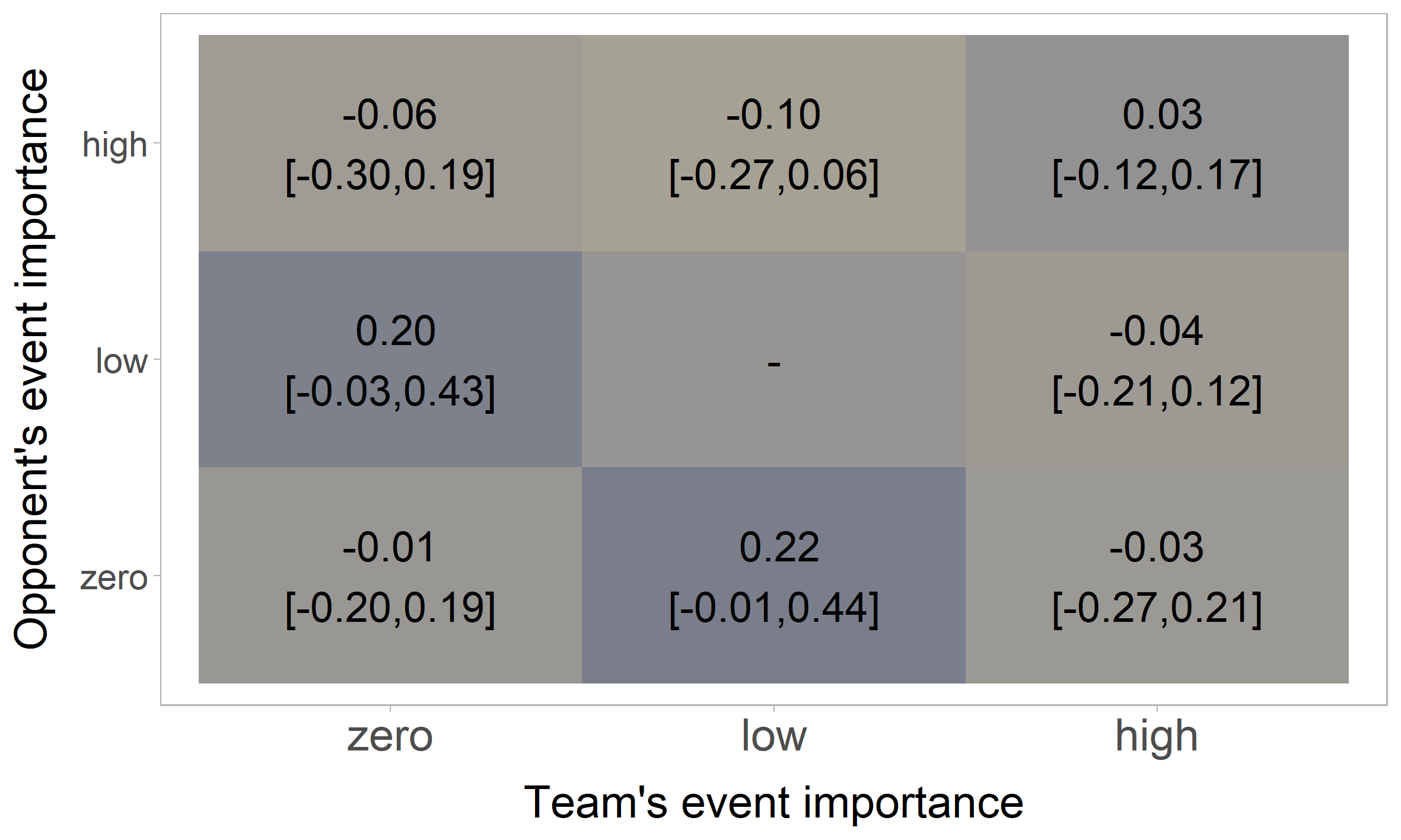}
      \caption{Dribbling attempts}
      \label{fig:gg_lm_est_contest_pmin}
    \end{subfigure}%
\end{figure}
\begin{figure}\ContinuedFloat
    \begin{subfigure}{.48\textwidth}
      \centering
      \includegraphics[width=\textwidth]{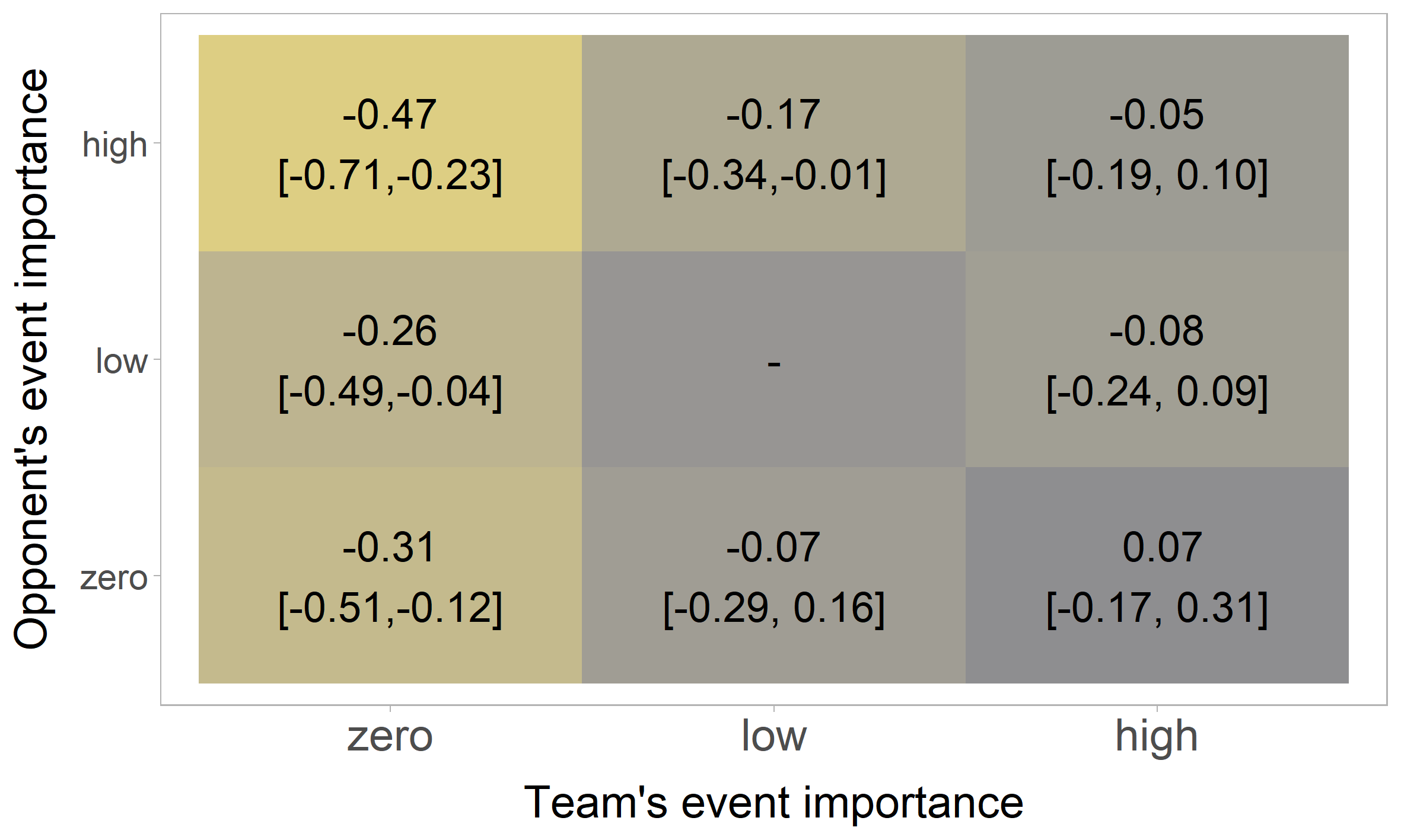}
      \caption{Final third entries}
      \label{fig:gg_lm_est_final_third_entries_pmin}
    \end{subfigure}
    \hfill
    \begin{subfigure}{.48\textwidth}
      \centering
      \includegraphics[width=\textwidth]{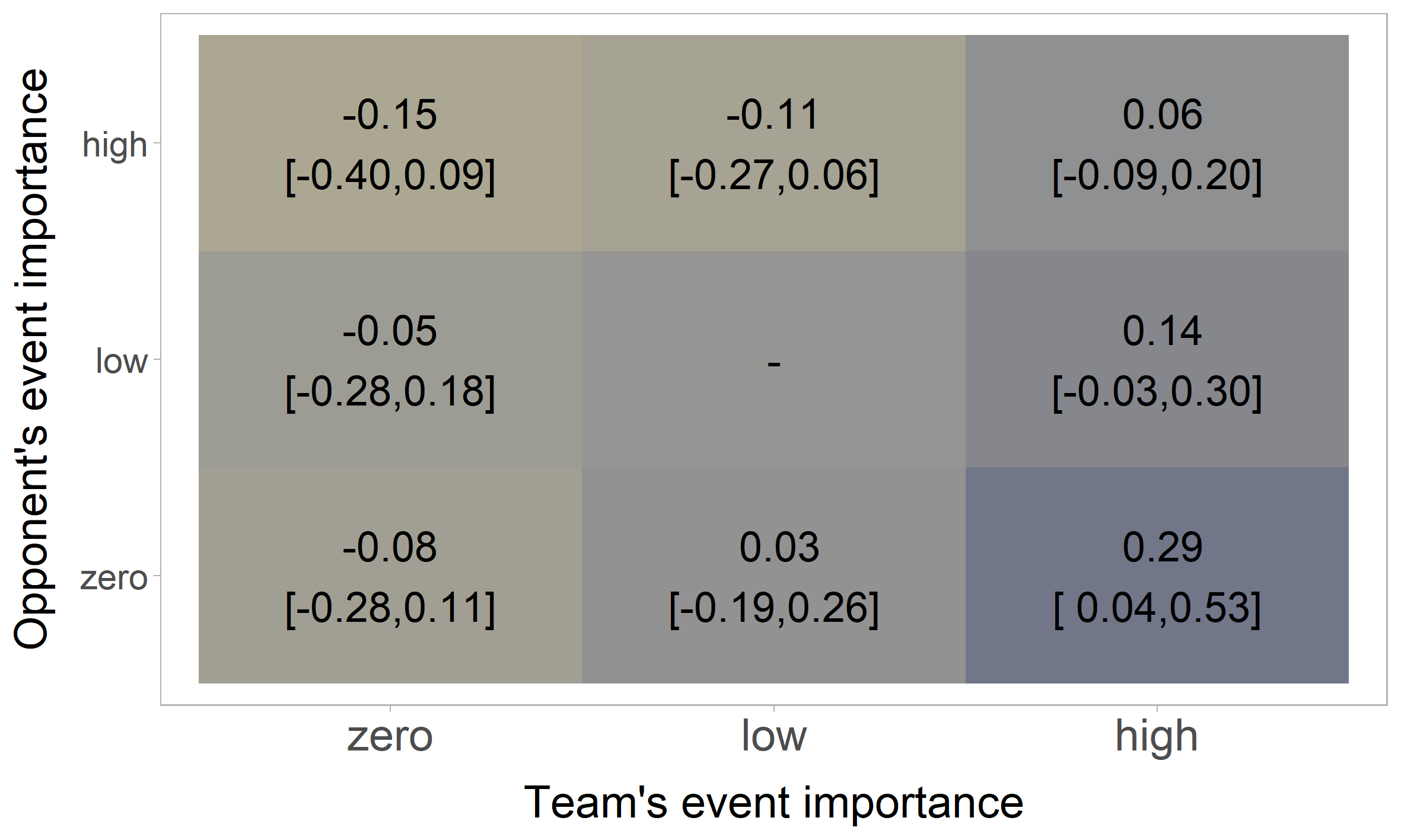}
      \caption{Penalty area entries}
      \label{fig:gg_lm_est_pen_area_entries_pmin}
    \end{subfigure}
\end{figure}
\begin{figure}\ContinuedFloat
    \begin{subfigure}{.48\textwidth}
      \centering
      \includegraphics[width=\textwidth]{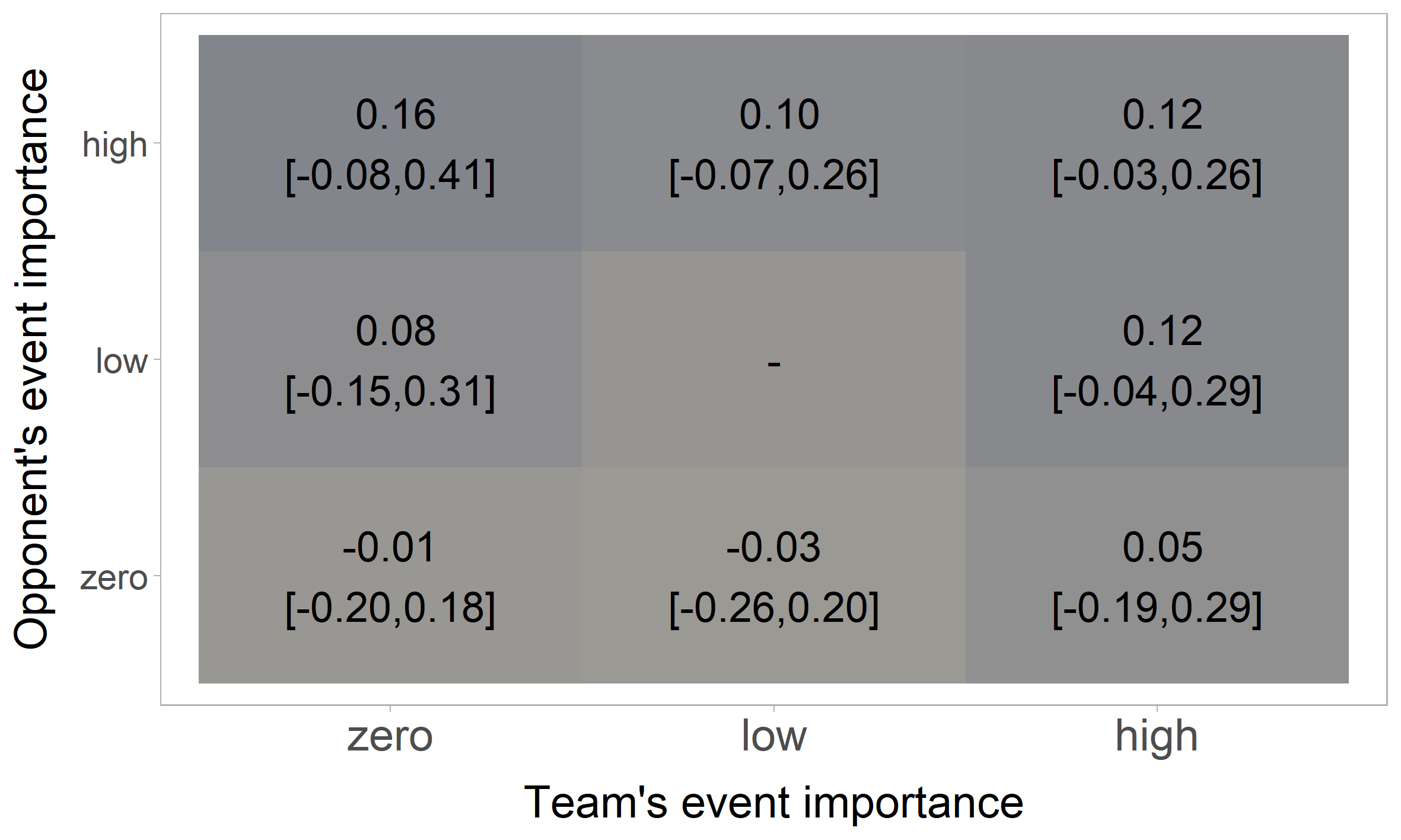}
      \caption{Errors leading to shot}
      \label{fig:gg_lm_est_error_lead_to_shot_pmin}
    \end{subfigure}
    \hfill
    \begin{subfigure}{.48\textwidth}
      \centering
      \includegraphics[width=\textwidth]{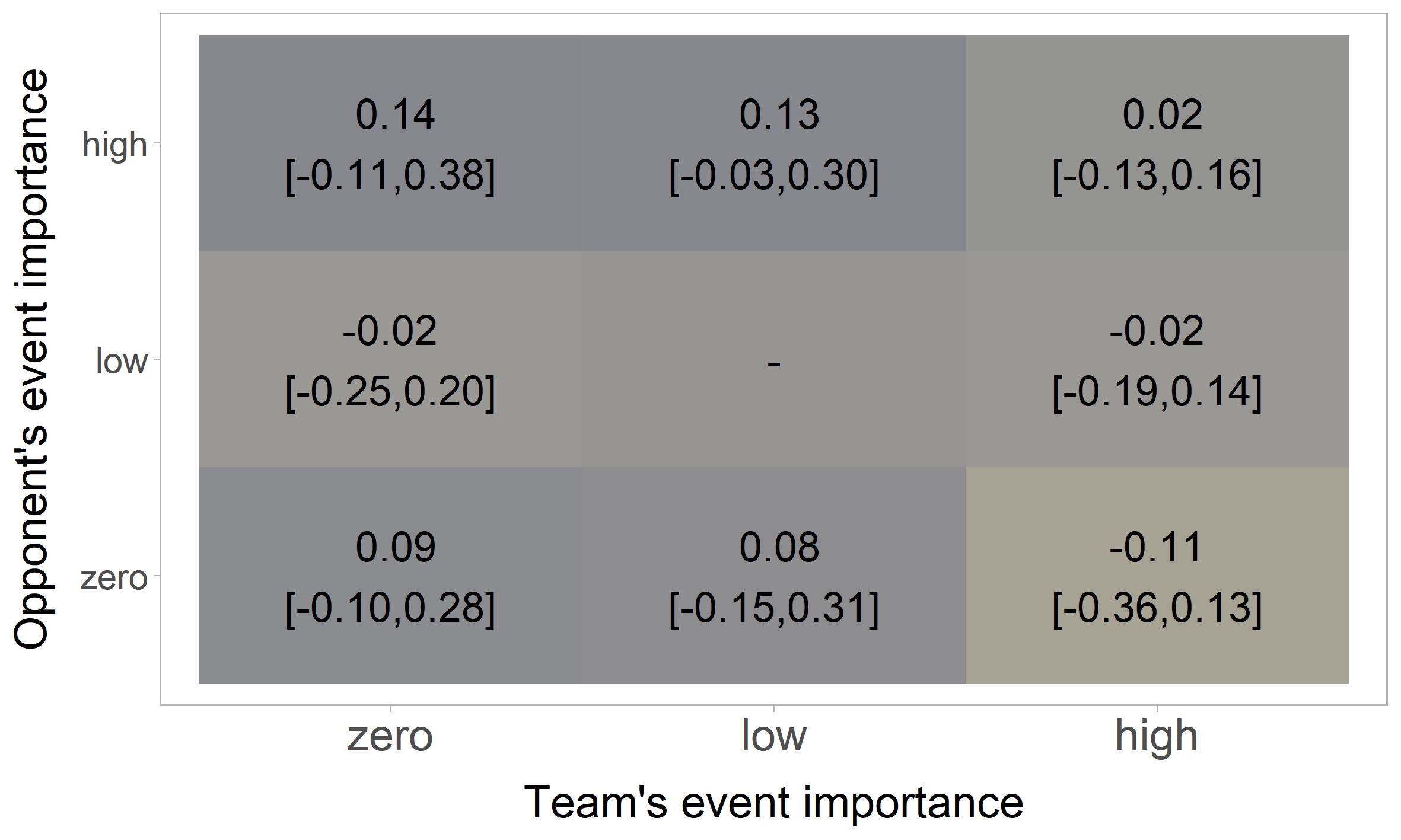}
      \caption{Errors leading to goal}
      \label{fig:gg_lm_est_error_lead_to_goal_pmin}
    \end{subfigure}%
    \caption{Estimates of the FE regression residuals of different outcomes on the event importance categories. 95\% confidence intervals in parentheses. The baseline is the low by low category.}
    \label{fig:gg_lm_est_opta_app}
\end{figure}
\end{document}